\documentclass[11pt]{article}
\usepackage{jheppub}
\usepackage{lmodern}
\usepackage[T1]{fontenc}   
\usepackage[utf8]{inputenc}

\usepackage{graphicx, epsfig} %include figure files
\usepackage{caption}
\usepackage{subcaption}
\usepackage{ragged2e}
\usepackage{amsmath,amssymb,amsfonts,dsfont,mathrsfs,amsthm,mathtools}
\usepackage{bm} %include bold math: \bm{} creates bold letters in math mode
\usepackage{color}
\usepackage[usenames]{xcolor}
\usepackage{hyperref}
\usepackage{siunitx}
\hypersetup{colorlinks=true,urlcolor=blue,linkcolor=magenta,citecolor=blue,filecolor=blue}
\usepackage[normalem]{ulem}
\usepackage{array}
\usepackage{hyperref}
\usepackage{booktabs}
\usepackage{blindtext}
\usepackage{upgreek}
\usepackage{subfloat}
\usepackage{braket}
\usepackage[normalem]{ulem}

\newcommand{\stkout}[1]{\ifmmode\text{\sout{\ensuremath{#1}}}\else\sout{#1}\fi}

\DeclareMathOperator{\sech}{sech}

\newcommand{\diff}[1]{\text{d}#1}

\newcommand{\GN}{G_\text{N}}

\newcommand{\Ls}{L_\star}
\newcommand{\RT}{{\Sigma_{A}}}

\newcommand{\ext}{{\Sigma_{\text{ext}}}}

\newcommand{\IEren}{I_{\text E}^{\text{ren}}}

\newcommand{\Aren}{\mathbf A_\text{ren}}

\newcommand{\tE}{t_\text{E}}
\newcommand{\iu}{\text i}
\newcommand{\tauE}{\tau_\text{E}}
\newcommand{\dtE}{\dot t_{\text E}}
\newcommand{\Tau}{\mathrm{T}}

\newcommand{\TauEl}{\mathrm{T}_{\text E,\ell}}

\makeatletter
\newcommand{\dal}{\mathop{\mathpalette\dal@\relax}}
\newcommand{\dal@}[2]{%
  \begingroup
  \sbox\z@{$\m@th#1\square$}%
  \dimen0=\fontdimen8
    \ifx#1\displaystyle\textfont\else
    \ifx#1\textstyle\textfont\else
    \ifx#1\scriptstyle\scriptfont\else
    \scriptscriptfont\fi\fi\fi3
  \makebox[\wd\z@]{%
    \hbox to \ht\z@{%
      \vrule width \dimen0
      \kern-\dimen0
      \vbox to \ht\z@{
        \hrule height \dimen0 width \ht\z@
        \vss
        \hrule height 2\dimen0
      }%
      \kern-2.5\dimen0
      \vrule width 2.5\dimen0
    }%
  }%
  \endgroup
}
\makeatother

\begin{document}
%%%%%%%
\noindent\flushright
{YITP-26-18} 
\noindent\flushleft\justifying
\vspace{-2cm}

\title{\huge Renormalized pseudoentropy in dS/CFT}
%\title{\huge Holographic pseudoentropy and renormalisation in dS/CFT}

%%%%%%%
\author[a]{Giorgos Anastasiou,}
\author[b]{Ignacio J. Araya,}
\author[c, d]{Avijit Das,}
\author[e,f]{Javier Moreno}

\affiliation[a]{Departamento de Ciencias, Facultad de Artes Liberales, Universidad Adolfo Iba\~nez,\\ Av. Diagonal Las Torres, 2640 Pe\~nalolen, Chile \vspace{0.1cm}}
\affiliation[b]{Departamento de F\'isica y Astronom\'ia, Facultad de Ciencias Exactas, Universidad Andres Bello,\\ Sazi\'e 2212, Piso 7, Santiago, Chile \vspace{0.1cm}}
\affiliation[c]{Instituto de Ciencias Exactas y Naturales, Universidad Arturo Prat,\\
Playa Brava 3256, 1111346, Iquique, Chile \vspace{0.1cm}}
\affiliation[d]{Facultad de Ciencias, Universidad Arturo Prat,\\
Avenida Arturo Prat Chacón 2120, 1110939, Iquique, Chile \vspace{0.1cm}}
\affiliation[e]{Center for Gravitational Physics and Quantum Information,\\
Yukawa Institute for Theoretical Physics, Kyoto University,\\
Kitashirakawa Oiwakecho, Sakyo-ku, Kyoto 606-8502, Japan
\vspace{0.1cm}
}
\affiliation[f]{Departament de F\'isica Qu\'antica i Astrof\'isica, Institut de Ci\'encies del Cosmos,\\
Universitat de Barcelona, Mart\'i i Franqu\'es 1, E-08028 Barcelona, Spain
}

\vspace{0.3cm}
\emailAdd{georgios.anastasiou@uai.cl} 
\emailAdd{ignacio.araya@unab.cl}
\emailAdd{adas@estudiantesunap.cl}
\emailAdd{javier.moreno@yukawa.kyoto-u.ac.jp}

\abstract{
We study holographic pseudoentropy for subregions in non-unitary Euclidean conformal field theories (CFTs) within the framework of the de Sitter/conformal field theory (dS/CFT) correspondence. Pseudoentropy, defined as the von Neumann entropy of a transition matrix, is computed holographically from codimension-two extremal surfaces in dS space and is divergent due to the asymptotic bulk volume at future infinity. We show that a finite and regulator-independent definition follows from the on-shell action of conformal gravity in four and six dimensions, implemented through the replica construction.
We illustrate the formalism for spherical entangling surfaces and small shape deformations thereof. The renormalized pseudoentropy isolates the universal contribution, which for a spherical entangling surface is proportional to the complex-valued central charge $a^\star$ of the non-unitary CFT. On an equal footing, for infinitesimal deformations away from the sphere, we recover, at quadratic order in the deformation parameter, an analytic continuation of the Mezei-like formula in its anti-de Sitter counterpart.
}

\maketitle
%%%%%%%%%%%%%%%%%%%
\section{Introduction}
\label{sec:intro}
%%%%%%%%%%%%%%%%%%

\begin{figure}
   \centering \includegraphics[width=0.7\linewidth]{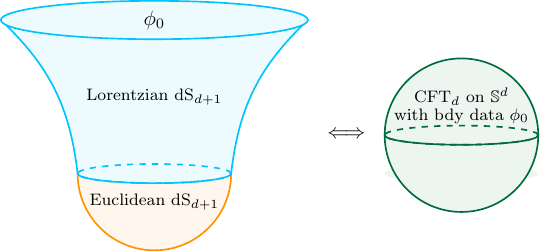}
\caption{A pictorial representation of the dS/CFT correspondence. The HH geometry is prepared by Wick-rotating the past half of dS space to a Euclidean section.}
\label{fig:dSCFT}
\end{figure}

The de Sitter/conformal field theory (dS/CFT) correspondence \cite{Strominger:2001pn,Strominger:2001gp,Spradlin:2001pw,Anninos:2012qw,Anninos:2011ui} proposes a duality between quantum gravity in dS$_{d+1}$ space and a Euclidean CFT$_d$ living on the future conformal boundary. This conjecture offers a promising avenue to address foundational questions in quantum cosmology and quantum gravity, particularly in understanding the wavefunction of the universe and the nature of cosmological observables \cite{Skenderis:2006jq,Skenderis:2006fb,McFadden:2009fg,McFadden:2010na,Hertog:2011ky,McFadden:2011kk,Bzowski:2011ab,Banerjee:2013mca}. However, compared to the well-established anti-de Sitter/conformal field theory (AdS/CFT) correspondence \cite{Maldacena:1997re,Witten:1998qj}, the dS/CFT dictionary remains less developed, partly due to the subtle nature of the dual CFT, which is Euclidean and non-unitary \cite{Strominger:2001pn,Strominger:2001gp,Spradlin:2001pw,Anninos:2012qw,Anninos:2011ui}. In this framework, the partition function of the non-unitary Euclidean CFT is identified with the Hartle--Hawking (HH) wavefunction of the universe in dS space \cite{Hartle:1983ai,Maldacena:2002vr},
\begin{equation}\label{eq:dSCFT}
\Psi_{\text{dS}}[\phi_{0}]
=
Z_{\text{CFT}}[\phi_{0}] \, ,
\end{equation}
where $\phi_{0}$ denotes the asymptotic profile of the bulk fields $\phi$, including the metric, evaluated at future infinity $\mathcal{I}^+$. 
The HH initial condition is prepared by a path integral over half of Euclidean dS space with the future boundary data fixed to $\phi_{0}$. 
The corresponding wavefunction takes the form \cite{Hartle:1983ai,Witten:2001kn}
\begin{equation} \label{eq:saddleds}
\Psi_{\text{dS}}[\phi_{0}]
\simeq \left. \exp\!\left(\iu\, I[\phi]\right)\right|_{\phi(\mathcal{I}^+)\simeq\phi_0}\,,
\end{equation}
where $I[\phi]$ is the on-shell Lorentzian dS action evaluated on histories satisfying the HH no-boundary condition with boundary data $\phi_0$ at $\mathcal{I}^+$; the approximation holds at the usual saddle point.\footnote{
As in AdS, the future conformal boundary $\mathcal{I}^+$ admits a conformal structure such that the fields obey the standard Fefferman--Graham (or Starobinsky) expansion \cite{AST_1985__S131__95_0,1983ZhPmR..37...55S,Anderson:2004wj}. 
The coefficient $\phi_0$ corresponds to the leading term in this expansion, and boundary data are specified up to a conformal equivalence class compatible with the boundary CFT symmetry.}
A sketch of the dS/CFT correspondence is presented in Fig.~\ref{fig:dSCFT}.

Within this framework, CFT correlators are obtained by functional differentiation of the wavefunction \cite{Maldacena:2002vr,Anninos:2012qw,Bzowski:2023nef},
\begin{equation}
\langle \mathcal O( x_1) \cdots \mathcal O(x_n) \rangle_{\text{CFT}}
= \frac{\delta^n \Psi_{\text{dS}}}{\delta \phi_0( x_1) \cdots \delta \phi_0( x_n)}\Bigg|_{\phi_0=0}\,,
\end{equation}
and should be interpreted as \emph{amplitudes}, namely coefficients in the late--time wavefunctional rather than bulk physical observables. 
Physical late--time observables in dS space are instead computed using the in--in (Schwinger--Keldysh) formalism \cite{Hartle:1983ai,Maldacena:2002vr,Witten:2001kn,Anninos:2012qw,Bzowski:2023nef},
\begin{equation}
\langle \phi_0( x_1) \cdots \phi_0( x_n) \rangle_{\text{dS}}
= \int \mathcal D\phi_0 \, |\Psi_{\text{dS}}[\phi_0]|^2 \, \phi_0( x_1) \cdots \phi_0( x_n)\,,
\end{equation}
which yields genuine bulk expectation values. Since they depend only on $|\Psi_{\text{dS}}|^{2}$, these correlators are insensitive to the overall phase of the wavefunction and are therefore free of infrared divergences and independent of local boundary counterterms \cite{Bzowski:2023nef,Poole:2025cmv}. Despite the physical significance of in--in correlators in the dS bulk, our analysis will be restricted to boundary CFT correlators obtained from the late--time wavefunction.

Having established the holographic framework, we now turn our attention to entanglement entropy \cite{Calabrese:2004eu,Calabrese:2009qy,Headrick:2019eth}, a key probe of quantum correlations and a central element of the bulk--boundary dictionary \cite{Witten:1998qj,Gubser:1998bc,Harlow:2011ke}. In holography, it provides a powerful diagnostic of the quantum structure of the boundary theory and offers a window into the emergence of spacetime from quantum information \cite{VanRaamsdonk:2010pw,Penington:2019npb,Takayanagi:2018pml,Takayanagi:2025ula}. As we now review, this connection admits a precise geometric realization within the correspondence.

In AdS/CFT, the Ryu--Takayanagi (RT) prescription \cite{Ryu:2006bv,Ryu:2006ef} provides a geometric formula for the entanglement entropy: the holographic entanglement entropy (HEE) of a boundary subregion $A$ is given by the area of a minimal extremal surface $(\Sigma_A)$ in the bulk anchored to the boundary of $A$, i.e.,
\begin{equation}\label{eq:RTformula}
    S(A)=\frac{\text{Area}(\Sigma_A)}{4G}\,.
\end{equation}

In dS/CFT, a similar prescription has been proposed \cite{Doi:2022iyj,Doi:2023zaf}, but with crucial distinctions: the bulk extremal surfaces anchored to the future boundary have mixed signature; near the boundary they are timelike, while deep in the interior they become spacelike \cite{Doi:2022iyj,Doi:2023zaf}%\rc{as} depicted in Fig.~\ref{fig:undefds}
. Hence the area functional in dS space contains both timelike and spacelike segments, i.e., $\Sigma_A=\Sigma_A^{(\text t)}\cup\Sigma_A^{(\text s)}$, and consequently $S(A)=S^{(\text t)}(A)+S^{( \text s)}(A)$, therefore the dual quantity is not standard entanglement entropy but rather \emph{pseudoentropy} \cite{Nakata:2020luh,Doi:2022iyj,Doi:2023zaf}.\footnote{For more recent developments on pseudoentropy in dS holography and related settings, see ~\cite{Guo:2022jzs,Aalsma:2022swk,Narayan:2022afv,He:2023eap,Alshal:2023kcd,Chen:2023prz,Narayan:2023ebn,Jiang:2023loq,Kawamoto:2023nki,Chen:2023eic,Guo:2023aio,Aguilar-Gutierrez:2023tic,Narayan:2023zen,Shinmyo:2023eci,Kanda:2023jyi,Yadav:2024ray,Doi:2024nty,Fareghbal:2024lqa,Caputa:2024gve,Goswami:2024vfl,Nanda:2025tid,Fujiki:2025rtx,Anastasiou:2025rvz,Huang:2025gmq}.} This new quantum information measure provides a generalization of entanglement entropy and it is defined as the von Neumann entropy of a transition matrix $\tau_A$ instead of a reduced density matrix, i.e., ~\cite{Nakata:2020luh}
\begin{equation}
S(A)=-\text{tr}(\tau_A\log\tau_A)\,,\quad\tau_A=\text{tr}_B\left(\frac{|\psi\rangle\langle\varphi|}{\braket{\varphi|\psi}}\right)\,,
\end{equation}
where $|\psi\rangle$ and $|\varphi\rangle$ are pure states in the total Hilbert space $\mathcal{H}_{\text{tot}}=\mathcal{H}_{A}\otimes \mathcal{H}_{B}$. 
A natural question concerns the physical interpretation of the states $|\psi\rangle$ and $|\varphi\rangle$ entering the definition of the transition matrix $\tau_A$ in the dS/CFT context. In contrast to unitary CFTs, the dual CFT to dS space is non-unitary, and the Euclidean path integral prepares states that are not related by Hermitian conjugation. The transition matrix $\tau_A$ is constructed by performing the path integral on the sphere with different boundary conditions in the upper and lower hemispheres, preparing the states $|\psi\rangle$ and $|\varphi\rangle$, respectively, and then gluing along the complement of the subregion $A$ \cite{Nakata:2020luh,Doi:2022iyj,Doi:2023zaf}.

\begin{figure}
    \centering
    \includegraphics[width=0.325\linewidth]{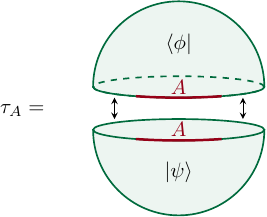}
   \caption{Transition matrix $\tau_A$ for a subsystem $A$ on the equator of $\mathbb S^d$, where the Euclidean CFT dual to dS is defined.}
 \label{fig:tau_A}
\end{figure}

In this sense, $|\psi\rangle$ and $|\varphi\rangle$ should be viewed as independent states defined by distinct Euclidean preparations, rather than related by conjugation. This structure is explicitly realized in the $\mathrm{dS}_3/\mathrm{CFT}_2$ correspondence, where the dual theory is given by a non-unitary Liouville CFT with complex action \cite{Hikida:2022ltr,Boruch:2021hqs}. In such cases, the path integral is inherently complex, leading to a non-Hermitian reduced operator $\tau_A$ and hence $\tau_A \neq \tau_A^\dagger$. A pictorial representation of this construction for a subsystem $A$ on the equator of $\mathbb S^d$ is shown in Fig.~\ref{fig:tau_A}.

A crucial difference between reduced density matrices and transition matrices lies in the non-Hermiticity of the latter, which implies that the associated pseudoentropy $S(A)$ can acquire complex values \cite{Doi:2022iyj,Doi:2023zaf}. This feature makes pseudoentropy a useful information measure for non-Hermitian quantum systems in condensed matter physics \cite{Couvreur:2016mbr,Herviou:2019yfb,Chang:2019jcj,Jian:2020byi}.

Additionally, related complex-valued measures  have been explored in the context of timelike entanglement entropy (TEE) 
\cite{Doi:2022iyj,Chen:2023gnh,Nakata:2020luh,Das:2023yyl,Doi:2023zaf,Guo:2024lrr,Grieninger:2023knz,Heller:2024whi,Jiang:2023ffu,Li:2022tsv,Xu:2024yvf,Anegawa:2024kdj,Narayan:2023ebn,Takayanagi:2025ula,Nunez:2025gxq,Katoch:2025bnh,Chu:2025sjv,Zhao:2025zgm,Harper:2025lav,Li:2025tud,Giataganas:2025ize,Afrasiar:2025eam,Jena:2024tly}, 
which is defined for timelike-separated regions in the dual unitary Lorentzian CFT within the AdS/CFT correspondence.
From the perspective adopted here, TEE can be regarded as a pseudoentropy-like observable that generalizes conventional entanglement measures beyond spacelike configurations. 
Holographically, the extremal surfaces associated with TEE are not constrained in the same manner as standard RT surfaces and can extend across black hole horizons, thereby probing regions deeper in the bulk spacetime \cite{Doi:2023zaf,Das:2023yyl}. With potential applications ranging from sub-horizon bulk reconstruction to the exploration of near-singularity physics \cite{Das:2023yyl,Anegawa:2024kdj}. It would be interesting to explore whether similar effects persist in dS space.

Like its AdS counterpart, holographic pseudoentropy in dS is ultraviolet (UV) divergent due to the infinite bulk volume near the conformal boundary. For a smooth entangling region $A$, the  pseudoentropy admits an expansion of the form\footnote{$\lfloor x \rfloor$ is the integer floor of $x$.} \cite{Doi:2022iyj,Doi:2023zaf}
\begin{equation}\label{eq:Sexp}
    S(A)=
    \sum_{k=1}^{\lfloor\frac{d-1}{2}\rfloor}
    c_{d-2k}\left(\frac{H}{\delta}\right)^{d-2k}
    +
    \begin{cases}
     (-1)^{\frac{d-1}{2}}\,\mathsf{S}_{\text{u}}\text{(A)}\,, & \text{odd } d\,,\\[4pt]
     (-1)^{\frac{d-2}{2}}\,\mathsf{S}_\text{u}\text{(A)}\!\left(\log\frac{H}{\delta}-\frac{\iu\pi}{2}\right), & \text{even } d\,,
    \end{cases}
\end{equation}
where $\delta$ is a UV cutoff, $H$ denotes a characteristic length scale of $A$, and the coefficients $c_{d-2k}$ are scheme-dependent. The quantity $\mathsf{S}_\text{u}\text{(A)}$ is finite, universal and captures the intrinsic pseudoentropy structure of the dual CFT. Extracting $\mathsf{S}_\text{u}\text{(A)}$ requires a systematic  renormalization
procedure, directly analogous to the AdS case. Holographic renormalization for bulk AdS spacetimes is well established 
\cite{Skenderis:2002wp,Henningson:1998gx,deHaro:2000vlm}, and closely related methods have also been developed in dS space \cite{Anninos:2013rza,Castro:2012gc,Banks:2006rx,Ghezelbash:2001vs,Poole:2025cmv}. 
The extension of these techniques to codimension-two extremal surface areas in AdS, and hence to HEE, has likewise been explored in the literature~\cite{Taylor:2016aoi,Anastasiou:2017xjr,Nishioka:2018khk,Anastasiou:2018rla,Anastasiou:2018mfk,Anastasiou:2019ldc,Taylor:2020uwf,Anastasiou:2021swo,Anastasiou:2022ljq}. 
In contrast, an analogous analysis for dS space---in particular for pseudoentropy---has not yet been studied in the literature.

In this paper, we address this gap by developing a framework for the renormalization of pseudoentropy in dS$_4$/CFT$_3$ and dS$_6$/CFT$_5$, guided by bulk Weyl invariance. Our approach is based on the conformal renormalization prescription 
\cite{Anastasiou:2018mfk,Anastasiou:2018rla,Anastasiou:2019ldc}, which relies on conformal gravity (CG), a higher-derivative theory invariant under local conformal transformations and admitting Einstein-(A)dS spacetimes as a consistent subsector of its solution space 
\cite{Maldacena:2011mk,Grumiller:2013mxa,Anastasiou:2016jix}. A key observation is that the CG action, when evaluated on Einstein-AdS backgrounds---or equivalently after imposing appropriate boundary conditions---reproduces the renormalized Einstein--AdS action 
\cite{Maldacena:2011mk,Anastasiou:2016jix,Anastasiou:2018mfk,Anastasiou:2018rla,Anastasiou:2019ldc,Hell:2023rbf}. 
This equivalence reflects the finiteness of the CG action for any asymptotically locally AdS spacetime~\cite{Anastasiou:2023oro}, and it is expected to hold as well for Einstein-dS spacetimes through analytic continuation.\footnote{As pointed out in \cite{Maldacena:2011mk,Hell:2023rbf}, because of the independence of CG to the value of the cosmological constant, both the AdS partition function and the HH wavefunction in dS can be obtained through Neuman boundary conditions in the tree-level CG theory.}

Remarkably, an analogous finiteness property holds for the corresponding HEE functional obtained via the replica construction \cite{Calabrese:2004eu,Lewkowycz:2013nqa,Barrella:2013wja,Solodukhin:2008dh}, which provides a general framework for computing entanglement entropy in quantum field theories \cite{Calabrese:2004eu}.
In holography, this construction is implemented by considering an $n$-fold replicated geometry endowed with a $\mathbb{Z}_n$ symmetry. 
For $n\neq 1$, the fixed locus of this symmetry introduces a conical defect that can be described as the backreaction of a codimension-two cosmic brane, and the quotient geometry acquires an orbifold structure with the entangling surface located at the defect. 
The HEE is then obtained by analytically continuing the replicated partition function to $n\to 1$ \cite{Lewkowycz:2013nqa,Fursaev:2006ih}. 
For Einstein gravity, this procedure reproduces the RT prescription \cite{Ryu:2006bv}, while in higher-curvature theories, additional subtleties such as the splitting problem arise \cite{Miao:2014nxa,Camps:2014voa}. 
Notable exceptions which are free from the splitting ambiguity include Lovelock and quadratic curvature gravities; in particular, CG in four bulk dimensions belongs to the latter class, for which the resulting entropy functional is conformally invariant~\cite{Solodukhin:2008dh} and free of infrared divergences \cite{Anastasiou:2022ljq,Anastasiou:2024rxe}.\footnote{Note that a specific combination of the six dimensional conformal invariants \cite{Lu:2011ks} is also free from the splitting ambiguity, as shown in \cite{Miao:2014nxa}.}

In the dS/CFT setting, the replica construction has been revisited recently in Ref.~\cite{Nanda:2025tid} (see also \cite{Chandrasekaran:2022eqq}). 
Rather than acting on a Euclidean partition function as in AdS/CFT, the replica trick is applied to the HH wavefunction $\Psi_{\mathrm{dS}}$. 
Taking the $n\to 1$ limit yields a pseudoentropy associated with the area of the brane $\Sigma_{A,n}$,
\begin{equation} \label{eq:SAds}
S(A)\;=\; \lim_{n \to 1} \left( 1 - n \partial_n \right)
\log \Psi_{\mathrm{dS},n}
\;=\; \frac{\mathbf{A}(\Sigma_{A,n})}{4 \GN} \, .
\end{equation}
The cosmic brane generically acquires a complex worldvolume, reflecting the mixed Euclidean--Lorentzian character of extremal surfaces in de Sitter space. As a consequence, its area---and hence the associated pseudoentropy---is, in general, complex. In particular, the divergent contributions arising near future infinity are purely imaginary.

Motivated by this construction, we develop a systematic framework for defining and renormalizing the area of codimension-two extremal surfaces in dS space. We apply these ideas to dS/CFT in four and six bulk dimensions.\footnote{In this work we restrict attention to even-dimensional bulk spacetimes, where conformal renormalization is well established. While the construction can in principle be extended to arbitrary even dimensions, explicit realizations are technically simpler in four and six dimensions. 
In odd bulk dimensions, a conformal renormalization prescription is absent due to the impossibility of constructing local bulk conformal invariants. Nevertheless, the universal part of the pseudoentropy can be recovered by systematically discarding the divergent, non-universal contributions, as we have shown for arbitrary dimensions in our recent work \cite{Anastasiou:2025rvz}.} Starting from the  CG action, we evaluate it on replicated manifolds containing codimension-two conical defects along extremal surfaces. The contribution of the conical defect in the replicated geometry reproduces the Graham--Witten action \cite{Graham:1999pm} in four dimensions and the Graham--Reichert action \cite{Graham:2017bew} in six dimensions. These functionals lead, when evaluated for extremal surfaces on Einstein-dS backgrounds, to the renormalized area functional which we identify with the finite part of the holographic pseudoentropy.

We illustrate our formalism with two explicit examples in both dimensions: a spherical entangling region and small deformations thereof. For the sphere, the renormalized pseudoentropy is proportional to the complex-valued central charge $a^\star$ of the non-unitary CFT. For deformations, we solve the linearized extremality equation and compute the change in pseudoentropy at quadratic order in the deformation parameter. The result takes the form of a Mezei-like formula \cite{Mezei:2014zla,Allais:2014ata,Anastasiou:2025rvz}, which relates the shape dependence to the coefficient $C_T$ of the stress-tensor two-point function. This coefficient is a universal CFT datum, here understood as being analytically continued from AdS to dS, and is defined through
\begin{equation}
    \langle T_{\mu\nu}(x) T_{\rho\sigma}(0) \rangle
    =
    \frac{C_T}{x^{2d}}
    \left[
        I_{\mu(\rho} I_{\sigma)\nu}
        -
        \frac{1}{d}\, \delta_{\mu\nu} \delta_{\rho\sigma}
    \right] \, ,
\end{equation}
which holds for general CFTs. Here
$I_{\mu\nu} \equiv \delta_{\mu\nu} - 2 x_\mu x_\nu / x^2$
is a theory-independent tensorial structure \cite{Osborn:1993cr}. This provides a non-trivial check of the dS/CFT dictionary.

The paper is organized as follows. In section~\ref{sec:4D}, we review the relation between CG and Einstein--(A)dS gravity in four dimensions, derive the renormalized pseudoentropy from the replica trick, and present the spherical and deformation examples. In section~\ref{sec:6D}, we extend the analysis to six dimensions. section~\ref{sec:conclusions} contains our conclusions and outlook. Notation, conventions, and additional technical details are provided in the appendices.

%%%%%%%%%%%%%%%%%%%%%%%
\section{Pseudoentropy from conformal renormalization: four-dimensional case}
\label{sec:4D}
%%%%%%%%%%%%%%%%%%%%%%%
In this section, we aim on developing  a systematic renormalization prescription for holographic pseudoentropy in four-dimensional de Sitter space based on bulk Weyl invariance. Our starting point is conformal gravity (CG), the unique four-dimensional purely metric theory that is invariant under local Weyl transformations.  The theory is defined by the action\footnote{For simplicity, from now on we omit in all integrals the volume element and the determinant of the metric, which in this case amounts to $\int_\mathcal M\equiv\int_\mathcal M\diff^4x\sqrt{|\mathcal G}|$.}
\begin{equation}\label{eq:CGaction4D}
I_{\text{CG}} = \alpha_{\text{CG}} \int_{\mathcal{M}}W^2\, ,
\end{equation}
where
\begin{equation}
W^2\equiv W^{\alpha \beta \gamma \delta} W_{\alpha \beta \gamma \delta} \,,
\end{equation}
denotes the Weyl-squared scalar, with the Weyl tensor expressed in terms of the Riemann tensor and the Schouten
tensor as
\begin{equation}
W^{\alpha\beta}_{\gamma \delta}
=R^{\alpha\beta}_{\gamma \delta}
-4\,S^{[\alpha}_{[\gamma}\delta^{\beta]}_{\delta]}\,,
\qquad
S^\alpha_{\beta}
=\frac{1}{D-2}
\left(
R^\alpha_{\beta}
-\frac{R}{2(D-1)}\delta^\alpha_{\beta}
\right)\,.
\end{equation}
Here $D=d+1$ is the bulk dimension of the spacetime manifold $\mathcal M$.
CG was initially introduced as an alternative to general relativity with better  properties in the ultraviolet~\cite{Stelle:1976gc,Capper:1975ig,Fradkin:1983tg,Julve:1978xn} and which allows for predictions of the rotational velocity curves of galaxies in astrophysics~\cite{Mannheim:1988dj,Mannheim:2005bfa,Mannheim:2010ti,Mannheim:2011ds}. However, the theory carries ghosts~\cite{Riegert:1984hf}, even though in certain backgrounds this problem can be evaded~\cite{Mannheim:2021oat}. 

Despite the ghost problem, CG admits desirable features which are relevant for our study. In particular, Einstein spacetimes belong to the solution space of the theory while also being free of infrared divergences for asymptotically locally (A)dS (Al(A)dS) spacetimes~\cite{Grumiller:2013mxa}. Interestingly, imposing Neumann boundary conditions allows to both eliminate the ghost mode and isolate the Einstein-(A)dS class of solutions.~\cite{Maldacena:2011mk,Hell:2023rbf}.

Equivalently, one may directly evaluate the action~\eqref{eq:CGaction4D} for Einstein-(A)dS spacetimes~\cite{Anastasiou:2016jix,Anastasiou:2020mik} imposing the condition $R_{\alpha\beta}=\Lambda \mathcal{G}_{\alpha\beta}\,, \Lambda=-3\sigma/\Ls^2$.\footnote{An equivalent definition of the Einstein condition is obtained by
requiring the vanishing of the trace-free part of the Ricci tensor,
$H^\alpha_{\ \beta}=0$.}
In this case, the Weyl tensor reduces to the (A)dS curvature tensor
\begin{equation}
F^{\alpha\beta}_{\gamma \delta}
=R^{\alpha\beta}_{\gamma \delta}
+\frac{\sigma}{\Ls^2}\delta^{\alpha\beta}_{\gamma \delta}\,,
\label{eq:AdScurv}
\end{equation}
where $\sigma=+1$ for AdS , $\sigma=-1$ for dS and $\Ls$ is the (A)dS radius. Overall, by imposing these conditions and fixing the coupling constant to $\alpha_{\text{CG}}=\sigma \Ls^2/(64\pi \GN)$, Eq.~\eqref{eq:CGaction4D} can be cast in the form
\begin{equation}
I_{\text{CG}}(E)= \frac{\sigma\Ls^2}{64\pi\GN} \int_{\mathcal M} F^2 -   \frac{\pi \sigma \Ls^2}{2\GN} \mathcal{\chi}\left(\mathcal{M}\right) \,,
\label{CGEinstein}
\end{equation}
where 
$F^2 \equiv F_{\alpha\beta}^{\gamma\delta} F^{\alpha\beta}_{\gamma\delta}$. We hereby added a topological term, being indifferent to local Weyl rescalings, that does not affect the asymptotics of the action; however, it is necessary to correctly reproduce BH entropy~\cite{Anastasiou:2023oro} and HEE~\cite{Anastasiou:2017xjr} in the AdS case. After some algebraic manipulation the latter can be rewritten as
\begin{align}
I_{\text{CG}}(E)= \frac{1}{16\pi\GN}
\int_{\mathcal M}\left(R+\frac{6\sigma}{\Ls^2}\right)
+\frac{\sigma\Ls^2}{64\pi\GN}
\int_{\mathcal M}\mathcal X_4-  \frac{\pi \sigma \Ls^2}{2\GN} \mathcal{\chi}(\mathcal{M}) \,,
\label{eq:RenEH4d}
\end{align}
where
\begin{equation}
\mathcal X_4 =\frac{1}{4} \delta_{\alpha_1 \alpha_2 \alpha_3 \alpha_4}^{\beta_1 \beta_2 \beta_3 \beta_4} R_{\beta_1 \beta_2}^{\alpha_1 \alpha_2} R_{\beta_3 \beta_4}^{\alpha_3 \alpha_4}
\end{equation}
is the Gauss-Bonnet density. This expression is manifestly finite for Al(A)dS spacetimes and corresponds to the renormalized Einstein-(A)dS action, i.e. $I_{\text{CG}}(E)=\IEren$.\footnote{In AdS, the conformal boundary is the usual timelike hypersurface given by conformal compactification at radial infinity, whereas in dS, there are two conformal boundaries corresponding to future and past time infinity. Because we are considering the HH no-boundary geometry to the past, we will only consider the conformal boundary at future time infinity ($\mathcal{I}^+$).} Indeed, in the $\sigma=+1$ case, the latter has been shown to be equivalent to standard holographic renormalization counterterm~\cite{Miskovic:2009bm}.

Since Al(A)dS spacetimes are conformally compact manifolds with boundaries \cite{Penrose:1962ij}, one may trade the bulk term for a  boundary contribution using the Gauss--Bonnet theorem,
\begin{equation}
\int_{\mathcal M}\mathcal X_4
=
32\pi^2\chi(\mathcal M)
+\int_{\partial\mathcal M}\mathcal B_3\,,
\label{eq:GBt}
\end{equation}
 where $\sqrt{h}\,\mathcal B_3 = -4\, \sqrt{h}\, \delta^{\mu_1 \mu_2 \mu_3}_{\nu_1 \nu_2 \nu_3}\,
k^{\nu_1}_{\mu_1}
\left(
\frac{1}{2}\, r^{\nu_2 \nu_3}_{\mu_2 \mu_3}
- \frac{1}{3}\,
k^{\nu_2}_{\mu_2}
k^{\nu_3}_{\mu_3}
\right)$ is the second Chern form, which depends explicitly on the intrinsic and extrinsic geometry of the boundary. As a result, the renormalized Einstein-(A)dS action reads

\begin{equation}
\IEren
=\frac{1}{16\pi\GN}
\int_{\mathcal M}\left(R+\frac{6\sigma}{\Ls^2}\right)
+\frac{\sigma\Ls^2}{64\pi\GN}
\int_{\partial\mathcal M}\mathcal B_3\,.
\label{eq:IEren}
\end{equation}
This result highlights the power of the CG approach: the intricate counterterm structure required for Einstein gravity emerges automatically from the bulk Weyl symmetry of the action \eqref{eq:CGaction4D}.

CG therefore provides a systematic renormalization scheme in which one starts from an already finite action, restricts to the Einstein sector, and obtains finite boundary observables without introducing additional counterterms. This framework, dubbed \textit{conformal renormalization}, will be particularly useful for defining and computing renormalized pseudoentropy, as we show in the following sections.

\subsection{Codimension-two local conformal invariant in four dimensions}
\label{subsec:Lsigma4D}
In both AdS/CFT and dS/CFT correspondence, the holographic entanglement entropy and pseudoentropy associated with a boundary subregion $A$ are derived from a codimension-two bulk surface $\Sigma$ anchored on $\partial A$. The bare area of $\Sigma$ diverges due to the infinite volume of the bulk spacetime and must therefore be renormalized. 

Taking advantage of the finiteness of the CG action, we are aiming on constructing codimension-2 functionals which are free of IR divergences, extending the paradigm of Refs.~\cite{Anastasiou:2022ljq,Anastasiou:2024rxe,Anastasiou:2025dex} to dS. To do so, we consider a replicated bulk manifold $\mathcal M^{(\vartheta)}$ that develops a conical singularity of opening angle $2\pi\vartheta$ along $\Sigma$ where $\vartheta$ is related to the replica index $n$ by $\vartheta = 1/n$. Considering the definition of the pseudoentropy $S(A)$ in terms of the dS wavefunction in Eq.~\eqref{eq:SAds} and replacing the saddle-point approximation of Eq.~\eqref{eq:saddleds}, one recovers the Lewkowycz--Maldacena (LM) prescription \cite{Lewkowycz:2013nqa} which states that both holographic entanglement entropy in AdS and pseudoentropy in dS are given by
\begin{equation}
S(A)= -\lim_{\vartheta \to 1}
\partial_\vartheta
I[\mathcal M^{(\vartheta)}]\,,
\label{eq:LMdS}
\end{equation}
where $I[\mathcal M^{(\vartheta)}]$ denotes the on-shell action evaluated on the orbifold geometry.

Based on these considerations, our starting point is the  CG action
\begin{equation}
  I_{\text{CG}} =\frac{\sigma \Ls^2}{64\pi \GN} \int_{\mathcal M}W^2 -  \frac{\pi \sigma  \Ls^2}{2\GN} \mathcal{\chi}(M) \label{eq:CGaction4Dren}
\end{equation}
which as already discussed is finite without additional counterterms (see Appendix \ref{App:BulkFinite}). We now implement the LM prescription \eqref{eq:LMdS} by evaluating the action \eqref{eq:CGaction4Dren} on a replicated bulk geometry $\mathcal M^{(\vartheta)}$. The replica construction introduces a codimension-two conical singularity of opening angle $2\pi\vartheta$ along the surface $\Sigma$, while the geometry remains smooth away from $\Sigma$. Evaluating the action on this orbifold geometry leads to
\begin{equation}
  I^{(\vartheta)}_{\text{CG}} =\frac{\sigma \Ls^2}{64\pi \GN} \int_{\mathcal{ M}^{(\vartheta)}} \left(W^{(\vartheta)}\right)^2  - \frac{\pi \sigma  \Ls^2}{2\GN} \mathcal{\chi}\left(\mathcal{ M}^{(\vartheta)}\right).
\end{equation}
As a consequence of the conical singularity, the Euler characteristic of the replicated manifold acquires an additional localized contribution supported on $\Sigma$ \cite{Fursaev:2013fta},
\begin{equation}
    \chi \left(\mathcal{ M}^{(\vartheta)}\right) =\chi (\mathcal{ M}) + (1-\vartheta) \chi(\Sigma),
\end{equation}
where $\Sigma$ denotes the codimension-two locus of the conical defect, equipped with the induced metric $\gamma_{ab}$, with $(a,b)$ labeling coordinates $y^a$ intrinsic to $\Sigma$---see appendix~\ref{app:A} for more details on indices notation.

Similarly, the Weyl-squared term evaluated on the orbifold geometry admits a systematic expansion in powers of the deficit angle
\cite{Fursaev:2013fta,Solodukhin:2008dh},
\begin{equation}
   \int_{\mathcal{ M}^{(\vartheta)}} {W^{{(\vartheta)}}}^2 = \int_{\mathcal M} W^2  + 8\pi(1-\vartheta) \int_\Sigma\mathcal {K}_{\Sigma} + \mathcal{O}[(1-\vartheta)^2], \label{eq:WeylSquaredExpansion}
\end{equation}
where $\mathcal{K}_{\Sigma}$ is a local, conformally covariant scalar, built from the intrinsic and extrinsic geometry of the codimension-two surface $\Sigma$,
\begin{equation}\label{eq:Ksigma}
\mathcal{K}_{\Sigma}= W_{ab}^{ab} - P_{ab}^{A} P^{ab}_{A}\,,
\qquad
P_{ab}^{A}= K_{ab}^{A} - \frac{1}{2} K^{A}\,\gamma_{ab}.
\end{equation}
with $P_{ab}^{A}$ being the traceless part of the extrinsic curvature along the normal direction labeled by $A$. Combining the localized contributions arising from the Euler characteristic and
the Weyl-squared term, the CG action evaluated on the orbifold
takes the form
\begin{equation}
     I^{(\vartheta)}_{\text{CG}}=I_{\text{CG}} +\frac{\sigma \Ls^2(1-\vartheta)}{8\GN}\int_\Sigma\mathcal{K}_{\Sigma}
     -\frac{\pi\sigma \Ls^2(1-\vartheta)}{2\GN}
     \chi(\Sigma)+\mathcal{O}[(1-\vartheta)^2],
\end{equation}
or, equivalently,
\begin{equation}
   I^{(\vartheta)}_{\text{CG}} =I_{\text{CG}}+\frac{\left(1-\vartheta\right)}{4\GN}\,\mathbf{L}(\Sigma)+\mathcal{O}[(1-\vartheta)^2],
\end{equation}
where the coefficient of the conical defect defines the codimension-two functional
\begin{equation}
    \mathbf{L}(\Sigma) = \frac{\sigma \Ls^2}{2}
    \int_\Sigma\mathcal{K}_{\Sigma}
    - 2\pi\sigma \Ls^2\,\chi(\Sigma).\label{eq:Lsigma}
\end{equation}
The functional $\mathbf{L}(\Sigma)$, known as Graham-Witten action \cite{Graham:1999pm}, is manifestly finite and invariant under bulk Weyl transformations \cite{Anastasiou:2025dex}. Within the LM framework, it captures the contribution of the conical defect to the on-shell action, analogous to the area of the RT surface for Einstein-(A)dS gravity. Since we are aiming for pseudoentropy in dS spacetimes, which is derived for extremal surfaces, in what follows we provide the link between $\textbf{L}\left(\Sigma\right)$ and renormalized area functionals, extending the results derived before in the literature for AdS~\cite{Anastasiou:2022ljq,Anastasiou:2024rxe}.

%%%%%%%%%%%%%%%%%%%%%%%%%%
\subsection{Renormalized pseudoentropy for dS$_{4}$/CFT$_{3}$}\label{subsec:pseudo4D}
%%%%%%%%%%%%%%%%%%%%%%%%%

Starting from the codimension-two conformal invariant $\mathbf{L}(\Sigma)$ derived in the previous subsection, we now construct the renormalized geometric functional governing holographic pseudoentropy in dS$_4$/CFT$_3$. In this framework, the pseudoentropy associated with a boundary region $A$ is obtained from the renormalized area of an extremal codimension-two surface $\Sigma$ anchored on $\partial A$.

Our starting point is the $\mathcal{O}\left[\left(1-\vartheta\right)\right]$  contribution to the CG action, encoded in the functional $\mathbf{L}(\Sigma)$ given in Eq.~\eqref{eq:Lsigma}. Restricting to an ambient Einstein--dS geometry---i.e., setting $\sigma=-1$, the evaluation of $\mathcal{K}_{\Sigma}$ defined in Eqs.~\eqref{eq:Ksigma} reduces to replacing the Weyl tensor by the dS curvature tensor $F_{cd}^{ab}$ introduced in
Eq.~\eqref{eq:AdScurv}. As a result,
\begin{equation}
    \mathcal{K}_{\Sigma}\big|_{\text E}=
F_{ab}^{ab}- P_{ab}^{A} P^{ab}_{A}.
    \label{eq:KsigmaE}
\end{equation}
Substituting this expression into Eq.~\eqref{eq:Lsigma}, the codimension-two functional evaluated on an Einstein--dS background takes the form
\begin{equation}
    \mathbf{L}(\Sigma)\big|_{\text E}=-\frac{\Ls^2}{2}\int_\Sigma\left(F_{ab}^{ab}-P_{ab}^{A} P^{ab}_{A}
\right)+2\pi \Ls^2\,\chi(\Sigma). \label{eq:LSE1}
\end{equation}
Using Eq.~\eqref{eq:AdScurv}, the definition of the traceless extrinsic curvature \eqref{eq:Ksigma}, and the Gauss-Codazzi relation $R_{ab}^{ab}=\mathcal{R}+K^{A}_{ab} K^{ab}_{A}-(K^{A})^2$, this expression can be rewritten as
\begin{equation}
    \mathbf{L}(\Sigma)\big|_{\text E} = - \frac{\Ls^2}{2} \int_\Sigma \left[\mathcal{R}
    -\frac{2}{\Ls^2}-\frac{1}{2}(K^{A})^2\right] +2\pi \Ls^2\,\chi(\Sigma).
    \label{eq:LSE2}
\end{equation}
We now specialize to extremal codimension-two surfaces, denoted by $\ext$. Extremality implies the vanishing of the trace of the extrinsic curvature,
\begin{equation}
   \Sigma = \ext \Rightarrow K^{A} = 0 \,.
    \label{eq:ext0}
\end{equation}
Under this condition, the functional $\mathbf{L}(\Sigma)\big|_{\text E}$ reduces to a purely intrinsic geometric quantity, which we identify with the renormalized area of the extremal surface,\footnote{See Appendix~\ref{app:finiteness} for the proof of finiteness of the renormalized area for generic codimension-two hypersurfaces.}
\begin{equation}
    \Aren(\ext)= \mathbf{L}(\ext)\big|_{\text E} = - \frac{\Ls^2}{2}
    \int_{\ext}\left( \mathcal{R}
    -\frac{2}{\Ls^2}\right) + 2\pi \Ls^2\,\chi(\ext).
    \label{eq:LSAren}
\end{equation}
Applying the two-dimensional Gauss--Bonnet theorem to the surface $\ext$,
\begin{equation}
     \int_{\ext}\mathcal{R} =
     \int_{\partial \ext} \mathcal{B}_1^{\ext}
     + 4\pi\,\chi(\ext),
\end{equation}
where $\sqrt{\mathfrak h }\mathcal{B}_1^{\ext}= -2\sqrt{\mathfrak h }\mathfrak K $ 
denotes the first Chern form evaluated on $\partial\ext$ and $\mathfrak K$ is the trace of the extrinsic curvature of $\partial \Sigma \subset \Sigma$. Then, the renormalized area simplifies to
\begin{equation}
    \Aren(\ext)=\mathbf{A}(\ext) - \frac{\Ls^2}{2} \int_{\partial \ext} \mathcal{B}_1^{\ext},
    \label{eq:RenA4D}
\end{equation}
with $\mathbf{A}(\ext)=\int_{\ext} \diff^2y\sqrt{\gamma}$ the bare area.

Finally, if the extremal surface $\ext$ satisfies the usual homology and anchoring conditions appropriate to a RT surface $\Sigma_A$ in the dual non-unitary CFT$_3$, the finite holographic pseudoentropy associated with the region $A$ is given by
\begin{equation}
\mathsf{S}_\text{u}\text{(A)} =-\frac{\Aren(\Sigma_A)}{4\GN}= -\frac{1}{4\GN}\mathbf{L}(\Sigma_A)\big|_{\text E}\,, \label{FAL4D}
\end{equation}
This expression applies to Einstein–dS backgrounds and provides a finite definition of holographic pseudoentropy which originates from a conformal invariant and in which all power-law divergences are removed by construction, as we show explicitly in the examples below.

%%%%%%%%%%%%%
\subsection{Explicit examples}\label{sec:examples4D}
%%%%%%%%%%%%%%%%%%%%%%%
\subsubsection{Spherical entangling surface}\label{sec:sphere4D}

We now evaluate the holographic pseudoentropy for a spherical subregion in the Euclidean CFT$_3$ dual to global dS$_4$.
Using the global patch with coordinates $x^\alpha=\{\tau,\tE,\theta,\phi\}$, the Lorentzian dS metric takes
the form
\begin{equation}\label{eq:GdS1} 
\diff s^2_\mathcal{G}=\mathcal{G}_{\alpha\beta}\diff x^\alpha\diff x^\beta=\Ls^2 \left(
-\diff \tau^2 + \cosh ^2 \tau \left[ \diff \tE^2
 +\sin^2 \tE( \diff \theta^2 + \cos^2 \theta \,\diff\phi^2)\right]\right)\,,
\end{equation}
with $\tau>0$ a timelike coordinate. The region $\tau<0$ is replaced by the Euclidean geometry obtained via the Wick rotation $\tau=\iu\tauE$, yielding
\begin{equation}\label{eq:GdT1} 
\diff s^2_\mathcal{G}=\Ls^2 \left(
\diff \tauE^2 + \cos ^2 \tauE \left[ \diff \tE^2
 +\sin^2 \tE( \diff \theta^2 + \cos^2 \theta \,\diff\phi^2)\right]\right)\,,
\end{equation}
with the two geometries glued smoothly at $\tau=\tauE=0$.\footnote{The mixed nature of the bulk manifold is a consequence of the HH geometry associated with the `no-boundary proposal' in dS/CFT holography \cite{Strominger:2001pn,Maldacena:2002vr}.}

\begin{figure*}
\centering
\begin{subfigure}{.5\textwidth}
  \centering
  \includegraphics[height=0.9\linewidth]{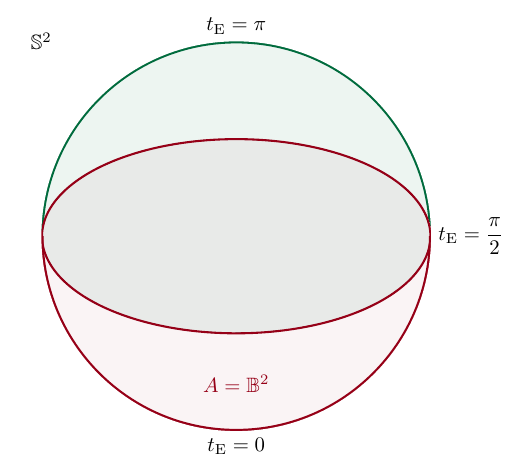}
\end{subfigure}%
\begin{subfigure}{.5\textwidth}
  \centering
  \includegraphics[height=0.9\linewidth]{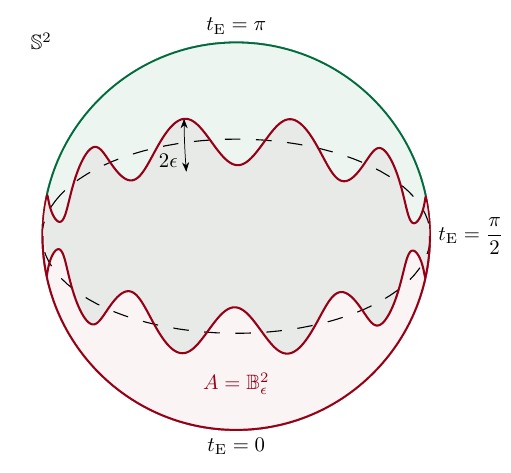}
\end{subfigure}
\caption{\justifying(Left) We show pictorially the disk entangling region $A=\mathbb B^2$ of unit radius defined in $\mathbb S^2$, which is in the equator---this is, at $\theta=0$ following \eqref{eq:GdS1}---of $\mathbb S^3$. (Right) We show the same disk-like entangling region of unit radius perturbed harmonically, $\mathbb B_\epsilon^2$, following \eqref{eq:tET}.}
  \label{fig:undefdefreg}
\end{figure*}

We consider an entangling region $A=\mathbb B^{2}$, with entangling surface $\partial A=\mathbb S^{1}$, defined at future infinity $\tau\to\infty$ by fixing the angular coordinate $\tE=T_0$ on the $\theta=0$ slice.\footnote{Under stereographic projection to $\mathbb R^{2}$, the region $A$ is mapped to a disk of radius $r=\tan(\tE/2)$.\label{foot:str}}
 In Fig.~\ref{fig:undefdefreg} (Left) we sketch the shape of the disk-like subregion $\mathbb B^{2}$ in the Euclidean CFT.

\begin{figure*}
\centering
\begin{subfigure}{.5\textwidth}
  \centering
  \includegraphics[height=0.65\linewidth]{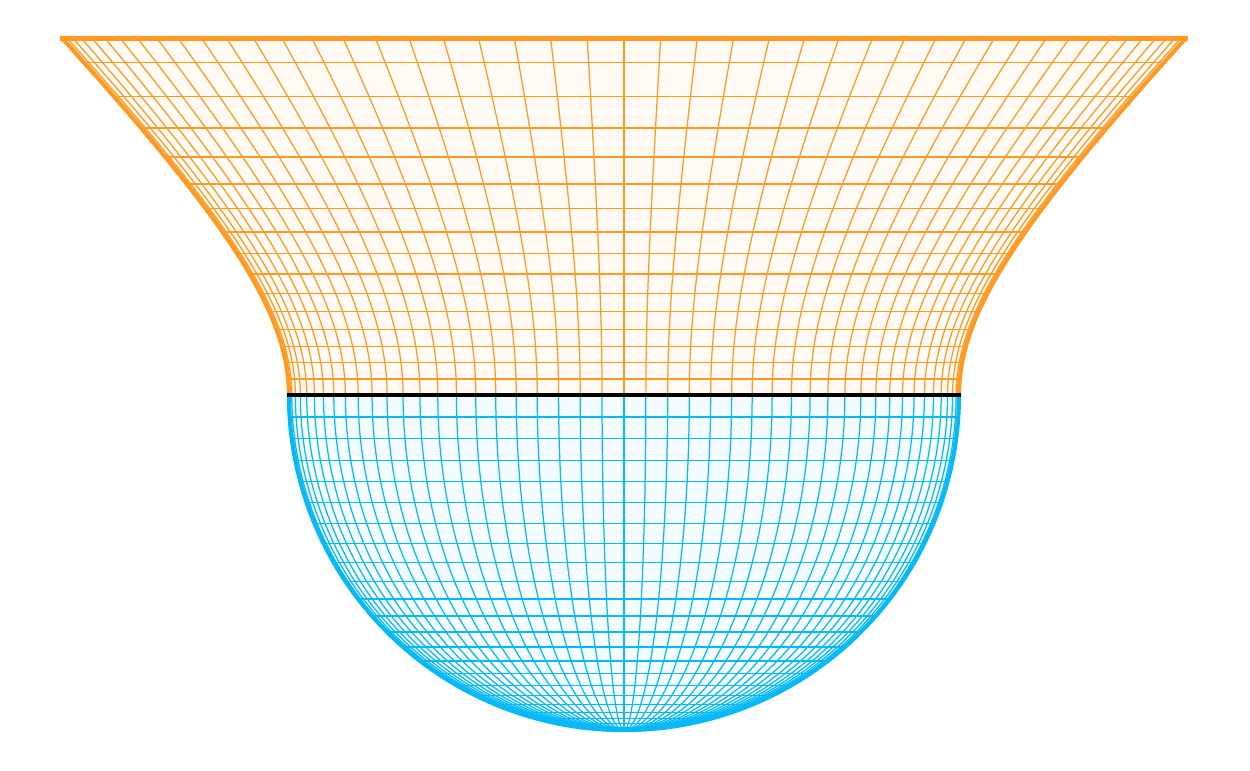}
\end{subfigure}%
\begin{subfigure}{.5\textwidth}
  \centering
  \includegraphics[height=0.65\linewidth]{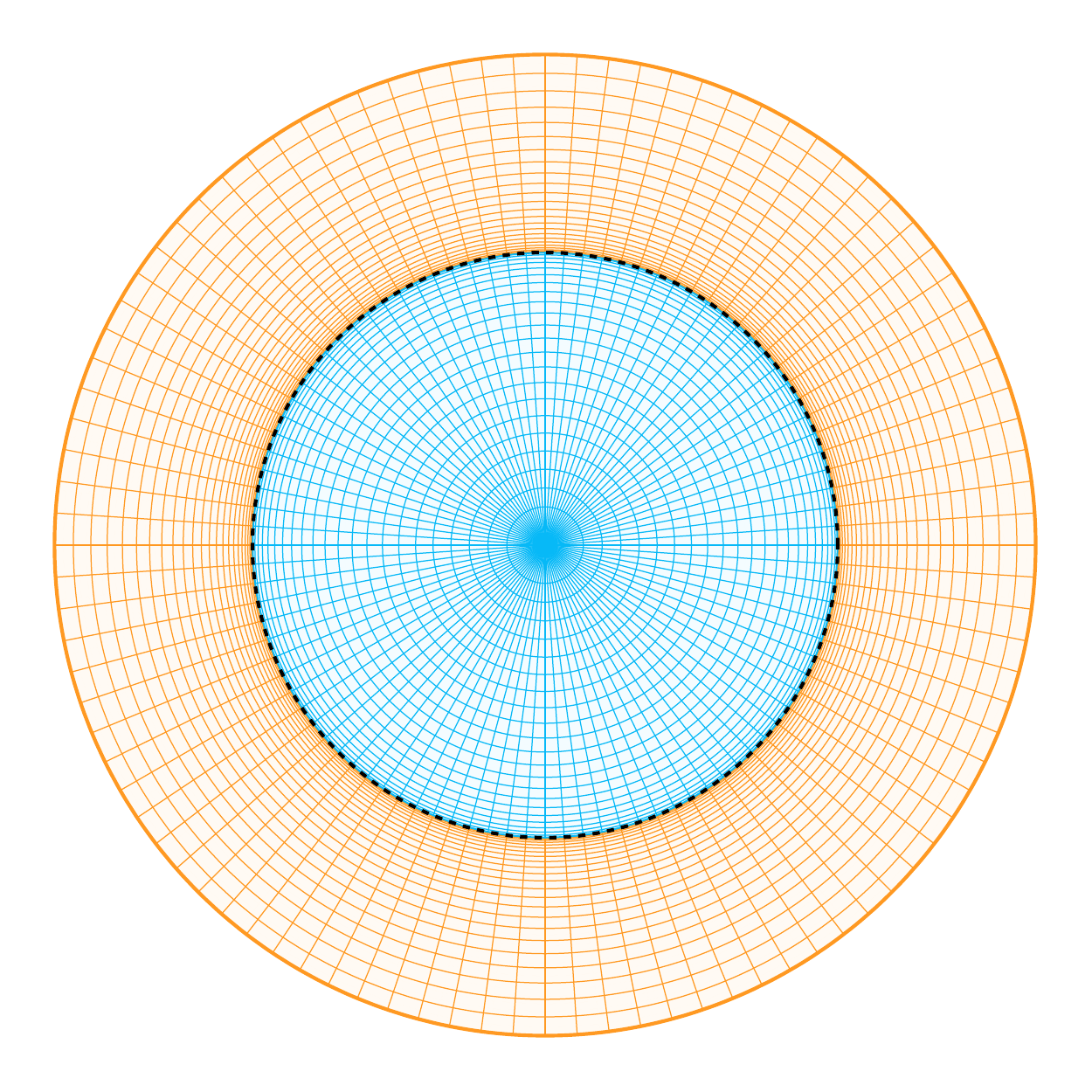}
\end{subfigure}
\caption{\justifying The RT surface associated with $\mathbb B^2$ subregion. The timelike (Lorentzian) section is shown in orange, while the spacelike (Euclidean) section is in blue. The curves are smoothly matched at $\tau = 0$, representing the junction between Lorentzian and Euclidean parts of the extremal surface. The plot is schematic and not drawn to scale. (Left) The RT surface front view (Right) The RT surface top view.}
\label{fig:unperturbedRT}
\end{figure*}

In the bulk, the associated codimension-two extremal surface $\Sigma_A$ is parameterized by coordinates $y^a=\{\tau,\phi\}$ in the Lorentzian region $\tau>0$, and by $y^a=\{\tauE,\phi\}$ in the Euclidean cap $0\le\tauE<\tan^{-1}(1/C)$. In particular, the corresponding embedding of $\Sigma_A$ is parametrized as \cite{Doi:2023zaf},
\begin{equation}\label{eq:SigmaA}
    \Sigma_A=\left\{\Sigma^{(\text t)}_A\cup\Sigma^{(\text s)}_A,\theta=0\right\}\,,\ \text{with }\begin{cases} 
\Sigma^{(\text t)}_A:\cos{\tE} =  \cos T_0 \tanh\tau\, & 0<\tau<\infty\,, \\
\Sigma^{(\text s)}_A\,:\cos{\tE} = C \tan\tauE\,\ & 0\leq\tauE < \tan^{-1} \frac{1}{C}\,,
\end{cases}
\end{equation}
where $\Sigma^{(\text t)}_A$ and $\Sigma^{(\text s)}_A$ denote the timelike and spacelike segments of the surface $\Sigma_A$, respectively, and $C$ is a constant.\footnote{While continuity at the junction does not fix $C$, smoothness across
the gluing surface uniquely determines it \cite{Anastasiou:2025rvz}.} For this embedding, the corresponding induced metric reads
\begin{equation}\label{eq:indgdS}
     \frac{\diff s^2_{\gamma}}{\Ls^2}=  \begin{cases} \displaystyle
\frac{\sin^2{T_0}\,\diff\tau^2}{\cos^2{T_0}\tanh^2{\tau}-1} +  ({\cosh^2{\tau}-\cos^2{T_0} \sinh^2{\tau}}) \diff\phi^2\,,  &  0 <  \tau < \infty\,,  \\\displaystyle
\frac{(1+C^2)\diff\tauE^2}{1-C^2 \tan^2{\tauE}}  + (\cos^2{\tauE}- C^2 \sin^2{\tauE})  \diff\phi^2,\,  & 0\leq\tauE < \tan^{-1} \frac{1}{C}\, .
\end{cases} 
\end{equation}
In Fig.~\ref{fig:unperturbedRT} we show pictorially this embedding.

To extract the finite holographic pseudoentropy, we employ the general expression~\eqref{FAL4D}, which relates pseudoentropy to the renormalized area functional~\eqref{eq:LSAren}. For both the timelike and spacelike segments of $\Sigma_A$, the intrinsic Ricci scalar of the induced metric is constant, i.e., $\mathcal R=2/\Ls^2$. As a consequence, the local curvature contribution to the renormalized area vanishes identically, and only the topological term governed by the Euler characteristic of $\Sigma_A$ contributes. Since $\Sigma_A$ is topologically a disk, $\chi(\Sigma_A)=1$, yielding 
\begin{equation}\label{eq:su04D}
\mathsf{S}_\text{u}(\mathbb B^2) = -\frac{\pi \Ls^2}{2\GN} \,.
\end{equation}
The resulting renormalized holographic pseudoentropy receives contributions solely from the spacelike segment of $\Sigma_A$. The timelike portion carries the usual area-law divergence, which is precisely removed by the renormalized area prescription, leaving a finite and real result \cite{Doi:2023zaf,Anastasiou:2025rvz}.

%%%%%%%%%%%%%%%%%%%%%%%%%%%% 
\subsubsection{Small deformations of the sphere}\label{sec:defs3d}
%%%%%%%%%%%%%%%%%%%%%%%%%

In odd-dimensional unitary CFTs the finite piece of  entanglement entropy 
associated to a spherical entangling surface coincides with the free energy on $\mathbb S^d$, namely $\mathsf{S}_\text{u,0} = -\log Z_{\mathbb{S}^d}$ \cite{Dowker:2010yj,Casini:2011kv}, while in even dimensions it is governed by the type-A conformal anomaly $a^\star$ \cite{Duff:1977ay,Bonora:1985cq,Deser:1993yx}. Different universal CFT quantities appear as the geometry of the entangling surface is deformed. A prominent example is the case of small perturbations around the $n$-ball, whose leading correction is controlled by the coefficient of the stress-energy tensor two-point function $C_T$ \cite{Osborn:1993cr}. This relation was first discovered holographically \cite{Allais:2014ata,Mezei:2014zla} and subsequently established by purely field-theoretic methods \cite{Faulkner:2015csl}. Here, we extend the result for small perturbations to
the case of holographic pseudoentropy in dS$_4$/CFT$_3$.

As reviewed in subsection~\ref{sec:sphere4D}, for a disk-shaped subregion of unit radius one has
$T_0=\pi/2$, so that the undeformed surface lies at constant $\tE=\pi/2$---see the discussion of footnote~\ref{foot:str}. To model the perturbed disk region, $\mathbb B_\epsilon^2$, of our interest we induce the Fourier (harmonic) expansion, i.e.,
\begin{equation}\label{eq:phiF}
    \mathbb B_\epsilon^2:T_0=\frac{\pi}{2}+\epsilon\Phi_\ell(\phi)\,,\quad \Phi(\phi)=\sum_\ell\left[\frac{a_\ell}{\sqrt{\pi}} \cos{(\ell \phi)}+\frac{b_\ell}{\sqrt{\pi}} \sin{(\ell \phi)}\right]\,,
\end{equation}
where $\epsilon\ll1$ is the perturbation parameter, $\ell$ labels the Fourier mode number and $a_\ell$, $b_\ell$ determined the amplitudes of the corresponding cosine and sine components of the deformation. In Fig.~\ref{fig:undefdefreg} (Right) we show a sketch of the subregion described by \eqref{eq:phiF} in the Euclidean CFT.

In this case, in the bulk side, the embedding of the timelike part of the RT surface in \eqref{eq:GdS1} is described by a function $\tE(\tau,\phi)$, which now depends explicitly on the angular coordinate $\phi$ as opposed to the expression in \eqref{eq:SigmaA}. Because of this, the induced metric reads
\begin{equation}\label{eq:GdefLT}
\frac{ds^2_{\gamma}}{\Ls^2}=  
\left({\tE'}^2  \cosh^2{\tau}-1\right)\diff\tau^2 + \cosh^2{\tau} \left(\sin^2{\tE} + {\dtE}^2\right)\diff\phi^2  +2\tE'\dtE \cosh^2{\tau}  \,\diff\tau \diff\phi\,,   
\end{equation}
where we denoted $\tE'= \partial_{\tau} \tE(\tau,\phi) $ and $\dtE = \partial_{\phi} \tE(\tau, \phi) $. Following \eqref{eq:phiF}, the embedding function must follow the form
\begin{equation}\label{eq:tET}
    \tE(\tau, \phi) =\frac{\pi}{2} + \epsilon f^{(\text t)} (\tau,\phi)\,.
\end{equation}
We assume that the function $f^{(\text t)}(\tau,\phi)$ allows for separation of variables $f^{(\text t)}(\tau,\phi)=\Tau_\ell(\tau)\Phi_\ell(\phi)$, with $\Phi_\ell(\phi)$ given in \eqref{eq:phiF} and satisfies the boundary conditions $\Tau_\ell(\tau\rightarrow\infty) = 1$ and $\Phi_\ell(\phi) = \Phi_\ell(\phi + 2\pi)$. These conditions are consequence of the homologous constraint on the RT surface, as it is anchored to the conformal boundary $\tau \rightarrow \infty$. On the other hand, the equations of motion for the radial functions $\Tau_\ell(\tau)$ are obtained by imposing the extremality condition \eqref{eq:ext0}, yielding\footnote{For the surface $\tE(\tau,\Omega_{d-2})$ in arbitrary dimension $d$, the vanishing of the trace of the extrinsic curvature along the normal leads to the differential equation for the RT surface. In particular, the extrinsic curvature along the angular directions vanishes identically, $K^{\theta}{}_{ab} = \gamma^\alpha_a \gamma^\theta_b \nabla_\alpha N_{\theta}^{\theta} = 0$, since $\gamma^\theta_b = 0$ for any $b$ \cite{Anastasiou:2025rvz}.}
\begin{equation}\label{eq:diffeq1}
\left[(\ell^2-1) \sech^2{\tau} + 3 \tanh{\tau}\frac{\diff}{\diff \tau}+ \frac{\diff^2}{d \tau^2}\right]\Tau_\ell(\tau)=0\,.
\end{equation}

\begin{figure*}
\centering
\begin{subfigure}{.5\textwidth}
  \centering
  \includegraphics[height=0.65\linewidth]{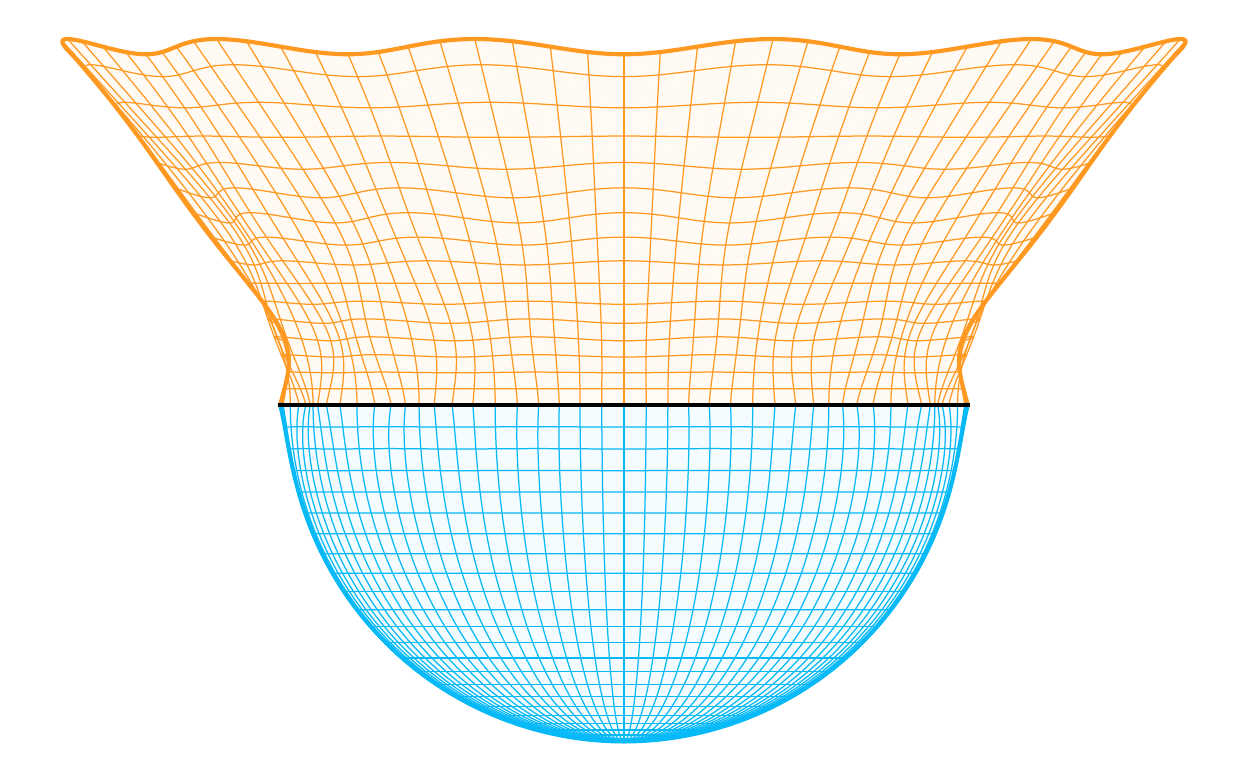}
\end{subfigure}%
\begin{subfigure}{.5\textwidth}
  \centering
  \includegraphics[height=0.65\linewidth]{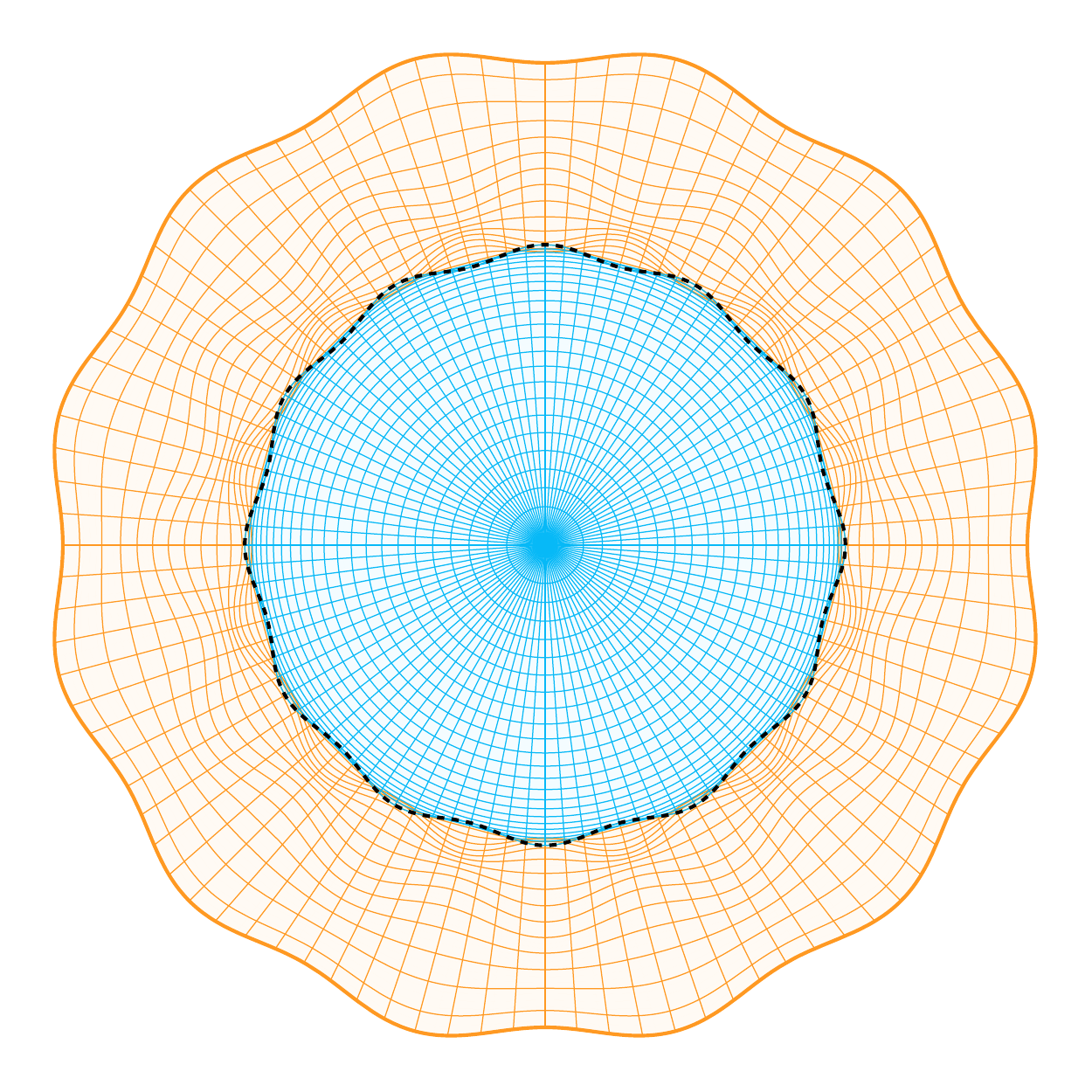}
  \label{fig:sub2}
\end{subfigure}
\caption{\justifying The RT surface for a small deformation of a ball-shaped subregion. The timelike (Lorentzian) section is shown in orange, while the spacelike (Euclidean) section is in blue. The curves are smoothly matched at $\tau = 0$, representing the junction between Lorentzian and Euclidean parts of the extremal surface. The plot is schematic and not drawn to scale. (Left) The RT surface front view (Right) The RT surface top view.}
\label{fig:perturbedRT}
\end{figure*}

We now turn our attention to the spacelike part. After Wick-rotating \eqref{eq:GdefLT}, we obtain
\begin{equation}\label{eq:GdefLS}
\frac{ds^2_{\gamma}}{\Ls^2}= 
\left(1- {\tE'}^2  \cos^2{\tauE}\right)\diff\tauE^2 + \cos^2{\tauE} \left(\sin^2{\tE} + {\dtE}^2\right)\diff\phi^2  +2\tE'\dtE \cos^2{\tauE} \,\diff\tauE \diff\phi\,,   
\end{equation}
where now we denoted $\tE'= \partial_{\tauE} \tE(\tauE,\phi) $ and $\dtE = \partial_{\phi} \tE(\tauE, \phi) $. We assume again an expansion for the Euclidean time of the type \eqref{eq:tET} with separation of variables with a polar part expanded in Fourier modes \eqref{eq:phiF} and a Wick-rotated global time, i.e., $\tE(\tauE, \phi) = \pi/2+ \epsilon f^{(\text s)} (\tauE,\phi)$, with $f^{(\text s)} (\tauE,\phi)=\TauEl(\tauE)\Phi_\ell(\phi)$. After these considerations, the differential equation for $\TauEl(\tauE)$ reads
\begin{equation}\label{eq:diffeq2}
\left[(\ell^2-1) \sec^2{\tauE} + 3 \tan{\tauE}\frac{\diff}{\diff \tauE}- \frac{\diff^2}{d \tauE^2}\right]\TauEl(\tauE)=0\,.
\end{equation}
Two out of the four undetermined coefficients from these second order differential equations are obtained using the boundary conditions
\begin{equation}\label{eq:bdycond}
\Tau_\ell(\tau\rightarrow\infty)=1\,,\quad \TauEl(\tauE=\pi/2)=0\,,
\end{equation}
which correspond to fixing the radius at infinite conformal time in the Lorentzian patch and imposing a regularity condition at the bulk turning point, respectively. The remaining two coefficients are obtained using junction conditions at $\tau = \tauE = 0$, namely
\begin{equation}\label{eq:juncon}
    \tE(\tau=0, \phi)=\tE(\tauE=0, \phi)\,, \quad \hspace{0.5cm}\partial_{\tau}\tE(\tau=0, \phi) =  \iu \partial_{\tauE}\tE(\tauE=0, \phi)\,.
\end{equation}
These conditions ensure continuity and differentiability as the RT surface transitions from the timelike to the spacelike region.\footnote{The continuity and differentiability conditions are well-motivated as essentially the minimal surface has to be the same RT surface in both the Lorentzian and Euclidean dS sections, and there is no localized matter or codimension-three defect at the junction which could induce a jump in the derivative. Also, the condition of \eqref{eq:juncon} considers that the $\tau = 0$ interface is a submanifold which is shared between both the Euclidean and Lorentzian sections of dS; therefore, the continuity of the derivative incorporates the ``change of coordinates'' between $\tau$ and $\tau_E$, which has a ``Jacobian'' of $\iu$.  Assuming the former, the surface is uniquely fixed for both Euclidean and Lorentzian sections.} Employing Eqs.~\eqref{eq:bdycond} and \eqref{eq:juncon} we obtain for the timelike part
\begin{equation}\label{eq:tauT} 
\Tau_\ell(\tau) =  \frac{(\sech{\tau})^{\frac{3}{2}}}{\sqrt{2\pi}} \left[- \pi P_{\ell-1/2}^{3/2}\left(\tanh\tau\right) +  2 \iu  Q_{\ell-1/2}^{3/2}\left(\tanh\tau\right)\right]\,,\quad \infty<\tau<0\,,
\end{equation}
where $P_m^l(x)$ and $Q_m^l(x)$ are the associated Legendre polynomials of first and second kind, respectively. On the other hand, the spacelike part reads 
\begin{equation}\label{eq:tauS} 
\TauEl(\tauE) =(-1)^{\frac{(\ell-1)(2 \ell -3)}{2}}\frac{\ell+\sin\tauE}{\cos\tauE} \tan^\ell\left(\frac{\pi }{4}-\frac{\tauE}{2}\right) \,,\quad 0\leq\tauE\leq\pi/2\,.
\end{equation}
In Fig.~\ref{fig:perturbedRT} we plot the profile of these perturbations.

Having determined the embedding functions for the timelike and spacelike segments of the deformed RT surface, given respectively by
Eqs.~\eqref{eq:tauT} and \eqref{eq:tauS}, we can explicitly evaluate the bare area functional. This is accomplished by substituting the corresponding induced metrics \eqref{eq:GdefLT} and \eqref{eq:GdefLS} into the area functional and expanding to second order in the deformation parameter $\epsilon$. For the timelike portion of the surface, one finds
\begin{align}\label{eq:eqnoninvder4}
\mathbf A\left(\Sigma^{(\text t)}_A\right)&= \epsilon^2\frac{\Ls^2 }{2} \sum_\ell\ell(\ell^2 -1)\left[1 +(-1)^\ell\right] (a_\ell^2 + b_\ell^2)\notag\\
&- \iu \frac{2\pi\Ls^2}{\delta}- \iu \epsilon^2 \sum_\ell\frac{\Ls^2(\ell^2-1)}{2\delta} (a_\ell^2 + b_\ell^2) +\mathcal{O}\left(\delta\right)\,,
\end{align}
where $\delta$ denotes the UV regulator associated with the approach to the conformal boundary. By contrast, the contribution from the spacelike segment is finite and given by
\begin{equation}\label{eq:SrenAsphbare4}
    \mathbf A\left(\Sigma^{(\text s)}_A\right)= 2 \pi \Ls^2-\epsilon^2\frac{\Ls^2 }{2} \sum_\ell (-1)^\ell \ell(\ell^2 -1)(a_\ell^2 + b_\ell^2)\ .
\end{equation}
The total bare area of the RT surface, $\mathbf A(\RT)=\mathbf A(\Sigma^{(\text t)}_A)+\mathbf A(\Sigma^{(\text s)}_A)$, is therefore obtained by adding the two contributions, yielding

\begin{align}\label{eq:ArenDef4}
   \mathbf A\left(\RT\right)& = 2 \pi \Ls^2+\epsilon^2\frac{\Ls^2 }{2} \sum_\ell\ell(\ell^2 -1)(a_\ell^2 + b_\ell^2) - \iu \frac{2\pi\Ls^2}{\delta}- \iu \epsilon^2 \sum_\ell\frac{\Ls^2(\ell^2-1)}{2\delta} (a_\ell^2 + b_\ell^2) +\mathcal{O}\left(\delta\right)\,.
\end{align}
The expression \eqref{eq:ArenDef4} contains a finite real contribution together with imaginary divergences proportional to the UV cutoff $\delta$, which originate from the timelike segment of the RT surface. In order to define a meaningful holographic pseudoentropy, these divergences must be removed. This is accomplished by implementing the codimension-two counterterm prescription introduced in section~\ref{subsec:pseudo4D}, which renormalizes the area functional by adding an appropriate boundary Chern form.

For the deformed embedding specified in Eq.~\eqref{eq:tET}, the relevant boundary Chern form $\mathcal B_1^{\partial\RT}$ reads
\begin{align}\label{eq:B1def4D}
\mathcal B_1^{\partial \RT}= -\iu\,\frac{2 }{\delta}+ \iu \epsilon^2 \sum_\ell \frac{(\Phi_\ell(\phi)^2-\Phi_\ell'(\phi)^2)}{\delta},
\end{align}
where $\Phi_\ell(\phi)$ is defined in Eq.~\eqref{eq:phiF}. Substituting Eq.~\eqref{eq:B1def4D} into the renormalized area functional \eqref{eq:RenA4D} and performing the $\phi$ integration yields
\begin{align}
\mathbf A^\text{ren}\left(\RT\right)&=\mathbf A\left(\RT\right)+ \iu \frac{2\pi\Ls^2}{\delta}+ \iu \epsilon^2 \sum_\ell\frac{\Ls^2(\ell^2-1)}{2\delta} (a_\ell^2 + b_\ell^2) +\mathcal{O}\left(\delta\right)\, .
\end{align}
These terms precisely cancel the divergent contributions appearing in Eq.~\eqref{eq:ArenDef4}, leaving a finite and universal result, i.e.,
\begin{align}\label{eq:RenAdef4D}
\mathbf A^\text{ren}\left(\RT\right)=2 \pi \Ls^2+\epsilon^2\frac{\Ls^2 }{2} \sum_\ell\ell(\ell^2 -1)(a_\ell^2 + b_\ell^2) \, .
\end{align}
The renormalized pseudoentropy for a deformed disk-shaped $\mathbb B^{2}_\epsilon$ subregion  then follows from Eq.~\eqref{FAL4D}, giving
\begin{align}\label{eq:FGdeform4D}
      \mathsf{S}_\text{u}(\mathbb B^{2}_\epsilon)=-\frac{\pi\Ls^2}{2\GN}&-\frac{\Ls^2}{8\GN}\epsilon^2 \sum_\ell\ell(\ell^2 -1)(a_\ell^2 + b_\ell^2) \,.
\end{align}
This result allows for a natural extrapolation to a Mezei-like formula \cite{Allais:2014ata,Mezei:2014zla,Faulkner:2015csl} in dS$_4$/CFT$_3$. Firstly, the absence of the linear term in $\epsilon$ indicates that the sphere is a local extremum of $\mathsf{S}_\text{u}(A)$ ---as observed in AdS/CFT \cite{Fonda:2015nma,Anastasiou:2020smm,Anastasiou:2022pzm} and in unitary CFTs \cite{Bueno:2021fxb,Bueno:2023gey}. Furthermore, the leading shape-dependent correction, is controlled by the analytic continuation ($\left.\Ls\right|_{\text{AdS}}\rightarrow-\iu \left.\Ls\right|_{\text{dS}}$) of the coefficient of the stress tensor two-point function $C_T$ in AdS/CFT, such that $\left.C_T\right|_{\text{AdS}}\rightarrow\left.C_T\right|_{\text{dS}}$. This indicates that $\left.C_T\right|_{\text{dS}}$ plays the same role for CFTs dual to Einstein-dS gravity.

By direct analogy with the AdS case, Eq.~\eqref{eq:FGdeform4D} can be rewritten as
\begin{align}\label{eq:FAdef4D}
   \mathsf{S}_\text{u}(\mathbb B^{2}_\epsilon) =-\frac{\pi\Ls^2}{2\GN}&+\frac{\pi^3C_T^{\text{dS}}}{24}\epsilon^2 \sum_\ell\ell(\ell^2 -1)(a_\ell^2+ b_\ell^2)\,,
\end{align}
where
\begin{equation}
C_{T}^{\text{dS}}=-\frac{3\Ls^2}{\pi^3\GN} \, .
\end{equation}
Following the universality relation for holographic theories found in Refs.~\cite{Mezei:2014zla} and extended to general (unitary) CFTs~\cite{Faulkner:2015csl}, the holographic pseudoentropy for shape deformations in dS$_4$/CFT$_3$ is expected to point out a universal behavior for a universality class of three-dimensional non-unitary Euclidean CFTs \cite{Anastasiou:2025dex}.

%%%%%%%%%%%%%%%%%%%%%
\section{Pseudoentropy from conformal renormalization: six-dimensional case}
\label{sec:6D}
%%%%%%%%%%%%%%%%%%%%%%

In this section we extend our analysis to six bulk dimensions. CG in six dimensions is substantially richer than in four dimensions, as it is constructed from three independent local conformal invariants~\cite{Bonora:1985cq,Deser:1993yx,Erdmenger:1997gy},
\begin{align}
I_1 &\equiv W_{\alpha\beta\gamma\delta}
      W^{\alpha\lambda\eta\beta}
      W_{\lambda}{}^{\gamma\delta}{}_{\eta}, \label{eq:I1} \\
I_2 &\equiv W_{\alpha\beta\gamma\delta}
      W^{\gamma\delta\lambda\eta}
      W_{\lambda\eta}{}^{\alpha\beta}, \label{eq:I2} \\
I_3 &\equiv
W_{\alpha\gamma\delta\lambda}
\left(
\delta^\alpha_\beta \dal
+ 4 R^\alpha{}_\beta
- \frac{6}{5}\delta^\alpha_\beta R
\right)
W^{\beta\gamma\delta\lambda}
+ \nabla_\alpha J^\alpha ,
\label{eq:I3}
\end{align}
where the total derivative involves
\begin{equation}
J_\alpha \equiv
R_\alpha{}^{\beta\gamma\delta}\nabla^\lambda R_{\lambda\beta\gamma\delta}
+3 R_{\beta\gamma\delta\lambda}\nabla_\alpha R^{\beta\gamma\delta\lambda}
- R_{\beta\gamma}\nabla_\alpha R^{\beta\gamma}
+\frac{1}{2}R\nabla_\alpha R
- R_\alpha^\beta \nabla_\beta R
+2 R_{\beta\gamma}\nabla^\beta R^\gamma_\alpha .
\end{equation}
The divergence term in \eqref{eq:I3} does not affect the bulk equations of motion, but it plays a crucial role in rendering the variational principle well defined.

Among all possible linear combinations of these invariants, Lü, Pang and Pope identified a unique one that admits Schwarzschild-AdS spacetimes as exact solutions \cite{Lu:2011ks}. This distinguished combination,
\begin{equation}
\mathcal{L}_{\rm CG}
= 4 I_1 + I_2 - \frac{1}{3} I_3 ,
\label{eq:L6D}
\end{equation}
defines what is commonly referred to as LPP CG. It was later shown that the same combination admits Einstein spacetimes in general~\cite{Anastasiou:2020mik,Anastasiou:2023oro}. Following conformal renormalization, this property makes it possible to construct the renormalized Einstein--(A)dS action directly from CG, without invoking the standard holographic counterterm prescription \cite{Anastasiou:2020mik,Anastasiou:2023oro,Anastasiou:2024rxe}. We exploit this structure to identify the conformally invariant codimension-two counterterms relevant for entanglement entropy and pseudoentropy.

The full six--dimensional CG action can be written as \cite{Anastasiou:2020mik}
\begin{align}
I_{\rm LPP}
&= \alpha_{\rm CG} \int_{\mathcal M} \left[W^3
\nonumber
+\frac{1}{2}
\delta^{\beta_1\cdots\beta_5}_{\alpha_1\cdots\alpha_5}
W^{\alpha_1\alpha_2}_{\beta_1\beta_2}
W^{\alpha_3\alpha_4}_{\beta_3\beta_4}
S^{\alpha_5}_{\beta_5}
+8 C^{\alpha\beta\lambda} C_{\alpha\beta\lambda}
\right]
\nonumber\\
&\quad
+ \alpha_{\rm CG} \int_{\partial\mathcal M}n_\alpha
\left(
8 W^{\alpha\kappa\lambda\beta} C_{\kappa\lambda\beta}
- W^{\kappa\lambda}_{\beta\sigma}
\nabla^\alpha W^{\beta\sigma}_{\kappa\lambda}
\right)\,,
\label{6DCGaction}
\end{align}
where now $W^3=
\delta^{\beta_1\cdots\beta_6}_{\alpha_1\cdots\alpha_6}
W^{\alpha_1\alpha_2}{}_{\beta_1\beta_2}
W^{\alpha_3\alpha_4}{}_{\beta_3\beta_4}
W^{\alpha_5\alpha_6}{}_{\beta_5\beta_6}/4!$ and $C_{\alpha\beta\lambda}$ is the Cotton tensor. Evaluating the action on Einstein spacetimes, such that
\begin{equation}
R_{\alpha\beta} = -\frac{5\sigma}{\Ls^2}\,\mathcal{G}_{\alpha\beta},
\qquad
W_{\alpha\beta}^{\gamma\delta} = F_{\alpha\beta}^{\gamma\delta},
\end{equation}
yields\footnote{Here we added a topological contribution to correctly reproduce the thermodynamic properties of Einstein spaces \cite{Anastasiou:2023oro}.} \cite{Anastasiou:2020mik,Anastasiou:2023oro,Anastasiou:2024rxe}
\begin{align}
I_{\rm LPP}\left(E\right)
&=
-4!\alpha_{\rm CG} \int_{\mathcal M}P_6\!\left(F\right)-\frac{\alpha_{\rm CG}}{2}
\int_{\partial\mathcal M}n^\alpha J_\alpha\big|_{\rm E}
-2(4\pi)^3 \alpha_{\rm CG} \chi(\mathcal M)\,,
\end{align}
where
\begin{equation}
P_6 \left(F\right)
=\frac{\sigma}{2(4!)\Ls^2}
\delta_{\alpha_{1} \ldots \alpha_{4}}^{\beta_{1}\ldots \beta_{4}}
F_{\beta_{1} \beta_{2}}^{\alpha_{1} \alpha_{2}}
F_{\beta_{3} \beta_{4}}^{\alpha_{3} \alpha_{4}}
-\frac{1}{(4!)^2}
\delta_{\alpha_{1} \ldots \alpha_{6}}^{\beta_{1}\ldots \beta_{6}}
F_{\beta_{1} \beta_{2}}^{\alpha_{1} \alpha_{2}}
F_{\beta_{3} \beta_{4}}^{\alpha_{3} \alpha_{4}}
F_{\beta_{5} \beta_{6}}^{\alpha_{5} \alpha_{6}} ,
\label{P6WE}
\end{equation}
and
\begin{equation}
J_\alpha\big|_{\rm E}
= \frac{1}{2}\nabla_\alpha
\big(
F^{\beta\gamma}_{\delta\kappa}
F^{\delta\kappa}_{\beta\gamma}
\big).
\end{equation}
As in four dimensions, requiring a Neumann boundary condition on the induced metric in CG removes the non-Einstein degrees of freedom, which has the dual CFT interpretation of setting the other sources beyond that of the stress-tensor to zero \cite{Anastasiou:2020mik,Anastasiou:2023oro,Hell:2023rbf}.

After evaluating the LPP CG action on Einstein manifolds, and choosing the coupling $\alpha_{\rm CG} = -\Ls^4/(384\pi \GN)$, one recovers the renormalized Einstein--(A)dS action, including the Euler term and a boundary counterterm quadratic in the boundary Weyl tensor, i.e. $I_{\text{LPP}} \left(E\right)=\IEren$. In particular, for Al(A)dS spacetimes one finds
\begin{equation}
\small
\IEren=\frac{1}{16\pi \GN}\int_{\mathcal{M}}\left( R+\frac{20 \sigma}{\Ls^{2}}-\frac{\Ls ^{4}}{72}\mathcal{X}_{6} +\frac{\Ls ^{4}}{48} \dal F^2 \right)+ \frac{\pi^2\Ls^4}{3\GN}\chi \left(\mathcal M\right) \,,
\end{equation}
where $\mathcal{X}_{6}=\frac{1}{8}\delta^{\alpha_1\ldots\alpha_{6}}_{\beta_1\ldots\beta_{6}}R^{\beta_1\beta_2}_{\alpha_1\alpha_2} \ldots R^{\beta_{5}\beta_{6}}_{\alpha_{5}\alpha_{6}}$ is the six-dimensional Euler density. In the following, we exploit this structure to construct codimension-two counterterms relevant for entanglement entropy and pseudoentropy.

\subsection{Codimension-two local conformal invariants in six dimensions}\label{sec:cod2-6D}

We now derive the codimension-two functional associated with six-dimensional CG, extending the construction presented in four dimensions. The starting point is the unique six-dimensional CG action that admits Einstein--(A)dS solutions as an exact subsector, and which can be written as \cite{Anastasiou:2020mik}
\begin{equation}
I_{\text{LPP}} = -\frac{\Ls^4}{384 \pi \GN} \int_{\mathcal M}\mathcal{C} + \frac{\pi^{2} \Ls^4}{3 \GN} \, \chi(\mathcal M), \label{eq:LPP2}
\end{equation}
where $\mathcal{C}=4I_1+I_2-\frac{1}{3}I_3$ is a specific combination of the conformal invariants of Eqs.~\eqref{eq:I1}-\eqref{eq:I3}. To extract the associated codimension-two invariant, we evaluate the action on an orbifold geometry $\mathcal M^{(\vartheta)}$ which then decomposes as
\begin{equation}
I_{\text{LPP}}\!\left(\mathcal M^{(\vartheta)}\right)
= I_{\text{LPP}}(\mathcal M) + \frac{1-\vartheta}{4\GN}\,\mathbf F(\Sigma),
\end{equation}
where $\Sigma$ denotes the locus of the conical singularity. The functional $\mathbf F(\Sigma)$ captures the universal contribution associated with the defect and is given by
\begin{equation}\label{eq:FS}
\mathbf F(\Sigma) = -\frac{\Ls^4}{48} \int_\Sigma\mathcal C_\Sigma + \frac{4}{3}\pi^2 \Ls^4 \chi(\Sigma).
\end{equation}
The defect density $\mathcal C_\Sigma$ arises from the expansion
\begin{equation}
\mathcal C\!\left(\mathcal M^{(\vartheta)}\right)
= \mathcal C(\mathcal M) + 2\pi(1-\vartheta)\mathcal C_\Sigma,
\qquad \mathcal C_\Sigma = 4F_1 + F_2 - \frac{1}{3}F_3 ,
\end{equation}
where the codimension-two invariants $F_i$ are induced by the corresponding bulk conformal invariants $I_i$. Their explicit expressions were firstly computed in Ref.~\cite{Miao:2020oey} and are presented in appendix~\ref{app:cod2-6D}. In deriving this result, we have also employed the decomposition of the Euler characteristic in the presence of conical defects, $\chi(\mathcal M^{(\vartheta)})=\chi(\mathcal M)+(1-\vartheta)\chi(\Sigma)$ 
\cite{Fursaev:1995ef,Fursaev:2013fta}. 

The functional $\mathbf F(\Sigma)$ constitutes the natural four-dimensional generalization of the Graham--Witten action $\mathbf L(\Sigma)$ and is often referred to as the Graham--Reichert action\footnote{In AdS space ($\sigma=1$), this term corresponds to the so-called Graham-Reichert anomaly, which is related to holographic entanglement entropy and a higher-dimensional version of Willmore energy \cite{Anastasiou:2024rxe}---see also~\cite{guven2005conformally,Gover:2016buc,Graham:2017bew,zhang2017graham,Blitz:2021qbp,Olanipekun:2021htq,2023arXiv230811433M,bernard2024analysis,Boulanger:2025oli,Wu:2025zbf,Lan:2025evd}.} \cite{Graham:2017bew}. When evaluated on Einstein backgrounds, it can be decomposed as
\begin{equation}\label{eq:IfromF}
\mathbf F(\Sigma)\big|_{\rm E} = \Ls^4\,\mathbf I(\Sigma)
+ \frac{4 \pi^2 \Ls^4}{3}\,\chi(\Sigma),
\end{equation}
such that\footnote{The details of this calculation are presented in Appendix~\ref{app:cod2-6D}.}
\begin{align}\label{eq:RHM5}
\mathbf I(\Sigma) =&\frac{1}{48}\int_{\Sigma}\bigg[ \frac{48}{\Ls^{4}}
-\mathcal X_4^{\Sigma} + (\partial K)^2 - K K^{ab} K K_{ab}
+ \frac{7}{16}K^4 - \frac{6\sigma}{\Ls^{2}}K^2 \nonumber\\
&\hspace{1.2cm}
- K_{A}K^{B} \Big(\mathcal R_{iB}^{iA} + \frac{\sigma}{\Ls^2}\delta_{iB}^{iA}\Big) \bigg] - \frac{1}{6\sqrt{\sigma}\Ls}
\int_{\partial\Sigma}\mathcal K_{\partial\Sigma}\big|_{\rm E},
\end{align}
where $\mathcal X_4^{\Sigma}=\frac{1}{4}\delta_{abcd}^{efgh}\mathcal R_{ef}^{ab}\mathcal R_{gh}^{cd}$ denotes the Euler density intrinsic to $\Sigma$, $(\partial K)^2\equiv\partial^a K^A\partial_a K_A$,
$K K^{ab} K K_{ab}\equiv K^{A}K_{A}{}^{ab}K_{B}K^{B}{}_{ab}$, $K^4\equiv (K^{A}K_{A})^{2}$ and $K_{\partial\Sigma}\big|_{\text E}=w_{ij}^{ij} -\kappa^{I}{}_{\langle ij\rangle}\kappa_{I}{}^{\langle ij\rangle}$, as defined in Eq.~\eqref{eq:btE}. $\mathbf F(\Sigma)$ provides the four-dimensional codimension-two conformal invariant analogous to $\mathbf L(\Sigma)$ in (A)dS$_4$ and will play a central role in our construction of renormalized pseudoentropy in the dS$_6$/CFT$_5$ correspondence.

%%%%%%%%%%%%%%%%%%%%%%%%%%%%%%%
\subsection{Renormalized pseudoentropy for dS$_6$/CFT$_5$}
%%%%%%%%%%%%%%%%%%%%%%%%%%%%%%%%%%%

Having identified $\mathbf F(\Sigma)$, we now particularize to the case in which the bulk is an Einstein--dS manifold---this is, with $\sigma=-1$---and $\Sigma$ is an extremal surface. This restriction is of direct relevance for holographic pseudoentropy, since RT surfaces form a distinguished subset of extremal codimension-two submanifolds.

When $\Sigma$ is extremal (i.e., $K^{A}=0$), which we denote by $\Sigma_{\mathrm{ext}}$, the general expression for $\mathbf F(\Sigma)$ obtained in subsection~\ref{sec:cod2-6D} simplifies dramatically. In particular, upon evaluation on Einstein backgrounds one finds that the defect functional reduces to the renormalized area
\begin{equation}
\mathbf F(\Sigma_{\mathrm{ext}})\big|_{\mathrm E}
= \mathbf A^{\mathrm{ren}}(\Sigma_{\mathrm{ext}})
\end{equation}
where
\begin{equation}\label{eq:renA4}
\mathbf A^{\mathrm{ren}}(\Sigma_{\mathrm{ext}})
= \mathbf A(\Sigma_{\mathrm{ext}})
-\frac{\Ls^4}{24}\int_{\Sigma_{\mathrm{ext}}}
\mathcal X_4^{\Sigma}
+\frac{4\pi^2\Ls^4}{3}\chi(\Sigma_{\mathrm{ext}})
-\frac{\iu \Ls^3}{6}\int_{\partial\Sigma_{\mathrm{ext}}}\mathcal K_{\partial\Sigma_{\mathrm{ext}}}.
\end{equation}
Using the de Sitter curvature tensor $F_{\Sigma}$ of the extremal surface $\Sigma_{\mathrm{ext}}$, given in terms on the intrinsic Riemann tensor by
\begin{equation}\label{dS_curv_Sigma}
\left(F_{\Sigma}\right)_{ab}^{ cd}
=
\mathcal R_{ab}^{ cd}
-\frac{1}{\Ls^{2}}\,
\delta_{ab}^{cd},
\end{equation}
the Euler density of $\Sigma_{\mathrm{ext}}$ can be expressed as
\begin{equation}
\mathcal X_4^{\Sigma}
=
F_{\Sigma}^2
+\frac{24}{\Ls^{4}}.
\end{equation}
Substituting this relation into \eqref{eq:renA4}, and using $\mathbf A(\Sigma_{\mathrm{ext}})
=\int_{\Sigma_{\mathrm{ext}}}
\diff^4 y\,\sqrt{\gamma}$, we obtain

\begin{equation} \label{eq:renA4inF}
\mathbf A^{\mathrm{ren}}(\Sigma_{\mathrm{ext}})
=
-\frac{\Ls^4}{24}
\int_{\Sigma_{\mathrm{ext}}}
F_{\Sigma}^2
+\frac{4\pi^2\Ls^4}{3}\,
\chi(\Sigma_{\mathrm{ext}})
-\frac{\iu \Ls^3}{6}
\int_{\partial\Sigma_{\mathrm{ext}}}
\mathcal K_{\partial\Sigma_{\mathrm{ext}}}.
\end{equation}
For conformally compact extremal surfaces, such as RT surfaces $\Sigma_A$ anchored at the conformal boundary, one can exchange the Euler density and the Euler characteristic in \eqref{eq:renA4} by a boundary term using the Gauss--Bonnet theorem, yielding 
\begin{equation}\label{eq:renA4nc}
\mathbf A^{\mathrm{ren}}(\RT) = \mathbf A(\RT)
-\int_{\partial\RT}\left[ \frac{\Ls^4}{24}\mathcal B_3^{\RT} +\frac{\iu \Ls^3}{6}\mathcal K_{\partial\RT} \right],
\end{equation}
where $\sqrt{\mathfrak h}\mathcal B_3^{\RT}=-2/3\sqrt{\mathfrak h }\,
\delta^{lmn}_{ijk}\, \mathfrak K^i{}_l \left( 3\mathfrak R^{jk}{}_{mn} -2\mathfrak K^j{}_m\mathfrak K^k{}_n \right)$ denotes the second Chern form associated with the boundary  of the four-dimensional surface $\RT$.

Finally the finite contribution to the holographic pseudoentropy in five boundary dimensions follows directly from Eq.~\eqref{eq:renA4nc}, what reads
\begin{equation}\label{FAL5}
\mathsf{S}_\text{u}(A) =\frac{\mathbf A^{\mathrm{ren}}(\RT)}{4\GN}
=\frac{1}{4\GN}\,\mathbf F(\RT)\big|_{\mathrm E}.
\end{equation}
Proceeding similarly as before, we explicitly show the validity of this expression for spherical entangling surfaces and deformations thereof.

%%%%%%%%%%%%%%%%%
\subsection{Explicit examples}
\subsubsection{Spherical entangling surface}
We begin by evaluating the renormalized area of the RT surface, $\mathbf A^{\mathrm{ren}}(\RT)$, associated with a four-ball entangling region, $\mathbb B^4$, in the non-unitary CFT dual to dS$_6$. As in the four-dimensional case, spherical entangling regions can be treated analytically. 

The bulk geometry is pure dS$_6$, whose global Lorentzian metric is
\begin{equation}\label{eq:GdS1-6D} 
\diff s^2_\mathcal{G}=\mathcal{G}_{\alpha\beta}\diff x^\alpha\diff x^\beta=\Ls^2 \left(
-\diff \tau^2 + \cosh ^2 \tau \left[ \diff \tE^2
 +\sin^2 \tE( \diff \theta^2 + \cos^2 \theta \,\diff \Omega^2_3)\right]\right)\,,
\end{equation}
where $\diff\Omega_3^2=\diff\theta_1^2+\sin^2\theta_1\left(\diff\theta_2^2+\sin^2\theta_2\diff\theta_3^2\right)$ denotes the line element on the unit three-sphere, with angular coordinates $0\leq\theta_1\leq\pi$, $0\leq\theta_2\leq\pi$ and $0\leq\theta_3\leq2\pi$. Following the standard prescription, the region $-\infty<\tau<0$ is replaced by its Euclidean continuation obtained via the Wick rotation $\tau=\iu\tau_E$, yielding
\begin{equation}\label{eq:GdT1} 
\diff s^2_g=\Ls^2 \left(
\diff \tauE^2 + \cos ^2 \tauE \left[ \diff \tE^2
 +\sin^2 \tE( \diff \theta^2 + \cos^2 \theta \,\diff \Omega^2_3)\right]\right)\,,
\end{equation}
with the Lorentzian and Euclidean geometries glued at $\tau=\tau_E=0$.

For this choice of entangling region, the RT surface $\RT=\{\Sigma^{(\text t)}_A\cup\Sigma^{(\text s)}_A,\theta=0\}$, with $\Sigma^{(\text t)}_A$ and $\Sigma^{(\text s)}_A$ as defined in \eqref{eq:SigmaA}, is described by the induced metric
\begin{equation}\label{eq:sphind}
    \quad \frac{\diff s^2_{\gamma}}{\Ls^2}=  \begin{cases} \displaystyle
\frac{\sin^2{T_0}\,\diff\tau^2}{\cos^2{T_0}\tanh^2{\tau}-1} +  ({\cosh^2{\tau}-\cos^2{T_0} \sinh^2{\tau}}) \diff \Omega^2_3\,,  &  0 < \tau < \infty  \\\\\displaystyle
\frac{(1+C^2)\diff\tauE^2}{1-C^2 \tan^2{\tauE}}  + (\cos^2{\tauE}- C^2 \sin^2{\tauE})  \diff \Omega^2_3\,,  & 0\leq\tauE < \tan^{-1} \frac{1}{C} \, .
\end{cases}  
\end{equation}

For the embedding under consideration, the finite part of the holographic pseudoentropy follows from the renormalized area computed according to the general prescription \eqref{eq:renA4inF}.
Using the induced metric given in \eqref{eq:sphind}, we compute the dS curvature tensor \eqref{dS_curv_Sigma} associated with $\Sigma_A$ and find that it vanishes identically, i.e.
\begin{equation}
\left(F_{\Sigma_A}\right)_{ab}^{cd}=0 \, .
\end{equation}
Hence, there is no curvature contribution to the renormalized area. In addition, the boundary term vanishes for the geometry under consideration. Indeed, since $ w_{ij}^{ij}=0$ and 
$\kappa^{I}{}_{\langle ij\rangle}=0 \,,$ it follows directly that
\begin{equation}
    \mathcal K_{\partial\RT}=0.
\end{equation}
Because $\Sigma_A$ is topologically a four-dimensional ball, its Euler characteristic is $\chi(\Sigma_A)=1$. As a result, the renormalized area simplifies to
\begin{equation}
    \mathbf A^{\mathrm{ren}}(\RT)
    =
    \frac{4\pi^2 \Ls^4}{3} \, .
\end{equation}
The renormalized area is therefore entirely given by the topological contribution (for holographic entanglement entropy in AdS, the same is true as shown in Ref.~\cite{Anastasiou:2020smm}). Consequently, Eq.~\eqref{eq:renA4} reproduces the universal contribution to the
four-ball pseudoentropy \cite{Doi:2023zaf,Anastasiou:2025rvz}
\begin{equation}\label{eq:F0}
   \mathsf{S}_\text{u}\left(\mathbb B^4\right)=\frac{\pi^2\Ls^4}{3\GN}\, .
\end{equation}
We now turn to entangling surfaces with reduced symmetry in order to further test the general formalism.

%%%%%%%%%%%%%%%%%%%%%%%%%%%%%%%
\subsubsection{Small deformations of the sphere}\label{sec:defs}
%%%%%%%%%%%%%%%%%%%%%%%%%%%%%%%%%%
Small deformations of the spherical entangling surface induce a nontrivial shape dependence governed by the stress-tensor two-point function coefficient. The computation closely parallels the four-dimensional analysis, with the renormalized area functional now given by Eq.~\eqref{eq:renA4nc}.

We consider a four-ball entangling region of unit radius and introduce infinitesimal deformations in the angular direction $\theta_{1}$. We denote the resulting region by $\mathbb B^{4}_{\epsilon}$ and parametrize the deformation as
\begin{equation}\label{eq:RTdef}    \mathbb B_\epsilon^4:\ \tE(\theta_1)=\frac{\pi}{2} + \epsilon \sum_{\ell} a_{\ell}Y_\ell(\theta_1)\,, \quad Y_\ell(\theta_1)=\frac{1}{2\pi^2\sqrt{\sin\theta_1}}Q_{\ell+\frac{1}{2}}^{\frac{1}{2}}\left(\cos\theta_1\right)\, ,
\end{equation}
where $a_{\ell}$ controls the amplitude of the deformation. This choice corresponds to a restricted class of deformations considered in Refs.~\cite{Mezei:2014zla,Anastasiou:2025rvz}.\footnote{We restrict to this class of deformations for convenience. In arbitrary dimensions, the pseudoentropy for a deformed subregion depends only on the magnitude angular integer $\ell$ ~\cite{Anastasiou:2025rvz}. More generally, for $(d-2)$-dimensional spherical harmonics $Y_{\ell,\mathbf m}(\Omega_{d-2})$, the result is independent of the orientation angular numbers $\mathbf m$ due to rotational symmetry. Therefore, this restricted class suffices to determine the general result, as in the case of entanglement entropy in AdS ~\cite{Mezei:2014zla}.} The embedding of the corresponding RT surface---$\tE=\tE(\tau,\Omega_{3})$ for the timelike segment and $\tE=\tE(\tauE,\Omega_{3})$ for the spacelike segment---is given by
\begin{align}\label{eq:emb6D}
\tE=\frac{\pi}{2}+ \epsilon \sum_{\ell} a_{\ell}Y_\ell(\theta_1)\times\begin{cases}
        \Tau_{\ell}(\tau) & \text{for } \tau\in(\infty,0)\,,\\
        \TauEl(\tauE) & \text{for } \tauE\in(0,\frac{\pi}{2})\,.
    \end{cases}
\end{align}
The functions $\Tau_{\ell}(\tau)$ and $\TauEl(\tauE)$ satisfy second-order differential equations derived from extremality, subject to the appropriate boundary and junction conditions given previously in Eqs.~\eqref{eq:bdycond} and \eqref{eq:juncon}. For $d=5$, the timelike solution reads
\begin{equation}\label{eq:tauT6} 
 \Tau_{\ell}(\tau) =  \frac{(\sech{\tau})^{\frac{5}{2}}}{3\sqrt{2\pi}} \left[ \pi P_{\ell+1/2}^{5/2}\left(\tanh\tau\right) -  2 \iu  Q_{\ell+1/2}^{5/2}\left(\tanh\tau\right)\right]\,,\quad \infty<\tau<0\,,
\end{equation}
while the spacelike solution obtains the form
\begin{equation}\label{eq:tauS6} 
\TauEl(\tauE) =(-1)^{\frac{(\ell-1)(2 \ell -3)}{2}}\frac{(\ell+3\sin\tauE)(\ell+2+\sin\tauE)}{3\cos\tauE (1+\sin\tauE)} \tan^\ell\left(\frac{\pi }{4}-\frac{\tauE}{2}\right) \,,\quad 0\leq\tauE\leq\pi/2\,.
\end{equation}
Using these embeddings, the bare area of the deformed RT surface can be computed perturbatively. The contribution from the timelike component is
\begin{align}\label{eq:eqnoninvder6}
\mathbf A\left(\Sigma^{(\text t)}_A\right)&=\epsilon^2\frac{\Ls^4}{18\pi}\sum_\ell a_\ell^2\,  \left[1 +(-1)^\ell\right]   \left(\ell-1\right)_{5}+\iu \frac{\pi^2\Ls^4}{3\, \delta^3}- \iu \frac{5\pi^2\Ls^4}{4 \, \delta}\notag\\
& + \iu \epsilon^2 \sum_\ell a_\ell^2 \frac{(\ell-1)(\ell+2)\Ls^4}{36\pi}\left[\frac{3}{\delta^3}-\frac{\left(\ell(\ell+1)(2\ell-1)\right)}{\delta}\right]+\mathcal{O}\left(\delta\right),
\end{align}
where $(x)_n \equiv \Gamma(x+n)/\Gamma(x)$ denotes the Pochhammer symbol. The spacelike contribution is finite and given by
\begin{equation}\label{eq:SrenAsphbare6}
    \mathbf A\left(\Sigma^{(\text s)}_A\right)=\frac{4\pi^2\Ls^4}{3}-\epsilon^2\frac{\Ls^4}{18\pi}\sum_\ell a_\ell^2\,  (-1)^\ell  \left(\ell-1\right)_{5}+\mathcal{O}\left(\epsilon^4\right)\ .
\end{equation}
Combining the two contributions yields the total bare area
\begin{align}\label{eq:ArenDef}
 \mathbf A\left(\RT\right)& =\frac{4\pi^2\Ls^4}{3}+\epsilon^2\frac{\Ls^4}{18\pi}\sum_\ell a_\ell^2\,     \left(\ell-1\right)_{5}+\iu \frac{\pi^2\Ls^4}{3\, \delta^3}- \iu \frac{5\pi^2\Ls^4}{4 \, \delta}\notag\\
& + \iu \epsilon^2 \sum_\ell a_\ell^2\frac{(\ell-1)(\ell+2)\Ls^4}{36\pi}\left[\frac{3}{\delta^3}-\frac{\left(\ell(\ell+1)(2\ell-1)\right)}{\delta}\right]+\mathcal{O}\left(\delta\right)\,.
\end{align}
We now verify that the renormalized area functional of Eq.~\eqref{eq:renA4nc} cancels all UV divergences. For the embedding \eqref{eq:emb6D}, the associated Chern form evaluates to
\begin{align}\label{eq:B3def6D}
\small
\mathcal B_3^{\partial \RT}
=&\;
\frac{4 \iu \cos^2\theta_1\cos\theta_2}{\delta^3}
-\frac{15 \iu \cos^2\theta_1\cos\theta_2}{\delta}
+\iu \epsilon^2 \sum_{\ell} a_\ell^2
\Biggl[
\frac{2 \cos^2\theta_1\cos\theta_2\,
\bigl({Y'_\ell}^2-3Y_\ell^2\bigr)}{\delta^3}
\notag\\[2pt]
&\hspace{1.5em}
-\frac{\cos\theta_2}{12\,\delta}
\Biggl(
2\bigl(9+8(\ell-1)_2(\ell+2)_2\bigr)
\cos^2\theta_1\,Y_\ell^2
\notag\\
&\hspace{2.5em}
+32\bigl(3+\ell(\ell+2)\bigr)\cos\theta_1\,Y_\ell
\bigl(-2\sin\theta_1\,Y'_\ell
+\cos\theta_1\,Y''_\ell\bigr)
\notag\\
&\hspace{2.5em}
+Y'_\ell
\Bigl[
\bigl(69+8\ell(\ell+2)
+(-27+8\ell(\ell+2))\cos2\theta_1\bigr)Y'_\ell -96\sin2\theta_1\,Y''_\ell
\Bigr]
\Biggr)
\Biggr]
+\mathcal O(\delta)\, .
\end{align}
For the remaining boundary contribution in Eq.~\eqref{eq:renA4nc}, we require the partial trace of the Weyl tensor and the quadratic contraction of the traceless extrinsic curvature of $\partial\Sigma$, as given in Eq.~\eqref{eq:btE}. The Weyl contribution vanishes identically, i.e.,
$w_{ij}^{ij}=0$, while the extrinsic curvature term is given by
\begin{equation}\label{eq:TraceKdef6D}
\mathfrak{\kappa}^{I}{}_{\langle ij\rangle}\mathfrak{\kappa}_{I}{}^{\langle ij\rangle}= -\frac{4\delta^{2}\epsilon^{2}}{3\Ls^{2}}
\sum_{\ell}a_\ell^2\left(\tan\theta_{1}\,Y_\ell'(\theta_{1})+Y_\ell''(\theta_{1})\right)^{2}\,.   
\end{equation}
Substituting Eqs.~\eqref{eq:B3def6D} and \eqref{eq:TraceKdef6D} into Eq.~\eqref{eq:renA4nc}, we obtain
\begin{align}
\mathbf A^\text{ren}\left(\RT\right)&=\mathbf A\left(\RT\right)-\iu \frac{\pi^2\Ls^4}{3\, \delta^3}+ \iu \frac{5\pi^2\Ls^4}{4 \, \delta}\notag\\
& - \iu \epsilon^2 \sum_\ell a_\ell^2 \frac{(\ell-1)(\ell+2)\Ls^4}{36\pi}\left[\frac{3}{\delta^3}-\frac{\left(\ell(\ell+1)(2\ell-1)\right)}{\delta}\right]+\mathcal{O}\left(\delta\right)\, ,
\end{align}
which exactly cancels the divergences in Eq.~\eqref{eq:ArenDef}, leaving the finite contribution
\begin{align}\label{eq:RenAdef6D}
\mathbf A^\text{ren}\left(\RT\right)=\frac{4\pi^2\Ls^4}{3}+\epsilon^2\frac{\Ls^4}{18\pi}\sum_\ell a_\ell^2\,     \left(\ell-1\right)_{5}\, .
\end{align}
Using Eq.~\eqref{eq:RenAdef6D} in Eq.~\eqref{FAL5}, the universal part of the pseudoentropy becomes
\begin{equation}\label{eq:FGdeform6D}
\mathsf{S}_\text{u}(\mathbb B^{4}_\epsilon)=\frac{\pi^2\Ls^4}{3\GN}+\epsilon^2\frac{\Ls^4}{72\pi \GN}\sum_\ell a_\ell^2\,     \left(\ell-1\right)_{5}+\mathcal{O}(\epsilon^4)\, .
\end{equation}
This result can again be viewed as the dS analogue of Mezei's general formula governing shape deformations of spherical entangling surfaces \cite{Mezei:2014zla,Allais:2014ata,Anastasiou:2025rvz}. Indeed,
rewriting Eq.~\eqref{eq:FGdeform6D} in terms of the stress-tensor two-point function coefficient yields

\begin{equation}\label{eq:FAdef6D}
   \mathsf{S}_\text{u}(\mathbb B^{4}_\epsilon)=\frac{\pi^2\Ls^4}{3\GN}+\epsilon^2\frac{\pi^3}{2160}C_T\sum_\ell a_\ell^2(\ell-1)_5+\mathcal O \left(\epsilon^4\right)\, ,
\end{equation}
where for five-dimensional non-unitary CFTs dual to Einstein–dS gravity one has that $C_{T}=30\Ls^{4}/(\pi^{4}\GN)$. The dependence on the
deformation parameter, angular multipole number and overall normalization matches precisely the expected universal structure.

\section{Conclusions}\label{sec:conclusions}

In this work we developed a systematic framework for computing the finite and universal part of the holographic pseudoentropy in the context of the dS/CFT correspondence, for bulk manifolds of four and six dimensions, via conformal renormalization. By exploiting the fact that the renormalized Einstein–dS action can be obtained from a combination of bulk conformal invariants when evaluated on Einstein manifolds \cite{Lu:2013hx}, we derive the corresponding codimension-two functional resulting from evaluating the invariants on the replica orbifold. From there, following the same argument as in the AdS case discussed in Ref.~\cite{Anastasiou:2024rxe}, we obtain the renormalized area for the codimension-two RT surface, which inherits the bulk finiteness.

In this analysis we find that in four bulk dimensions, the renormalized pseudoentropy for a boundary subregion $A$ can be obtained from the Graham–Witten action $\mathbf L(\Sigma)$---Eq.~\eqref{eq:Lsigma}, which is a codimension-two conformal invariant and which reduces for Einstein–dS backgrounds and minimal surfaces to the renormalized area $\Aren (\Sigma_A)$. This functional, given in Eq.~\eqref{eq:LSAren}, is directly free from divergences and it corresponds to the finite universal part of the pseudoentropy.
In six bulk dimensions, the analogous construction is controlled by the Graham–Reichert action $\textbf{F}(\Sigma)$ of Eq.~\eqref{eq:FS}, which generalizes the four-dimensional result and provides a renormalized area for four-dimensional extremal surfaces in Einstein--dS$_6$ manifolds.

We applied this formalism to spherical entangling regions and to small shape deformations thereof. For the sphere, the renormalized pseudoentropy isolates the universal part which is proportional to the $a^\star$ charge according to Eqs.~\eqref{eq:su04D} and \eqref{eq:F0}. The leading correction in the deformation parameter, $\mathcal{O}\left( \epsilon^2 \right)$, is in precise agreement with the Mezei formula  analytically continued from AdS/CFT, as shown in Eqs.~\eqref{eq:FAdef4D} and \eqref{eq:FAdef6D}. It is worth emphasizing that, from the boundary perspective, the functional form of the shape dependence is largely fixed by conformal symmetry.\footnote{We thank the referee for this insightful comment.} In particular, as shown in ~\cite{Faulkner:2015csl}, shape deformations can be mapped to insertions of the stress-tensor, so that the leading $\mathcal{O}(\epsilon^2)$ correction is governed by the stress-tensor two-point function. The latter is completely determined by conformal symmetry up to the overall coefficient $C_T$. In this sense, the structure of the result does not rely on unitarity, which only imposes positivity constraints on $C_T$.

These findings, together with the results for dS holography in quadratic curvature gravity theories presented in Ref.~\cite{Anastasiou:2025rvz}, provide evidence on the universality of the Mezei relation, i.e., that the shape dependence of holographic pseudoentropy in dS/CFT mirrors that of entanglement entropy in AdS/CFT, reinforcing the interpretation that pseudoentropy in dS space captures universal data of the non-unitary dual CFT. Some of this data, corresponding to the bulk metric sector which is dual to the CFT stress-tensor, can be obtained from AdS by the direct analytic continuation of $L|_{\text{AdS}} \rightarrow \iu L |_{\text{dS}}$, as seen in our results.

In four bulk dimensions, the non-unitarity of the dual CFT is manifest in the negativity of $C_T$. In six dimensions, however, $C_T$ remains strictly positive, which might naively be interpreted as a sign of restored unitarity. This conclusion is premature: positivity of $C_T$ is only a necessary, but not sufficient, condition for unitarity, and violations may instead appear at higher orders in perturbation theory, which we have not explored. Notably, in our recent work \cite{Anastasiou:2025rvz}, we evaluated $C_T$ in arbitrary dimensions for CFTs dual to de Sitter space and found that, in odd bulk dimensions, $C_T$ is purely imaginary, unambiguously signaling non-unitarity.

Besides our findings, some issues remain unsolved. For example, there are other types of non-unitary CFTs which do not have holographic duals and which exhibit atypical properties such as having $C_T=0$ \cite{Gurarie:2004ce,dubail2010conformal}. It is unclear whether or not the universality of the shape deformation results would hold for this broader class of non-holographic non-unitary CFTs. Furthermore, the renormalized pseudoentropy defined here is intrinsically finite; however, its interpretation as a complex-valued information measure in non-Hermitian systems warrants further investigation. In particular, unlike the Willmore energy and the reduced Hawking mass---both of which can be derived from the Graham–Witten anomaly $\mathbf L(\Sigma)$ in AdS manifolds---the codimension-two functionals obtained in dS space do not obey any global bound or monotonicity property. This failure can be traced to their generically complex nature.

Future work could also extend the conformal renormalization approach to pseudo-Rényi entropies or study higher-order shape deformations around the sphere, which are possibly sensitive to three-point (or higher) stress-tensor coefficients. These explorations would allow the extraction of more information from holographic non-unitary CFTs, possibly leading to deeper insights into the holographic structure of dS space and its non-unitary dual theories.

\section*{Acknowledgements}
We thank Pablo Bueno and Tadashi Takayanagi for interesting discussions. This work is partially funded by ANID FONDECYT grants 11240059, 1240043, 1261016, 1231133, and 3230626. I.J.A. gratefully acknowledges support from the Simons Center for Geometry and Physics, Stony Brook University, at which some of the research for this work was performed. A.D. thanks the Departamento de Física y Astronomía at Universidad Andrés Bello and the Departamento de Ciencias at Universidad Adolfo Ibáñez for their hospitality during this work. The work of A.D. is supported by Becas de postgrado UNAP. The work of J.M. is supported by the Beatriu de Pinós fellowship BP 2024 00033 of the Agència de Gestió d'Ajuts Universitaris i de Recerca, Generalitat de Catalunya.

\appendix
\section{Notation and conventions}\label{app:A}

In this appendix we present the conventions used throughout the paper. In the first column of table~\ref{tab:notation}, we provide a list of objects defined on the different manifolds presented in the first line. The gravity theory is defined on the $D=(d+1)$-dimensional bulk manifold $\mathcal M$ and its dual CFT lives on its boundary $\partial \mathcal M$. We denote as $\Sigma$ the codimension-two bulk hypersurface which is cobordant with the entangling region $A$ (e.g. the RT surface $\Sigma_A$)  and its boundary as $\partial \Sigma=\partial A$. In Table \ref{tab:notation}, we also differentiate between various embeddings that can be defined for submanifolds, such as $\partial \Sigma$, which can be embedded in either $\Sigma$ or $\partial \mathcal M$.

\begin{table}[h!]
    \centering
    \begin{tabular}{|l|c|c|c|c|c|}
     \cline{2-6}
    \multicolumn{1}{c|}{}
 & $\mathcal{M}$ &  $\partial\mathcal{M} \subset \mathcal{M}$ & $\Sigma \subset \mathcal{M}$ & $\partial \Sigma \subset \Sigma$ & $\partial A = \partial \Sigma \subset \partial \mathcal{M}$ \\ \hline\hline
    Indices & $\alpha,\ldots,\lambda$ &$\mu,\ldots,\omega$& $a,\ldots,h$ & $i,\ldots,q$ & $i,\ldots,q$ \\
    Coordinates & $x^\alpha$ &$X^\mu$& $y^a$ & $Y^i$ & $Y^i$ \\
    Metric & $\mathcal{G}_{\alpha\beta}$ & $h_{\mu\nu}$ & $\gamma_{ab}$ & $\mathfrak h_{ij}$ & $\mathfrak h_{ij}$ \\
    Covariant derivative & $\nabla_\alpha$ & $\nabla_\mu^{\partial \mathcal M}$ & $\nabla_a^{\Sigma}$ & $\nabla_i^{\partial \Sigma}$ & $\nabla_i^{\partial \Sigma}$ \\
    Riemann tensor & $R_{\alpha\beta\gamma\delta}$ & $r_{\mu\nu\rho\sigma}$ & $\mathcal R_{abcd}$ & $\mathfrak R_{ijkl}$ & $\mathfrak R_{ijkl}$ \\
    Unit normal(s) &  & $n_{\alpha}$ & $N^{A}{}_{\alpha}$ & $\mathfrak{n}_{a}$ & $\mathfrak l^{I}{}_{\mu}$ \\
    Extrinsic curvature & & $k_{\mu\nu}$ &$K^{A}{}_{ab}$& $\mathfrak K_{ij}$ & $\kappa^{I}{}_{ij}$ \\\hline
    \end{tabular}
    \caption{Notation and conventions}
    \label{tab:notation}
\end{table}

\section{Finiteness of the bulk CG action}
\label{App:BulkFinite}

The four-dimensional conformal gravity (CG) action is
\begin{equation}
    I_{\rm CG}
    =
    \alpha_{\rm CG}
    \int_{\mathcal M} \diff^4 x \, \sqrt{\mathcal G}\,
    W_{\alpha\beta}^{\gamma\delta}
    W^{\alpha\beta}_{\gamma\delta} \, ,
    \label{eq:CGactionApp}
\end{equation}
where $W_{\alpha\beta}^{\gamma\delta}$ is the Weyl tensor of the bulk metric $\mathcal G_{\alpha\beta}$.
A key property of CG is that, for Al(A)dS spacetimes, the action is free of infrared divergences without the addition of boundary counterterms~\cite{Grumiller:2013mxa}. 
This can be established by a simple power-counting analysis in the holographic coordinate.

To make this explicit, we write the metric in Fefferman--Graham (FG) form (for AlAdS) or, equivalently, in the Starobinsky expansion (for AldS):
\begin{equation}
    ds^2
    =
    \frac{\Ls^2}{z^2}
    \left(
        \sigma\, dz^2
        +
        g_{\mu\nu}(z,X;\sigma)\, dX^\mu dX^\nu
    \right),
    \label{eq:FGmetricApp}
\end{equation}
where $\sigma=+1$ for AdS and $\sigma=-1$ for dS, and $\Ls$ denotes the (A)dS radius. 
The conformal boundary is located at $z=0$. 
For notational simplicity, we use $z$ for both the spatial radial coordinate in AdS and the timelike radial coordinate in dS.

The induced metric admits the near-boundary expansion
\begin{equation}
   g_{\mu\nu}(z,X;\sigma)
   =
   g^{(0)}_{\mu\nu}(X;\sigma)
   + \frac{z}{\Ls} g^{(1)}_{\mu\nu}
   + \frac{z^2}{\Ls^2} g^{(2)}_{\mu\nu}
   + \frac{z^3}{\Ls^3} g^{(3)}_{\mu\nu}
   + \mathcal O(z^4) .
   \label{eq:FGexpansionApp}
\end{equation}
Here $g^{(0)}_{\mu\nu}$ is the boundary metric and acts as the source for the holographic stress tensor.
The coefficient $g^{(1)}_{\mu\nu}$ arises from the higher-derivative nature of the CG equations of motion and is absent in Einstein gravity. 
In the holographic dictionary for CG, this mode sources a partially massless response~\cite{Grumiller:2013mxa}. 
More generally, a non-vanishing $g^{(1)}_{\mu\nu}$ modifies the asymptotic structure relative to the Einstein sector.

In FG gauge, the Weyl-squared density decomposes as
\begin{equation}
    W^{\alpha \beta}_{\gamma \delta} W^{\gamma \delta}_{\alpha \beta} = W^{\mu \nu}_{\rho \sigma} W_{\mu\nu}^{\rho \sigma} + 4 W^{\rho z}_{\mu \nu} W_{\rho z}^{\mu \nu} + 4 W^{\mu z}_{\nu z} W_{\mu z}^{\nu z} \, ,
\end{equation}
where $W^{\mu \nu}_{\rho \sigma}$, $W^{\rho z}_{\mu \nu}$, and $W^{\mu z}_{\nu z}$ are the independent projections of the Weyl tensor. A straightforward expansion of the Weyl tensor components near $z=0$ shows that each independent projection behaves as
\begin{align}
W^{\mu z}_{\nu z} = {}& \frac{z^2}{2\Ls^2} \left[
- {H}^{(0)\mu}_{\nu} - \left( g^{(2)\mu}_{\nu} - \frac{1}{3}g^{(2)} \delta^{\mu}_{\nu} \right)
+ \frac{1}{4}  \left( g^{(1)\mu}_{\nu} - \frac{1}{3} g^{(1)} \delta^{\mu}_{\nu} \right) \, {g}^{(1)}
\right] + {O}(z^3) \, ,  \\
W^{\rho z}_{\mu \nu} = {}& \frac{z^2}{2\Ls^2} \Bigg[
2 {D}^{(0)}_{[\nu} {g}^{(1)\rho}_{\mu]} 
+ \delta^\rho_\nu {D}^{(0)}_{[\mu} {g}^{(1)\sigma}_{\sigma]} - \delta^\rho_\mu {D}^{(0)}_{[\nu} {g}^{(1)\sigma}_{\sigma]}
\Bigg] + \mathcal{O}(z^3)\, ,  \\
W^{\mu \nu }_{\rho \sigma} = {}& \frac{z^2}{2\Ls^2} \Bigg[
\frac{1}{4} {g}^{(1)}_{\psi \omega} {g}^{(1)\psi \omega} \delta^{\mu \nu}_{\rho \sigma}
- \frac{1}{12} \left({{g}^{(1)}}\right)^{2} \delta^{\mu \nu}_{\rho \sigma}
+ {g}^{(1)} {g}^{(1)[\mu}_{[
\rho} \delta^{\nu]}_{\sigma]} 
- 2\delta^{[\mu}_{[\rho} {g}^{(1)\nu]}_{\omega} {g}^{(1)\omega}_{\sigma]} 
- \frac{1}{4} {g}^{(1)[\mu}_{[\mu} {g}^{\nu]}_{(1)\sigma]} \nonumber \\
{}& - \frac{2}{3} {g}^{(2)} \delta^{\mu \nu}_{\rho \sigma} + 4 \delta^{[\mu}_{[\rho} {g}^{(2)\nu]}_{\sigma]} - 4  \delta^{[\mu}_{[\rho}r^{(0)\nu]}_{\sigma]} + \frac{1}{3} r^{(0)} \delta^{\mu \nu}_{\rho \sigma} + 2 r^{(0)\mu \nu}_{\rho \sigma} \Bigg] 
+ \mathcal{O}(z^3) \,,
\end{align}
for a generic Al(A)dS geometry.\footnote{
In the Einstein sector, the leading $\mathcal O(z^2)$ term in 
$W_{\mu\nu}^{\rho\sigma}$ is proportional to the Weyl tensor of  $g^{(0)}_{\mu\nu}$, which vanishes identically when the boundary metric is conformally flat.}
Here, $D_{(0)\mu}$ and $r^{(0)\mu \nu}_{\rho \sigma}$ are the covariant derivative and the Riemann tensor  associated with ${g}_{\mu\nu}^{(0)}$, respectively. Indices are raised and lowered with the same metric. Consequently,
\begin{equation}
    W_{\alpha\beta}^{\gamma\delta}
    W^{\alpha\beta}_{\gamma\delta}
    \sim z^4 .
\end{equation}
On the other hand, the bulk volume element behaves as
\begin{equation}
    \sqrt{\mathcal G}
    =
    \frac{\Ls^4}{z^4}
    \sqrt{g(z,X)}
    \sim
    \frac{\Ls^4}{z^4}
    \sqrt{g^{(0)}} \left( 1 + \mathcal O(z) \right).
\end{equation}
The factor $z^{-4}$ from the determinant is therefore precisely canceled by the $z^4$ falloff of the Weyl-squared density. 
Near the boundary, the integrand behaves as
\begin{equation}
    \sqrt{\mathcal G}\, W^2
    \sim
    \sqrt{g^{(0)}} + \mathcal O(z).
\end{equation}
The radial integral is thus finite:
\begin{equation}
    I_{\rm CG}
    \sim
    \int_\epsilon \diff z \int_{\partial\mathcal M} \diff^3X\,
    \sqrt{g^{(0)}} \,
    \left( 1 + \mathcal O(z) \right)
    =
    \text{finite}
    + \mathcal O(\epsilon),
\end{equation}
with $\epsilon$ a cutoff regulating the location of the conformal boundary.

This power-counting argument establishes that the CG action is free of infrared divergences for asymptotically locally AdS and dS spacetimes. In particular, no additional boundary counterterms are required to render the on-shell action finite.

\section{Explicit covariant embedding and finiteness of the renormalized area}
\label{app:finiteness}
%%%%%%%%%%%%%%%%%%%%%%%%%%%%%%%%%%%%%%%%

We consider a codimension-two minimal surface $\Sigma$ embedded in a bulk manifold $\mathcal M$, with boundary $\partial\Sigma$ anchored at the conformal boundary $\partial \mathcal M$. Let $y^a=(\tau,Y^i)$ denote intrinsic coordinates on $\Sigma$, and $x^\alpha=(\eta,X^\mu)$ coordinates on $\mathcal M$.\footnote{Here, for de Sitter spacetime, we adopt the standard notation $\eta$ for the radial coordinate instead of $z$.} The embedding
$x^\alpha=x^\alpha(y^a)$ induces the metric on $\Sigma$,
\begin{equation}
    \gamma_{ab}
    =\frac{\partial x^\alpha}{\partial y^a}
     \frac{\partial x^\beta}{\partial y^b}\,
     \mathcal G_{\alpha\beta}\, .
    \label{b1}
\end{equation}
Fixing the diffeomorphism gauge by $\tau=\eta$ and $\gamma_{\tau i}=0$, the
induced metric on $\Sigma$ takes Gauss--normal form,
\begin{equation}
    ds^2=\gamma_{\eta\eta}\diff\eta^2+\mathfrak h_{ij}\diff Y^i \diff Y^j \, .
\end{equation}
Using~\eqref{b1} together with the Fefferman--Graham expansion of the bulk
metric, one finds
\begin{equation}
    \gamma_{\eta\eta}
    =\frac{\Ls^2}{\eta^2}\!\left(
    -1+\frac{\partial X^\mu}{\partial\eta}
         \frac{\partial X^\nu}{\partial\eta}\,
         g_{\mu \nu}\right), \qquad
    \mathfrak h_{ij}
    =\frac{\Ls^2}{\eta^2}\sigma_{i j},
    \label{b.3}
\end{equation}
where
\begin{equation}
    \sigma_{ij}(\eta,Y)
    =\frac{\partial X^\mu}{\partial Y^i}
     \frac{\partial X^\nu}{\partial Y^j}\,
     g_{\mu\nu}(\eta,X)\, .
\end{equation}
Near the conformal boundary, the embedding functions admit the expansion
\begin{equation}
  X^\mu(\eta,Y)
  =x^{(0)\mu}(Y)
   +\frac{\eta^2}{\Ls^2}x^{(2)\mu}(Y)
   +\mathcal O(\eta^3),
  \label{b.5}
\end{equation}
while the boundary metric has the standard Fefferman--Graham form
\begin{equation}
    g_{\mu \nu}(\eta,X)
    =g^{(0)}_{\mu\nu}(X)
     +\frac{\eta^2}{\Ls^2}g^{(2)}_{\mu\nu}(X)
     +\frac{\eta^3}{\Ls^3}g^{(3)}_{\mu\nu}(X)
     +\mathcal O(\eta^4).
    \label{b.6}
\end{equation}
Accordingly,
\begin{equation}
    \sigma_{ij}(\eta,Y)
    =\sigma^{(0)}_{ij}(Y)
     +\frac{\eta^2}{\Ls^2}\sigma^{(2)}_{ij}(Y)
     +\mathcal O(\eta^3),
\end{equation}
with
\begin{equation}
    \sigma^{(0)}_{ij}
    =\frac{\partial X^{(0)\mu}}{\partial Y^i}
     \frac{\partial X^{(0)\nu}}{\partial Y^j}\,
     g^{(0)}_{\mu\nu}(X^{(0)}),
\end{equation}
and
\begin{equation}
    \sigma^{(2)}_{ij}
    =\frac{\partial X^{(0)\mu}}{\partial Y^i}
     \frac{\partial X^{(0)\nu}}{\partial Y^j}\,
     g^{(2)}_{\mu\nu}(X^{(0)})
     +2\frac{\partial X^{(0)\mu}}{\partial Y^i}
        \frac{\partial X^{(2)\nu}}{\partial Y^j}\,
        g^{(0)}_{\mu\nu}(X^{(0)}).
\end{equation}
Substituting into~\eqref{b.3}, one obtains
\begin{equation}
    \gamma_{\eta\eta}
    =\frac{\Ls^2}{\eta^2}\!\left[
    -1+\frac{4\eta^2}{\Ls^4}
     X^{(2)\mu}X^{(2)\nu}g^{(0)}_{\mu\nu}
     +\mathcal O(\eta^4)\right],
    \label{b.10}
\end{equation}
\begin{equation}
    \mathfrak h_{ij}
    =\frac{\Ls^2}{\eta^2}\!\left[
    \sigma^{(0)}_{ij}
    +\frac{\eta^2}{\Ls^2}\sigma^{(2)}_{ij}
    +\mathcal O(\eta^3)\right].
    \label{b.11}
\end{equation}
Also, for $\dim(\Sigma)=2$, the determinant of the induced metric is given by
\begin{equation}
   \sqrt{\gamma}
   =\frac{\Ls^2}{\eta^2}\sqrt{-\sigma^{(0)}}\!\left[
   1+\frac{\eta^2}{2\Ls^2}
   \Bigl(\frac{\sigma^{(2)}}{\sigma^{(0)}}
         -\frac{4}{\ell^2}
          X^{(2)\mu}X^{(2)\nu}g^{(0)}_{\mu\nu}\Bigr)
   +\mathcal O(\eta^3)\right].
   \label{b.13}
\end{equation}
To establish the finiteness of~\eqref{eq:LSAren} and~\eqref{FAL4D}, we analyse the
asymptotic behaviour of the intrinsic Ricci scalar of $\Sigma$. Using
\eqref{b.10} and~\eqref{b.11}, one finds
\begin{equation}
    \mathcal R=\frac{2}{\Ls^2}+\mathcal O(\eta^2).
    \label{b.14}
\end{equation}
Substituting~\eqref{b.13} and~\eqref{b.14} into~\eqref{eq:LSAren}, we obtain
\begin{align}
    \Aren(\Sigma)
    &=-\frac{\Ls^2}{8G}
      \sqrt{\gamma}
      \left(\mathcal R-\frac{2}{\Ls^2}\right)
      +\frac{\pi\Ls^2}{2G}\chi(\Sigma) \nonumber\\
    &=
      \int_\Sigma \diff^2y\,
      \frac{\sqrt{-\sigma^{(0)}}}{\eta^2}
      \mathcal O(\eta^2)
      +\frac{\pi\Ls^2}{2G}\chi(\Sigma)
      =\text{finite}+\mathcal O(\eta).
\end{align}
This demonstrates the finiteness of the renormalized area \eqref{eq:LSAren}, and consequently of the renormalized dS
 pseudoentropy~\eqref{FAL4D}.

%%%%%%%%%%%%%%%%%%%%%%%%%%%
\section{Codimension-2 evaluation of the six-dimensional LPP invariant} \label{app:cod2-6D}
%%%%%%%%%%%%%%%%%%%%%%%%%%%%
In this appendix we  present the technical steps leading to the codimension-two functional $\mathbf F(\Sigma)$ introduced in
section~\ref{sec:cod2-6D}. The six-dimensional case exhibits additional subtleties compared to four dimensions. In particular, cubic curvature invariants are affected by the splitting problem \cite{Miao:2014nxa,Camps:2014voa,Miao:2015iba}, namely the
fact that distinct regularizations of the geometry near a conical defect lead to different expressions for the defect contribution and therefore to different holographic entanglement entropy functionals.\footnote{This ambiguity is absent in several classes of higher-curvature theories, including quadratic gravity \cite{Fursaev:2013fta}, $f(R)$ gravity, and Lovelock theories \cite{Camps:2013zua,Hung:2011xb,deBoer:2011wk}.}

For bulk conformal invariants, however, it was shown in Ref.~\cite{Miao:2015iba} that conformal symmetry  constrains the induced codimension-two functional. This allows one to parametrize a class of consistent splittings for which the universal defect contribution is independent of the specific regularization. Applying this prescription to the conformal
combination~\eqref{eq:L6D}, one obtains the following expressions
\cite{Miao:2015iba}:
\begin{align}
F_{1} =& 3 \left(W^{\alpha \beta \gamma \delta }W_{\beta \ \ \gamma }^{\ \lambda \eta }\varepsilon _{\lambda \alpha }\varepsilon _{\delta \eta } -\frac{1}{4}W^{\lambda \delta \eta \beta }W_{\ \delta \eta \beta }^{\alpha }g_{\alpha \lambda }^{\perp } +\frac{1}{20}W^{\alpha\beta\gamma\delta}W_{\alpha\beta\gamma\delta} \right)+3 K^{\iota }{}_{\langle \lambda \alpha\rangle} K_{\iota}{}^{\langle\beta \gamma\rangle}W^\lambda{}_\beta{}^\alpha{}_\gamma  \nonumber \\
&-3 K^{\iota }{}_{\ \langle\lambda \alpha\rangle}K_{\iota}{}_{\langle\beta \gamma\rangle} K_{\zeta }{}^{\langle\lambda [\gamma\rangle}K^{\zeta}{}^{\langle\alpha]\beta\rangle}+3\varepsilon ^{\iota \zeta } K_{\iota\langle\lambda \eta\rangle}K_{\zeta}{}^{\langle\alpha\eta \rangle}\varepsilon^{\gamma \delta }W^\lambda{}_{\alpha\gamma \delta} +\frac{3}{4}\left (K^{\iota }{}_{\langle\alpha \beta\rangle} K_{\iota }{}^{\langle\alpha \beta \rangle}\right )^{2} \nonumber \\
&+3\varepsilon ^{\iota \zeta}\varepsilon ^{\kappa \delta }K_{\iota\langle\lambda \eta\rangle}K_{\zeta}{}^{\langle\alpha \eta\rangle} K_{\kappa}{}^{\langle\gamma\lambda\rangle}K_{\delta}{}_{\langle\gamma \alpha \rangle}-\frac{3}{4} K^{\iota }{}_{\langle\lambda \eta\rangle} K_{\iota }{}^{\langle\lambda \eta\rangle}W^{\alpha \beta \gamma \delta }\varepsilon _{\alpha \beta }\varepsilon _{\gamma \delta },\\ \notag 
F_{2} =&3 \left(W^{\alpha \beta \gamma \delta }W_{\gamma \delta }^{\lambda \eta }\varepsilon _{\lambda \eta }\varepsilon _{\alpha\beta} -W^{\lambda \delta \eta \beta } W^{\alpha}{}_{\delta \eta \beta }g_{\alpha \lambda }^{\perp } +\frac{1}{5}W^{\alpha\beta\gamma\delta}W_{\alpha\beta\gamma\delta}\right)\\
&-6 K^{\iota }{}_{\langle\delta}{}^{(\gamma\rangle}K_{\iota }{}^{\langle\alpha )\delta\rangle}\left(2W_{\beta \gamma \lambda \alpha }g^{\perp \beta \lambda }+K^{\zeta }{}_{\langle\alpha \eta\rangle}K_{\zeta\langle\gamma}{}^{\eta\rangle}\right)\\&+6\varepsilon ^{\alpha\beta}\varepsilon ^{\gamma\delta}K_{\alpha}{}^{\langle\zeta\iota\rangle}K_{\beta}{}^{\langle\eta}{}_{\iota\rangle}\left(2W_{\gamma\eta\delta\zeta}+K_{\delta\langle\eta \lambda\rangle}K_{\delta}{}^{\langle\lambda}{}_{\zeta\rangle}\right), \nonumber \\
F_{3} =& -6\mathcal X_4^\Sigma+12F_{1} +3F_{2} +192 \left(\Upsilon _{a}^{a} -\frac{1}{2} S_{ab}S^{ab} +\frac{1}{4}\left (S_{a}^{a}\right )^{2} -\frac{1}{4} K^{A} K_{A cb}S^{cb}  \right. \nonumber\\
& \left. +\frac{3}{32}K^{A}K_{A } S_{b}^{b} -\frac{1}{16} K^{A }K^{B } S_{AB } -\frac{1}{32}K^{A }K_{A cb}K_{B }K^{B cb} +\frac{7}{1024}\left (K^{A }K_{A }\right )^{2}\right), \label{Ficonical}
\end{align}
where
\begin{align}
\Upsilon_{a b} &=\frac{1}{4}\left[\frac{1}{16}\left(\partial _{a}K^{A} \partial _{b}K_{A} +K_{A}K^{A}{}_{ac}K_{B}K^{B}{}_b^c -K^{A}K^{B} W_{aAbB}-K^{A}K_{A}S_{a b} \right.\right.\nonumber  \\
&\left.-K^{A}K^{B}S_{AB}\gamma_{a b}\right) +S_{a\alpha}S_{b}^{\alpha} -B_{a b}-\frac{1}{2}\left(S_{aA} \partial _{b}K^{A} -S_a^cK^{A}{}_{bc}K_{A} +K^{A}C_{a bA}\right. \nonumber  \\
&\left.\left.+\nabla_{a}^{\Sigma}\left (K^{A}S_{Ab}\right ) -K^{A}K^{B}{}_{ab}S_{AB} \right)\right]. \label{upsilonfin}
\end{align}
Here, $K^{\gamma}{}_{\alpha\beta}$ denotes the extrinsic curvature of $\Sigma$. For $F_1$ and $F_2$, we retain a fully covariant representation, introducing the binormal $\varepsilon_{\alpha\beta}=n_\alpha^A n_\beta^B\epsilon_{AB}$, with
$\epsilon_{AB}$ the Levi--Civita tensor in the normal bundle, and identifying $g^{\perp}_{\alpha\beta}$ as the induced metric on the  normal two-dimensional space. For later use, we perform an explicit normal decomposition in $F_3$ by splitting bulk indices as $\alpha=(A,a)$, where $a$ labels directions tangent to $\Sigma$ and $A$ labels normal directions. The tensors $C_{\alpha\beta\gamma}$
and $B_{\alpha\beta}$ correspond to the Cotton and Bach tensors of the ambient manifold $\mathcal M$, respectively, and $\nabla_a^{\Sigma}$ denotes the covariant derivative compatible with the induced metric $\gamma_{ab}$.

After several algebraic manipulations, the defect functional
$\mathbf F(\Sigma)$ can be recast into the compact form
\begin{align}
\mathbf{F}(\Sigma)&=-\frac{4\Ls^4}{3}\int _{\Sigma }\left[\frac{1}{32}\mathcal X_4^\Sigma+\Upsilon_a^a +\frac{1}{2} S_{ab}S^{ab} -\frac{1}{4}\left (S_a^a\right )^{2} +\frac{1}{4} K^A K_{A ab}S^{ab} -\frac{3}{32}K^A K_A S_a^a\right.\nonumber  \\
&\left.+\frac{1}{16} K^A K^B S_{AB}+\frac{1}{32}K^A K_{A ab}K_B K^{B ab}-\frac{7}{1024}\left (K^A K_A \right )^2\right]+\frac{4\pi ^{2}\Ls^4}{3}\chi \left (\Sigma \right ) +\text{b.\,t.}, \label{eq:FS2}
\end{align}
where $\mathcal X_4^\Sigma$ is the Euler density intrinsic to the four-dimensional surface $\Sigma$. The boundary terms, denoted by $\text{b.\,t.}$, originate from the total derivative $\nabla_\alpha J^\alpha$ contained in the invariant $I_3$. While such contributions were omitted in Ref.~\cite{Miao:2015iba}, their inclusion is required in order for $F_3$ to define a conformal invariant in the presence of boundaries.

The missing boundary contribution in Eq.~\eqref{eq:FS2} arises entirely from $I_3$ and can be fixed unambiguously. Evaluating the CG action \eqref{eq:LPP2} on Einstein backgrounds, the surviving boundary term reduces to

\begin{equation}\label{eq:J}
J\big|_{\text E}=\frac{\Ls^{4}}{384 \pi \GN}\int _{ \partial \mathcal M} n^{\alpha}J_{\alpha}\big|_{\text E}=-\frac{\sigma^{3/2}\Ls^{3}}{192 \pi \GN}\int _{ \partial \mathcal M} w^2,
\end{equation}
where $w^2=w_{\mu\nu}^{\rho\sigma}w^{\mu\nu}_{\rho\sigma}$ denotes the squared Weyl tensor at the asymptotic AdS boundary. Although the second equality holds only asymptotically, it is sufficient for canceling bulk divergences \cite{Anastasiou:2020mik}.

To extract the codimension-two contribution, Eq.~\eqref{eq:J} must be evaluated on the conically singular geometry. Using the decomposition of the Weyl-squared tensor introduced in Eq.~\eqref{eq:WeylSquaredExpansion}, adapted to the boundary submanifold $\partial\mathcal M$, one finds
\begin{equation}\label{eq:btE}
\text{b.\,t.}\big|_{\text E}= -\frac{\sigma^{3/2}\Ls^{3}}{6}\int_{ \partial \Sigma }\left (w_{ij}^{ij} -\kappa^{I}{}_{\langle ij\rangle}\kappa_{I}{}^{\langle ij\rangle}\right )=-\frac{\sigma^{3/2}\Ls^3}{6}\int_{\partial\Sigma}\mathcal K_{\partial\Sigma}\big|_{\text E}. 
\end{equation}
Here, $\kappa^{I}{}_{\langle ij\rangle}$ denotes the traceless extrinsic curvature of $\partial\Sigma$ embedded in $\partial\mathcal M$. The final equality makes it explicit that, when evaluated on Einstein backgrounds, the boundary contribution reproduces the functional introduced in Eq.~\eqref{eq:Ksigma}, but now evaluated on $\partial \Sigma$.

Combining the bulk contribution \eqref{eq:FS2}, the definition of
$\Upsilon_a^{a}$ in Eq.~\eqref{upsilonfin}, and the boundary term
\eqref{eq:btE}, and restricting to Einstein spacetimes where the Bach and Cotton tensors vanish and the Schouten tensor is proportional to the metric, the defect functional simplifies considerably. This procedure leads directly to Eq.~\eqref{eq:IfromF} for $\mathbf F(\Sigma)\big|_{\text E}$, as stated in section~\ref{sec:cod2-6D}.

\bibliographystyle{JHEP}
\bibliography{Biblio}

\providecommand{\href}[2]{#2}\begingroup\raggedright\begin{thebibliography}{100}

\bibitem{Strominger:2001pn}
A.~Strominger, \emph{{The dS / CFT correspondence}}, \href{http://dx.doi.org/10.1088/1126-6708/2001/10/034}{\emph{JHEP} {\bfseries 10} (2001) 034}, [\href{https://arxiv.org/abs/hep-th/0106113}{{\ttfamily hep-th/0106113}}].

\bibitem{Strominger:2001gp}
A.~Strominger, \emph{{Inflation and the dS / CFT correspondence}}, \href{http://dx.doi.org/10.1088/1126-6708/2001/11/049}{\emph{JHEP} {\bfseries 11} (2001) 049}, [\href{https://arxiv.org/abs/hep-th/0110087}{{\ttfamily hep-th/0110087}}].

\bibitem{Spradlin:2001pw}
M.~Spradlin, A.~Strominger and A.~Volovich, \emph{{Les Houches lectures on de Sitter space}},  in \emph{{Les Houches Summer School: Session 76: Euro Summer School on Unity of Fundamental Physics: Gravity, Gauge Theory and Strings}}, pp.~423--453, 10, 2001.
\newblock \href{https://arxiv.org/abs/hep-th/0110007}{{\ttfamily hep-th/0110007}}.

\bibitem{Anninos:2012qw}
D.~Anninos, \emph{{De Sitter Musings}}, \href{http://dx.doi.org/10.1142/S0217751X1230013X}{\emph{Int. J. Mod. Phys. A} {\bfseries 27} (2012) 1230013}, [\href{https://arxiv.org/abs/1205.3855}{{\ttfamily 1205.3855}}].

\bibitem{Anninos:2011ui}
D.~Anninos, T.~Hartman and A.~Strominger, \emph{{Higher Spin Realization of the dS/CFT Correspondence}}, \href{http://dx.doi.org/10.1088/1361-6382/34/1/015009}{\emph{Class. Quant. Grav.} {\bfseries 34} (2017) 015009}, [\href{https://arxiv.org/abs/1108.5735}{{\ttfamily 1108.5735}}].

\bibitem{Skenderis:2006jq}
K.~Skenderis and P.~K. Townsend, \emph{{Hidden supersymmetry of domain walls and cosmologies}}, \href{http://dx.doi.org/10.1103/PhysRevLett.96.191301}{\emph{Phys. Rev. Lett.} {\bfseries 96} (2006) 191301}, [\href{https://arxiv.org/abs/hep-th/0602260}{{\ttfamily hep-th/0602260}}].

\bibitem{Skenderis:2006fb}
K.~Skenderis and P.~K. Townsend, \emph{{Pseudo-Supersymmetry and the Domain-Wall/Cosmology Correspondence}}, \href{http://dx.doi.org/10.1088/1751-8113/40/25/S18}{\emph{J. Phys. A} {\bfseries 40} (2007) 6733--6742}, [\href{https://arxiv.org/abs/hep-th/0610253}{{\ttfamily hep-th/0610253}}].

\bibitem{McFadden:2009fg}
P.~McFadden and K.~Skenderis, \emph{{Holography for Cosmology}}, \href{http://dx.doi.org/10.1103/PhysRevD.81.021301}{\emph{Phys. Rev. D} {\bfseries 81} (2010) 021301}, [\href{https://arxiv.org/abs/0907.5542}{{\ttfamily 0907.5542}}].

\bibitem{McFadden:2010na}
P.~McFadden and K.~Skenderis, \emph{{The Holographic Universe}}, \href{http://dx.doi.org/10.1088/1742-6596/222/1/012007}{\emph{J. Phys. Conf. Ser.} {\bfseries 222} (2010) 012007}, [\href{https://arxiv.org/abs/1001.2007}{{\ttfamily 1001.2007}}].

\bibitem{Hertog:2011ky}
T.~Hertog and J.~Hartle, \emph{{Holographic No-Boundary Measure}}, \href{http://dx.doi.org/10.1007/JHEP05(2012)095}{\emph{JHEP} {\bfseries 05} (2012) 095}, [\href{https://arxiv.org/abs/1111.6090}{{\ttfamily 1111.6090}}].

\bibitem{McFadden:2011kk}
P.~McFadden and K.~Skenderis, \emph{{Cosmological 3-point correlators from holography}}, \href{http://dx.doi.org/10.1088/1475-7516/2011/06/030}{\emph{JCAP} {\bfseries 06} (2011) 030}, [\href{https://arxiv.org/abs/1104.3894}{{\ttfamily 1104.3894}}].

\bibitem{Bzowski:2011ab}
A.~Bzowski, P.~McFadden and K.~Skenderis, \emph{{Holographic predictions for cosmological 3-point functions}}, \href{http://dx.doi.org/10.1007/JHEP03(2012)091}{\emph{JHEP} {\bfseries 03} (2012) 091}, [\href{https://arxiv.org/abs/1112.1967}{{\ttfamily 1112.1967}}].

\bibitem{Banerjee:2013mca}
S.~Banerjee, A.~Belin, S.~Hellerman, A.~Lepage-Jutier, A.~Maloney, D.~Radicevic et~al., \emph{{Topology of Future Infinity in dS/CFT}}, \href{http://dx.doi.org/10.1007/JHEP11(2013)026}{\emph{JHEP} {\bfseries 11} (2013) 026}, [\href{https://arxiv.org/abs/1306.6629}{{\ttfamily 1306.6629}}].

\bibitem{Maldacena:1997re}
J.~M. Maldacena, \emph{{The Large N limit of superconformal field theories and supergravity}}, \href{http://dx.doi.org/10.1023/A:1026654312961}{\emph{Int. J. Theor. Phys.} {\bfseries 38} (1999) 1113--1133}, [\href{https://arxiv.org/abs/hep-th/9711200}{{\ttfamily hep-th/9711200}}].

\bibitem{Witten:1998qj}
E.~Witten, \emph{{Anti-de Sitter space and holography}}, {\emph{Adv. Theor. Math. Phys.} {\bfseries 2} (1998) 253--291}, [\href{https://arxiv.org/abs/hep-th/9802150}{{\ttfamily hep-th/9802150}}].

\bibitem{Hartle:1983ai}
J.~B. Hartle and S.~W. Hawking, \emph{{Wave Function of the Universe}}, \href{http://dx.doi.org/10.1103/PhysRevD.28.2960}{\emph{Phys. Rev. D} {\bfseries 28} (1983) 2960--2975}.

\bibitem{Maldacena:2002vr}
J.~M. Maldacena, \emph{{Non-Gaussian features of primordial fluctuations in single field inflationary models}}, \href{http://dx.doi.org/10.1088/1126-6708/2003/05/013}{\emph{JHEP} {\bfseries 05} (2003) 013}, [\href{https://arxiv.org/abs/astro-ph/0210603}{{\ttfamily astro-ph/0210603}}].

\bibitem{Witten:2001kn}
E.~Witten, \emph{{Quantum gravity in de Sitter space}},  in \emph{{Strings 2001: International Conference}}, 6, 2001.
\newblock \href{https://arxiv.org/abs/hep-th/0106109}{{\ttfamily hep-th/0106109}}.

\bibitem{AST_1985__S131__95_0}
C.~Fefferman and C.~R. Graham, \emph{{Conformal invariants}},  in \emph{\'Elie Cartan et les math\'ematiques d'aujourd'hui - Lyon, 25-29 juin 1984}, no.~S131 in Ast\'erisque, pp.~95--116.
\newblock Soci\'et\'e math\'ematique de France, 1985.

\bibitem{1983ZhPmR..37...55S}
A.~A. {Starobinskii}, \emph{{Isotropization of arbitrary cosmological expansion given an effective cosmological constant}}, {\emph{ZhETF Pisma Redaktsiiu} {\bfseries 37} (Jan., 1983) 55--58}.

\bibitem{Anderson:2004wj}
M.~T. Anderson, \emph{{On the structure of asymptotically de Sitter and anti-de Sitter spaces}}, \href{http://dx.doi.org/10.4310/ATMP.2004.v8.n5.a4}{\emph{Adv. Theor. Math. Phys.} {\bfseries 8} (2004) 861--893}, [\href{https://arxiv.org/abs/hep-th/0407087}{{\ttfamily hep-th/0407087}}].

\bibitem{Bzowski:2023nef}
A.~Bzowski, P.~McFadden and K.~Skenderis, \emph{{Renormalisation of IR divergences and holography in de Sitter}}, \href{http://dx.doi.org/10.1007/JHEP05(2024)053}{\emph{JHEP} {\bfseries 05} (2024) 053}, [\href{https://arxiv.org/abs/2312.17316}{{\ttfamily 2312.17316}}].

\bibitem{Poole:2025cmv}
A.~Poole, K.~Skenderis and M.~Taylor, \emph{{Gravitational charges and radiation in asymptotically locally de Sitter spacetimes}},  \href{https://arxiv.org/abs/2512.14243}{{\ttfamily 2512.14243}}.

\bibitem{Calabrese:2004eu}
P.~Calabrese and J.~L. Cardy, \emph{{Entanglement entropy and quantum field theory}}, \href{http://dx.doi.org/10.1088/1742-5468/2004/06/P06002}{\emph{J. Stat. Mech.} {\bfseries 0406} (2004) P06002}, [\href{https://arxiv.org/abs/hep-th/0405152}{{\ttfamily hep-th/0405152}}].

\bibitem{Calabrese:2009qy}
P.~Calabrese and J.~Cardy, \emph{{Entanglement entropy and conformal field theory}}, \href{http://dx.doi.org/10.1088/1751-8113/42/50/504005}{\emph{J. Phys.} {\bfseries A42} (2009) 504005}, [\href{https://arxiv.org/abs/0905.4013}{{\ttfamily 0905.4013}}].

\bibitem{Headrick:2019eth}
M.~Headrick, \emph{{Lectures on entanglement entropy in field theory and holography}},  \href{https://arxiv.org/abs/1907.08126}{{\ttfamily 1907.08126}}.

\bibitem{Gubser:1998bc}
S.~S. Gubser, I.~R. Klebanov and A.~M. Polyakov, \emph{{Gauge theory correlators from noncritical string theory}}, \href{http://dx.doi.org/10.1016/S0370-2693(98)00377-3}{\emph{Phys. Lett.} {\bfseries B428} (1998) 105--114}, [\href{https://arxiv.org/abs/hep-th/9802109}{{\ttfamily hep-th/9802109}}].

\bibitem{Harlow:2011ke}
D.~Harlow and D.~Stanford, \emph{{Operator Dictionaries and Wave Functions in AdS/CFT and dS/CFT}},  \href{https://arxiv.org/abs/1104.2621}{{\ttfamily 1104.2621}}.

\bibitem{VanRaamsdonk:2010pw}
M.~Van~Raamsdonk, \emph{{Building up spacetime with quantum entanglement}}, \href{http://dx.doi.org/10.1142/S0218271810018529}{\emph{Gen. Rel. Grav.} {\bfseries 42} (2010) 2323--2329}, [\href{https://arxiv.org/abs/1005.3035}{{\ttfamily 1005.3035}}].

\bibitem{Penington:2019npb}
G.~Penington, \emph{{Entanglement Wedge Reconstruction and the Information Paradox}}, \href{http://dx.doi.org/10.1007/JHEP09(2020)002}{\emph{JHEP} {\bfseries 09} (2020) 002}, [\href{https://arxiv.org/abs/1905.08255}{{\ttfamily 1905.08255}}].

\bibitem{Takayanagi:2018pml}
T.~Takayanagi, \emph{{Holographic Spacetimes as Quantum Circuits of Path-Integrations}}, \href{http://dx.doi.org/10.1007/JHEP12(2018)048}{\emph{JHEP} {\bfseries 12} (2018) 048}, [\href{https://arxiv.org/abs/1808.09072}{{\ttfamily 1808.09072}}].

\bibitem{Takayanagi:2025ula}
T.~Takayanagi, \emph{{Essay: Emergent Holographic Spacetime from Quantum Information}}, \href{http://dx.doi.org/10.1103/pg4r-fy8n}{\emph{Phys. Rev. Lett.} {\bfseries 134} (2025) 240001}, [\href{https://arxiv.org/abs/2506.06595}{{\ttfamily 2506.06595}}].

\bibitem{Ryu:2006bv}
S.~Ryu and T.~Takayanagi, \emph{{Holographic derivation of entanglement entropy from AdS/CFT}}, \href{http://dx.doi.org/10.1103/PhysRevLett.96.181602}{\emph{Phys. Rev. Lett.} {\bfseries 96} (2006) 181602}, [\href{https://arxiv.org/abs/hep-th/0603001}{{\ttfamily hep-th/0603001}}].

\bibitem{Ryu:2006ef}
S.~Ryu and T.~Takayanagi, \emph{{Aspects of Holographic Entanglement Entropy}}, \href{http://dx.doi.org/10.1088/1126-6708/2006/08/045}{\emph{JHEP} {\bfseries 08} (2006) 045}, [\href{https://arxiv.org/abs/hep-th/0605073}{{\ttfamily hep-th/0605073}}].

\bibitem{Doi:2022iyj}
K.~Doi, J.~Harper, A.~Mollabashi, T.~Takayanagi and Y.~Taki, \emph{{Pseudoentropy in dS/CFT and Timelike Entanglement Entropy}}, \href{http://dx.doi.org/10.1103/PhysRevLett.130.031601}{\emph{Phys. Rev. Lett.} {\bfseries 130} (2023) 031601}, [\href{https://arxiv.org/abs/2210.09457}{{\ttfamily 2210.09457}}].

\bibitem{Doi:2023zaf}
K.~Doi, J.~Harper, A.~Mollabashi, T.~Takayanagi and Y.~Taki, \emph{{Timelike entanglement entropy}}, \href{http://dx.doi.org/10.1007/JHEP05(2023)052}{\emph{JHEP} {\bfseries 05} (2023) 052}, [\href{https://arxiv.org/abs/2302.11695}{{\ttfamily 2302.11695}}].

\bibitem{Nakata:2020luh}
Y.~Nakata, T.~Takayanagi, Y.~Taki, K.~Tamaoka and Z.~Wei, \emph{{New holographic generalization of entanglement entropy}}, \href{http://dx.doi.org/10.1103/PhysRevD.103.026005}{\emph{Phys. Rev. D} {\bfseries 103} (2021) 026005}, [\href{https://arxiv.org/abs/2005.13801}{{\ttfamily 2005.13801}}].

\bibitem{Guo:2022jzs}
W.-z. Guo, S.~He and Y.-X. Zhang, \emph{{Constructible reality condition of pseudo entropy via pseudo-Hermiticity}}, \href{http://dx.doi.org/10.1007/JHEP05(2023)021}{\emph{JHEP} {\bfseries 05} (2023) 021}, [\href{https://arxiv.org/abs/2209.07308}{{\ttfamily 2209.07308}}].

\bibitem{Aalsma:2022swk}
L.~Aalsma, S.~E. Aguilar-Gutierrez and W.~Sybesma, \emph{{An outsider{\textquoteright}s perspective on information recovery in de Sitter space}}, \href{http://dx.doi.org/10.1007/JHEP01(2023)129}{\emph{JHEP} {\bfseries 01} (2023) 129}, [\href{https://arxiv.org/abs/2210.12176}{{\ttfamily 2210.12176}}].

\bibitem{Narayan:2022afv}
K.~Narayan, \emph{{de Sitter space, extremal surfaces, and time entanglement}}, \href{http://dx.doi.org/10.1103/PhysRevD.107.126004}{\emph{Phys. Rev. D} {\bfseries 107} (2023) 126004}, [\href{https://arxiv.org/abs/2210.12963}{{\ttfamily 2210.12963}}].

\bibitem{He:2023eap}
S.~He, J.~Yang, Y.-X. Zhang and Z.-X. Zhao, \emph{{Pseudoentropy for descendant operators in two-dimensional conformal field theories}}, \href{http://dx.doi.org/10.1103/PhysRevD.109.025014}{\emph{Phys. Rev. D} {\bfseries 109} (2024) 025014}, [\href{https://arxiv.org/abs/2301.04891}{{\ttfamily 2301.04891}}].

\bibitem{Alshal:2023kcd}
H.~Alshal, \emph{{Einstein{\textquoteright}s equations and the pseudo-entropy of pseudo-Riemannian information manifolds}}, \href{http://dx.doi.org/10.1007/s10714-023-03130-7}{\emph{Gen. Rel. Grav.} {\bfseries 55} (2023) 86}, [\href{https://arxiv.org/abs/2301.13017}{{\ttfamily 2301.13017}}].

\bibitem{Chen:2023prz}
H.-Y. Chen, Y.~Hikida, Y.~Taki and T.~Uetoko, \emph{{Complex saddles of three-dimensional de Sitter gravity via holography}}, \href{http://dx.doi.org/10.1103/PhysRevD.107.L101902}{\emph{Phys. Rev. D} {\bfseries 107} (2023) L101902}, [\href{https://arxiv.org/abs/2302.09219}{{\ttfamily 2302.09219}}].

\bibitem{Narayan:2023ebn}
K.~Narayan and H.~K. Saini, \emph{{Notes on time entanglement and pseudo-entropy}}, \href{http://dx.doi.org/10.1140/epjc/s10052-024-12855-x}{\emph{Eur. Phys. J. C} {\bfseries 84} (2024) 499}, [\href{https://arxiv.org/abs/2303.01307}{{\ttfamily 2303.01307}}].

\bibitem{Jiang:2023loq}
X.~Jiang, P.~Wang, H.~Wu and H.~Yang, \emph{{Timelike entanglement entropy in dS$_{3}$/CFT$_{2}$}}, \href{http://dx.doi.org/10.1007/JHEP08(2023)216}{\emph{JHEP} {\bfseries 08} (2023) 216}, [\href{https://arxiv.org/abs/2304.10376}{{\ttfamily 2304.10376}}].

\bibitem{Kawamoto:2023nki}
T.~Kawamoto, S.-M. Ruan, Y.-k. Suzuki and T.~Takayanagi, \emph{{A half de Sitter holography}}, \href{http://dx.doi.org/10.1007/JHEP10(2023)137}{\emph{JHEP} {\bfseries 10} (2023) 137}, [\href{https://arxiv.org/abs/2306.07575}{{\ttfamily 2306.07575}}].

\bibitem{Chen:2023eic}
D.~Chen, X.~Jiang and H.~Yang, \emph{{Holographic TT{\textasciimacron} deformed entanglement entropy in dS3/CFT2}}, \href{http://dx.doi.org/10.1103/PhysRevD.109.026011}{\emph{Phys. Rev. D} {\bfseries 109} (2024) 026011}, [\href{https://arxiv.org/abs/2307.04673}{{\ttfamily 2307.04673}}].

\bibitem{Guo:2023aio}
W.-z. Guo and J.~Zhang, \emph{{Sum rule for the pseudo-R{\'e}nyi entropy}}, \href{http://dx.doi.org/10.1103/PhysRevD.109.106008}{\emph{Phys. Rev. D} {\bfseries 109} (2024) 106008}, [\href{https://arxiv.org/abs/2308.05261}{{\ttfamily 2308.05261}}].

\bibitem{Aguilar-Gutierrez:2023tic}
S.~E. Aguilar-Gutierrez, A.~K. Patra and J.~F. Pedraza, \emph{{Entangled universes in dS wedge holography}}, \href{http://dx.doi.org/10.1007/JHEP10(2023)156}{\emph{JHEP} {\bfseries 10} (2023) 156}, [\href{https://arxiv.org/abs/2308.05666}{{\ttfamily 2308.05666}}].

\bibitem{Narayan:2023zen}
K.~Narayan, \emph{{Further remarks on de Sitter space, extremal surfaces, and time entanglement}}, \href{http://dx.doi.org/10.1103/PhysRevD.109.086009}{\emph{Phys. Rev. D} {\bfseries 109} (2024) 086009}, [\href{https://arxiv.org/abs/2310.00320}{{\ttfamily 2310.00320}}].

\bibitem{Shinmyo:2023eci}
K.~Shinmyo, T.~Takayanagi and K.~Tasuki, \emph{{Pseudo entropy under joining local quenches}}, \href{http://dx.doi.org/10.1007/JHEP02(2024)111}{\emph{JHEP} {\bfseries 02} (2024) 111}, [\href{https://arxiv.org/abs/2310.12542}{{\ttfamily 2310.12542}}].

\bibitem{Kanda:2023jyi}
H.~Kanda, T.~Kawamoto, Y.-k. Suzuki, T.~Takayanagi, K.~Tasuki and Z.~Wei, \emph{{Entanglement phase transition in holographic pseudo entropy}}, \href{http://dx.doi.org/10.1007/JHEP03(2024)060}{\emph{JHEP} {\bfseries 03} (2024) 060}, [\href{https://arxiv.org/abs/2311.13201}{{\ttfamily 2311.13201}}].

\bibitem{Yadav:2024ray}
G.~Yadav, \emph{{Communicating multiverses in a holographic de Sitter braneworld}}, \href{http://dx.doi.org/10.1103/PhysRevD.110.026028}{\emph{Phys. Rev. D} {\bfseries 110} (2024) 026028}, [\href{https://arxiv.org/abs/2404.00763}{{\ttfamily 2404.00763}}].

\bibitem{Doi:2024nty}
K.~Doi, N.~Ogawa, K.~Shinmyo, Y.-k. Suzuki and T.~Takayanagi, \emph{{Probing de Sitter space using CFT states}}, \href{http://dx.doi.org/10.1007/JHEP02(2025)093}{\emph{JHEP} {\bfseries 02} (2025) 093}, [\href{https://arxiv.org/abs/2405.14237}{{\ttfamily 2405.14237}}].

\bibitem{Fareghbal:2024lqa}
R.~Fareghbal, \emph{{Flat-space limit of holographic pseudoentropy in (A)dS spacetimes}}, \href{http://dx.doi.org/10.1103/PhysRevD.110.066019}{\emph{Phys. Rev. D} {\bfseries 110} (2024) 066019}, [\href{https://arxiv.org/abs/2408.03061}{{\ttfamily 2408.03061}}].

\bibitem{Caputa:2024gve}
P.~Caputa, B.~Chen, T.~Takayanagi and T.~Tsuda, \emph{{Thermal pseudo-entropy}}, \href{http://dx.doi.org/10.1007/JHEP01(2025)003}{\emph{JHEP} {\bfseries 01} (2025) 003}, [\href{https://arxiv.org/abs/2411.08948}{{\ttfamily 2411.08948}}].

\bibitem{Goswami:2024vfl}
K.~Goswami, K.~Narayan and G.~Yadav, \emph{{No-boundary extremal surfaces in slow-roll inflation and other cosmologies}}, \href{http://dx.doi.org/10.1007/JHEP03(2025)193}{\emph{JHEP} {\bfseries 03} (2025) 193}, [\href{https://arxiv.org/abs/2409.14208}{{\ttfamily 2409.14208}}].

\bibitem{Nanda:2025tid}
K.~K. Nanda, K.~Narayan, S.~Porey and G.~Yadav, \emph{{$dS$ extremal surfaces, replicas, boundary Renyi entropies in $dS/CFT$ and time entanglement}},  \href{https://arxiv.org/abs/2509.02775}{{\ttfamily 2509.02775}}.

\bibitem{Fujiki:2025rtx}
K.~Fujiki, M.~Kohara, K.~Shinmyo, Y.-k. Suzuki and T.~Takayanagi, \emph{{Entropic Interpretation of Einstein Equation in dS/CFT}}, {\emph{unpublished} (11, 2025) }, [\href{https://arxiv.org/abs/2511.07915}{{\ttfamily 2511.07915}}].

\bibitem{Anastasiou:2025rvz}
G.~Anastasiou, I.~J. Araya, A.~Das and J.~Moreno, \emph{{Universality of pseudoentropy for deformed spheres in dS/CFT}},  \href{https://arxiv.org/abs/2512.02164}{{\ttfamily 2512.02164}}.

\bibitem{Huang:2025gmq}
X.~Huang and C.-T. Ma, \emph{{dS/CFT from Defect}},  \href{https://arxiv.org/abs/2512.11759}{{\ttfamily 2512.11759}}.

\bibitem{Hikida:2022ltr}
Y.~Hikida, T.~Nishioka, T.~Takayanagi and Y.~Taki, \emph{{CFT duals of three-dimensional de Sitter gravity}}, \href{http://dx.doi.org/10.1007/JHEP05(2022)129}{\emph{JHEP} {\bfseries 05} (2022) 129}, [\href{https://arxiv.org/abs/2203.02852}{{\ttfamily 2203.02852}}].

\bibitem{Boruch:2021hqs}
J.~Boruch, P.~Caputa, D.~Ge and T.~Takayanagi, \emph{{Holographic path-integral optimization}}, \href{http://dx.doi.org/10.1007/JHEP07(2021)016}{\emph{JHEP} {\bfseries 07} (2021) 016}.

\bibitem{Couvreur:2016mbr}
R.~Couvreur, J.~L. Jacobsen and H.~Saleur, \emph{{Entanglement in nonunitary quantum critical spin chains}}, \href{http://dx.doi.org/10.1103/PhysRevLett.119.040601}{\emph{Phys. Rev. Lett.} {\bfseries 119} (2017) 040601}, [\href{https://arxiv.org/abs/1611.08506}{{\ttfamily 1611.08506}}].

\bibitem{Herviou:2019yfb}
L.~Herviou, N.~Regnault and J.~H. Bardarson, \emph{{Entanglement spectrum and symmetries in non-Hermitian fermionic non-interacting models}}, \href{http://dx.doi.org/10.21468/scipostphys.7.5.069}{\emph{SciPost Phys.} {\bfseries 7} (2019) 069}, [\href{https://arxiv.org/abs/1908.09852}{{\ttfamily 1908.09852}}].

\bibitem{Chang:2019jcj}
P.-Y. Chang, J.-S. You, X.~Wen and S.~Ryu, \emph{{Entanglement spectrum and entropy in topological non-Hermitian systems and nonunitary conformal field theory}}, \href{http://dx.doi.org/10.1103/PhysRevResearch.2.033069}{\emph{Phys. Rev. Res.} {\bfseries 2} (2020) 033069}, [\href{https://arxiv.org/abs/1909.01346}{{\ttfamily 1909.01346}}].

\bibitem{Jian:2020byi}
C.-M. Jian, B.~Bauer, A.~Keselman and A.~W.~W. Ludwig, \emph{{Criticality and entanglement in nonunitary quantum circuits and tensor networks of noninteracting fermions}}, \href{http://dx.doi.org/10.1103/PhysRevB.106.134206}{\emph{Phys. Rev. B} {\bfseries 106} (2022) 134206}, [\href{https://arxiv.org/abs/2012.04666}{{\ttfamily 2012.04666}}].

\bibitem{Chen:2023gnh}
Z.~Chen, \emph{{Complex-valued Holographic Pseudo Entropy via Real-time AdS/CFT Correspondence}},  \href{https://arxiv.org/abs/2302.14303}{{\ttfamily 2302.14303}}.

\bibitem{Das:2023yyl}
A.~Das, S.~Sachdeva and D.~Sarkar, \emph{{Bulk reconstruction using timelike entanglement in (A)dS}}, \href{http://dx.doi.org/10.1103/PhysRevD.109.066007}{\emph{Phys. Rev. D} {\bfseries 109} (2024) 066007}, [\href{https://arxiv.org/abs/2312.16056}{{\ttfamily 2312.16056}}].

\bibitem{Guo:2024lrr}
W.-z. Guo, S.~He and Y.-X. Zhang, \emph{{Relation between time- and spacelike entanglement entropy}}, \href{http://dx.doi.org/10.1103/gmkp-lrh3}{\emph{Phys. Rev. D} {\bfseries 112} (2025) 086020}, [\href{https://arxiv.org/abs/2402.00268}{{\ttfamily 2402.00268}}].

\bibitem{Grieninger:2023knz}
S.~Grieninger, K.~Ikeda and D.~E. Kharzeev, \emph{{Temporal entanglement entropy as a probe of renormalization group flow}}, \href{http://dx.doi.org/10.1007/JHEP05(2024)030}{\emph{JHEP} {\bfseries 05} (2024) 030}, [\href{https://arxiv.org/abs/2312.08534}{{\ttfamily 2312.08534}}].

\bibitem{Heller:2024whi}
M.~P. Heller, F.~Ori and A.~Serantes, \emph{{Geometric Interpretation of Timelike Entanglement Entropy}}, \href{http://dx.doi.org/10.1103/PhysRevLett.134.131601}{\emph{Phys. Rev. Lett.} {\bfseries 134} (2025) 131601}, [\href{https://arxiv.org/abs/2408.15752}{{\ttfamily 2408.15752}}].

\bibitem{Jiang:2023ffu}
X.~Jiang, P.~Wang, H.~Wu and H.~Yang, \emph{{Timelike entanglement entropy and TT{\textasciimacron} deformation}}, \href{http://dx.doi.org/10.1103/PhysRevD.108.046004}{\emph{Phys. Rev. D} {\bfseries 108} (2023) 046004}, [\href{https://arxiv.org/abs/2302.13872}{{\ttfamily 2302.13872}}].

\bibitem{Li:2022tsv}
Z.~Li, Z.-Q. Xiao and R.-Q. Yang, \emph{{On holographic time-like entanglement entropy}}, \href{http://dx.doi.org/10.1007/JHEP04(2023)004}{\emph{JHEP} {\bfseries 04} (2023) 004}, [\href{https://arxiv.org/abs/2211.14883}{{\ttfamily 2211.14883}}].

\bibitem{Xu:2024yvf}
J.~Xu and W.-z. Guo, \emph{{Imaginary part of timelike entanglement entropy}}, \href{http://dx.doi.org/10.1007/JHEP02(2025)094}{\emph{JHEP} {\bfseries 02} (2025) 094}, [\href{https://arxiv.org/abs/2410.22684}{{\ttfamily 2410.22684}}].

\bibitem{Anegawa:2024kdj}
T.~Anegawa and K.~Tamaoka, \emph{{Black hole singularity and timelike entanglement}}, \href{http://dx.doi.org/10.1007/JHEP10(2024)182}{\emph{JHEP} {\bfseries 10} (2024) 182}, [\href{https://arxiv.org/abs/2406.10968}{{\ttfamily 2406.10968}}].

\bibitem{Nunez:2025gxq}
C.~Nunez and D.~Roychowdhury, \emph{{Timelike entanglement entropy: A top-down approach}}, \href{http://dx.doi.org/10.1103/vjyt-xc15}{\emph{Phys. Rev. D} {\bfseries 112} (2025) 026030}, [\href{https://arxiv.org/abs/2505.20388}{{\ttfamily 2505.20388}}].

\bibitem{Katoch:2025bnh}
G.~Katoch, D.~Sarkar and B.~Sen, \emph{{Holographic timelike entanglement in AdS3 Vaidya}}, \href{http://dx.doi.org/10.1103/lmsy-vs86}{\emph{Phys. Rev. D} {\bfseries 112} (2025) 046026}, [\href{https://arxiv.org/abs/2504.14313}{{\ttfamily 2504.14313}}].

\bibitem{Chu:2025sjv}
C.-S. Chu and H.~Parihar, \emph{{Timelike entanglement entropy with gravitational anomalies}}, \href{http://dx.doi.org/10.1007/JHEP08(2025)038}{\emph{JHEP} {\bfseries 08} (2025) 038}, [\href{https://arxiv.org/abs/2504.19694}{{\ttfamily 2504.19694}}].

\bibitem{Zhao:2025zgm}
Z.-X. Zhao, L.~Zhao and S.~He, \emph{{Timelike entanglement entropy in higher curvature gravity}}, \href{http://dx.doi.org/10.1007/JHEP12(2025)156}{\emph{JHEP} {\bfseries 12} (2025) 156}, [\href{https://arxiv.org/abs/2509.04181}{{\ttfamily 2509.04181}}].

\bibitem{Harper:2025lav}
J.~Harper, T.~Kawamoto, R.~Maeda, N.~Nakamura and T.~Takayanagi, \emph{{Non-hermitian Density Matrices from Time-like Entanglement and Wormholes}},  \href{https://arxiv.org/abs/2512.13800}{{\ttfamily 2512.13800}}.

\bibitem{Li:2025tud}
G.-Y. Li, M.-H. Xiao, S.~He and J.-R. Sun, \emph{{Entanglement first law for timelike entanglement entropy and linearized Einstein's equation}},  \href{https://arxiv.org/abs/2511.17098}{{\ttfamily 2511.17098}}.

\bibitem{Giataganas:2025ize}
D.~Giataganas, \emph{{Timelike Entanglement Entropy and Renormalization Group Flow Irreversibility}},  \href{https://arxiv.org/abs/2512.16499}{{\ttfamily 2512.16499}}.

\bibitem{Afrasiar:2025eam}
M.~Afrasiar, J.~K. Basak and K.-Y. Kim, \emph{{Aspects of holographic timelike entanglement entropy in black hole backgrounds}},  \href{https://arxiv.org/abs/2512.21327}{{\ttfamily 2512.21327}}.

\bibitem{Jena:2024tly}
S.~S. Jena and S.~Mahapatra, \emph{{A note on the holographic time-like entanglement entropy in Lifshitz theory}}, \href{http://dx.doi.org/10.1007/JHEP01(2025)055}{\emph{JHEP} {\bfseries 01} (2025) 055}, [\href{https://arxiv.org/abs/2410.00384}{{\ttfamily 2410.00384}}].

\bibitem{Skenderis:2002wp}
K.~Skenderis, \emph{{Lecture notes on holographic renormalization}}, \href{http://dx.doi.org/10.1088/0264-9381/19/22/306}{\emph{Class. Quant. Grav.} {\bfseries 19} (2002) 5849--5876}, [\href{https://arxiv.org/abs/hep-th/0209067}{{\ttfamily hep-th/0209067}}].

\bibitem{Henningson:1998gx}
M.~Henningson and K.~Skenderis, \emph{{The Holographic Weyl anomaly}}, \href{http://dx.doi.org/10.1088/1126-6708/1998/07/023}{\emph{JHEP} {\bfseries 07} (1998) 023}, [\href{https://arxiv.org/abs/hep-th/9806087}{{\ttfamily hep-th/9806087}}].

\bibitem{deHaro:2000vlm}
S.~de~Haro, S.~N. Solodukhin and K.~Skenderis, \emph{{Holographic reconstruction of space-time and renormalization in the AdS / CFT correspondence}}, \href{http://dx.doi.org/10.1007/s002200100381}{\emph{Commun. Math. Phys.} {\bfseries 217} (2001) 595--622}, [\href{https://arxiv.org/abs/hep-th/0002230}{{\ttfamily hep-th/0002230}}].

\bibitem{Anninos:2013rza}
D.~Anninos, F.~Denef, G.~Konstantinidis and E.~Shaghoulian, \emph{{Higher Spin de Sitter Holography from Functional Determinants}}, \href{http://dx.doi.org/10.1007/JHEP02(2014)007}{\emph{JHEP} {\bfseries 02} (2014) 007}, [\href{https://arxiv.org/abs/1305.6321}{{\ttfamily 1305.6321}}].

\bibitem{Castro:2012gc}
A.~Castro and A.~Maloney, \emph{{The Wave Function of Quantum de Sitter}}, \href{http://dx.doi.org/10.1007/JHEP11(2012)096}{\emph{JHEP} {\bfseries 11} (2012) 096}, [\href{https://arxiv.org/abs/1209.5757}{{\ttfamily 1209.5757}}].

\bibitem{Banks:2006rx}
T.~Banks, B.~Fiol and A.~Morisse, \emph{{Towards a quantum theory of de Sitter space}}, \href{http://dx.doi.org/10.1088/1126-6708/2006/12/004}{\emph{JHEP} {\bfseries 12} (2006) 004}, [\href{https://arxiv.org/abs/hep-th/0609062}{{\ttfamily hep-th/0609062}}].

\bibitem{Ghezelbash:2001vs}
A.~M. Ghezelbash and R.~B. Mann, \emph{{Action, mass and entropy of Schwarzschild-de Sitter black holes and the de Sitter / CFT correspondence}}, \href{http://dx.doi.org/10.1088/1126-6708/2002/01/005}{\emph{JHEP} {\bfseries 01} (2002) 005}, [\href{https://arxiv.org/abs/hep-th/0111217}{{\ttfamily hep-th/0111217}}].

\bibitem{Taylor:2016aoi}
M.~Taylor and W.~Woodhead, \emph{{Renormalized entanglement entropy}}, \href{http://dx.doi.org/10.1007/JHEP08(2016)165}{\emph{JHEP} {\bfseries 08} (2016) 165}, [\href{https://arxiv.org/abs/1604.06808}{{\ttfamily 1604.06808}}].

\bibitem{Anastasiou:2017xjr}
G.~Anastasiou, I.~J. Araya and R.~Olea, \emph{{Renormalization of Entanglement Entropy from topological terms}}, \href{http://dx.doi.org/10.1103/PhysRevD.97.106011}{\emph{Phys. Rev.} {\bfseries D97} (2018) 106011}, [\href{https://arxiv.org/abs/1712.09099}{{\ttfamily 1712.09099}}].

\bibitem{Nishioka:2018khk}
T.~Nishioka, \emph{{Entanglement entropy: holography and renormalization group}}, \href{http://dx.doi.org/10.1103/RevModPhys.90.035007}{\emph{Rev. Mod. Phys.} {\bfseries 90} (2018) 035007}, [\href{https://arxiv.org/abs/1801.10352}{{\ttfamily 1801.10352}}].

\bibitem{Anastasiou:2018rla}
G.~Anastasiou, I.~J. Araya and R.~Olea, \emph{{Topological terms, AdS$_{2n}$ gravity and renormalized Entanglement Entropy of holographic CFTs}}, \href{http://dx.doi.org/10.1103/PhysRevD.97.106015}{\emph{Phys. Rev.} {\bfseries D97} (2018) 106015}, [\href{https://arxiv.org/abs/1803.04990}{{\ttfamily 1803.04990}}].

\bibitem{Anastasiou:2018mfk}
G.~Anastasiou, I.~J. Araya, C.~Arias and R.~Olea, \emph{{Einstein-AdS action, renormalized volume/area and holographic R\'enyi entropies}}, \href{http://dx.doi.org/10.1007/JHEP08(2018)136}{\emph{JHEP} {\bfseries 08} (2018) 136}, [\href{https://arxiv.org/abs/1806.10708}{{\ttfamily 1806.10708}}].

\bibitem{Anastasiou:2019ldc}
G.~Anastasiou, I.~J. Araya, A.~Guijosa and R.~Olea, \emph{{Renormalized AdS gravity and holographic entanglement entropy of even-dimensional CFTs}}, \href{http://dx.doi.org/10.1007/JHEP10(2019)221}{\emph{JHEP} {\bfseries 10} (2019) 221}, [\href{https://arxiv.org/abs/1908.11447}{{\ttfamily 1908.11447}}].

\bibitem{Taylor:2020uwf}
M.~Taylor and L.~Too, \emph{{Renormalized entanglement entropy and curvature invariants}}, \href{http://dx.doi.org/10.1007/JHEP12(2020)050}{\emph{JHEP} {\bfseries 12} (2020) 050}, [\href{https://arxiv.org/abs/2004.09568}{{\ttfamily 2004.09568}}].

\bibitem{Anastasiou:2021swo}
G.~Anastasiou, I.~J. Araya, J.~Moreno, R.~Olea and D.~Rivera-Betancour, \emph{{Renormalized holographic entanglement entropy for quadratic curvature gravity}}, \href{http://dx.doi.org/10.1103/PhysRevD.104.086003}{\emph{Phys. Rev. D} {\bfseries 104} (2021) 086003}, [\href{https://arxiv.org/abs/2102.11242}{{\ttfamily 2102.11242}}].

\bibitem{Anastasiou:2022ljq}
G.~Anastasiou, I.~J. Araya and R.~Olea, \emph{{Energy functionals from Conformal Gravity}}, \href{http://dx.doi.org/10.1007/JHEP10(2022)123}{\emph{JHEP} {\bfseries 10} (2022) 123}, [\href{https://arxiv.org/abs/2209.02006}{{\ttfamily 2209.02006}}].

\bibitem{Maldacena:2011mk}
J.~Maldacena, \emph{{Einstein Gravity from Conformal Gravity}}, {\emph{unpublished} (2011) }, [\href{https://arxiv.org/abs/1105.5632}{{\ttfamily 1105.5632}}].

\bibitem{Grumiller:2013mxa}
D.~Grumiller, M.~Irakleidou, I.~Lovrekovic and R.~McNees, \emph{{Conformal gravity holography in four dimensions}}, \href{http://dx.doi.org/10.1103/PhysRevLett.112.111102}{\emph{Phys. Rev. Lett.} {\bfseries 112} (2014) 111102}, [\href{https://arxiv.org/abs/1310.0819}{{\ttfamily 1310.0819}}].

\bibitem{Anastasiou:2016jix}
G.~Anastasiou and R.~Olea, \emph{{From conformal to Einstein Gravity}}, \href{http://dx.doi.org/10.1103/PhysRevD.94.086008}{\emph{Phys. Rev.} {\bfseries D94} (2016) 086008}, [\href{https://arxiv.org/abs/1608.07826}{{\ttfamily 1608.07826}}].

\bibitem{Hell:2023rbf}
A.~Hell, D.~Lust and G.~Zoupanos, \emph{{On the ghost problem of conformal gravity}}, \href{http://dx.doi.org/10.1007/JHEP08(2023)168}{\emph{JHEP} {\bfseries 08} (2023) 168}, [\href{https://arxiv.org/abs/2306.13714}{{\ttfamily 2306.13714}}].

\bibitem{Anastasiou:2023oro}
G.~Anastasiou, I.~J. Araya, C.~Corral and R.~Olea, \emph{{Conformal Renormalization of topological black holes in AdS$_{6}$}}, \href{http://dx.doi.org/10.1007/JHEP11(2023)036}{\emph{JHEP} {\bfseries 11} (2023) 036}, [\href{https://arxiv.org/abs/2308.09140}{{\ttfamily 2308.09140}}].

\bibitem{Lewkowycz:2013nqa}
A.~Lewkowycz and J.~Maldacena, \emph{{Generalized gravitational entropy}}, \href{http://dx.doi.org/10.1007/JHEP08(2013)090}{\emph{JHEP} {\bfseries 08} (2013) 090}, [\href{https://arxiv.org/abs/1304.4926}{{\ttfamily 1304.4926}}].

\bibitem{Barrella:2013wja}
T.~Barrella, X.~Dong, S.~A. Hartnoll and V.~L. Martin, \emph{{Holographic entanglement beyond classical gravity}}, \href{http://dx.doi.org/10.1007/JHEP09(2013)109}{\emph{JHEP} {\bfseries 09} (2013) 109}, [\href{https://arxiv.org/abs/1306.4682}{{\ttfamily 1306.4682}}].

\bibitem{Solodukhin:2008dh}
S.~N. Solodukhin, \emph{{Entanglement entropy, conformal invariance and extrinsic geometry}}, \href{http://dx.doi.org/10.1016/j.physletb.2008.05.071}{\emph{Phys. Lett.} {\bfseries B665} (2008) 305--309}, [\href{https://arxiv.org/abs/0802.3117}{{\ttfamily 0802.3117}}].

\bibitem{Fursaev:2006ih}
D.~V. Fursaev, \emph{{Proof of the holographic formula for entanglement entropy}}, \href{http://dx.doi.org/10.1088/1126-6708/2006/09/018}{\emph{JHEP} {\bfseries 09} (2006) 018}, [\href{https://arxiv.org/abs/hep-th/0606184}{{\ttfamily hep-th/0606184}}].

\bibitem{Miao:2014nxa}
R.-X. Miao and W.-z. Guo, \emph{{Holographic Entanglement Entropy for the Most General Higher Derivative Gravity}}, \href{http://dx.doi.org/10.1007/JHEP08(2015)031}{\emph{JHEP} {\bfseries 08} (2015) 031}, [\href{https://arxiv.org/abs/1411.5579}{{\ttfamily 1411.5579}}].

\bibitem{Camps:2014voa}
J.~Camps and W.~R. Kelly, \emph{{Generalized gravitational entropy without replica symmetry}}, \href{http://dx.doi.org/10.1007/JHEP03(2015)061}{\emph{JHEP} {\bfseries 03} (2015) 061}, [\href{https://arxiv.org/abs/1412.4093}{{\ttfamily 1412.4093}}].

\bibitem{Anastasiou:2024rxe}
G.~Anastasiou, I.~J. Araya, P.~Bueno, J.~Moreno, R.~Olea and A.~Vilar~Lopez, \emph{{Higher-dimensional Willmore energy as holographic entanglement entropy}}, \href{http://dx.doi.org/10.1007/JHEP01(2025)081}{\emph{JHEP} {\bfseries 01} (2025) 081}, [\href{https://arxiv.org/abs/2409.19485}{{\ttfamily 2409.19485}}].

\bibitem{Lu:2011ks}
H.~Lu, Y.~Pang and C.~N. Pope, \emph{{Conformal Gravity and Extensions of Critical Gravity}}, \href{http://dx.doi.org/10.1103/PhysRevD.84.064001}{\emph{Phys. Rev. D} {\bfseries 84} (2011) 064001}, [\href{https://arxiv.org/abs/1106.4657}{{\ttfamily 1106.4657}}].

\bibitem{Chandrasekaran:2022eqq}
V.~Chandrasekaran, G.~Penington and E.~Witten, \emph{{Large N algebras and generalized entropy}}, \href{http://dx.doi.org/10.1007/JHEP04(2023)009}{\emph{JHEP} {\bfseries 04} (2023) 009}, [\href{https://arxiv.org/abs/2209.10454}{{\ttfamily 2209.10454}}].

\bibitem{Graham:1999pm}
C.~R. Graham and E.~Witten, \emph{{Conformal anomaly of submanifold observables in AdS / CFT correspondence}}, \href{http://dx.doi.org/10.1016/S0550-3213(99)00055-3}{\emph{Nucl. Phys. B} {\bfseries 546} (1999) 52--64}, [\href{https://arxiv.org/abs/hep-th/9901021}{{\ttfamily hep-th/9901021}}].

\bibitem{Graham:2017bew}
C.~R. Graham and N.~Reichert, \emph{{Higher-dimensional Willmore energies via minimal submanifold asymptotics}}, \href{http://dx.doi.org/10.4310/AJM.2020.v24.n4.a3}{\emph{Asian J. Math.} {\bfseries 24} (2020) 571--610}, [\href{https://arxiv.org/abs/1704.03852}{{\ttfamily 1704.03852}}].

\bibitem{Mezei:2014zla}
M.~Mezei, \emph{{Entanglement entropy across a deformed sphere}}, \href{http://dx.doi.org/10.1103/PhysRevD.91.045038}{\emph{Phys. Rev.} {\bfseries D91} (2015) 045038}, [\href{https://arxiv.org/abs/1411.7011}{{\ttfamily 1411.7011}}].

\bibitem{Allais:2014ata}
A.~Allais and M.~Mezei, \emph{{Some results on the shape dependence of entanglement and R\'enyi entropies}}, \href{http://dx.doi.org/10.1103/PhysRevD.91.046002}{\emph{Phys. Rev. D} {\bfseries 91} (2015) 046002}, [\href{https://arxiv.org/abs/1407.7249}{{\ttfamily 1407.7249}}].

\bibitem{Osborn:1993cr}
H.~Osborn and A.~C. Petkou, \emph{{Implications of conformal invariance in field theories for general dimensions}}, \href{http://dx.doi.org/10.1006/aphy.1994.1045}{\emph{Annals Phys.} {\bfseries 231} (1994) 311--362}, [\href{https://arxiv.org/abs/hep-th/9307010}{{\ttfamily hep-th/9307010}}].

\bibitem{Stelle:1976gc}
K.~S. Stelle, \emph{{Renormalization of Higher Derivative Quantum Gravity}}, \href{http://dx.doi.org/10.1103/PhysRevD.16.953}{\emph{Phys. Rev.} {\bfseries D16} (1977) 953--969}.

\bibitem{Capper:1975ig}
D.~Capper and M.~Duff, \emph{{Conformal Anomalies and the Renormalizability Problem in Quantum Gravity}}, \href{http://dx.doi.org/10.1016/0375-9601(75)90030-4}{\emph{Phys. Lett. A} {\bfseries 53} (1975) 361}.

\bibitem{Fradkin:1983tg}
E.~S. Fradkin and A.~A. Tseytlin, \emph{{Conformal Anomaly in Weyl Theory and Anomaly Free Superconformal Theories}}, \href{http://dx.doi.org/10.1016/0370-2693(84)90668-3}{\emph{Phys. Lett. B} {\bfseries 134} (1984) 187}.

\bibitem{Julve:1978xn}
J.~Julve and M.~Tonin, \emph{{Quantum Gravity with Higher Derivative Terms}}, \href{http://dx.doi.org/10.1007/BF02748637}{\emph{Nuovo Cim. B} {\bfseries 46} (1978) 137--152}.

\bibitem{Mannheim:1988dj}
P.~D. Mannheim and D.~Kazanas, \emph{{Exact Vacuum Solution to Conformal Weyl Gravity and Galactic Rotation Curves}}, \href{http://dx.doi.org/10.1086/167623}{\emph{Astrophys. J.} {\bfseries 342} (1989) 635--638}.

\bibitem{Mannheim:2005bfa}
P.~D. Mannheim, \emph{{Alternatives to dark matter and dark energy}}, \href{http://dx.doi.org/10.1016/j.ppnp.2005.08.001}{\emph{Prog. Part. Nucl. Phys.} {\bfseries 56} (2006) 340--445}, [\href{https://arxiv.org/abs/astro-ph/0505266}{{\ttfamily astro-ph/0505266}}].

\bibitem{Mannheim:2010ti}
P.~D. Mannheim and J.~G. O'Brien, \emph{{Impact of a global quadratic potential on galactic rotation curves}}, \href{http://dx.doi.org/10.1103/PhysRevLett.106.121101}{\emph{Phys. Rev. Lett.} {\bfseries 106} (2011) 121101}, [\href{https://arxiv.org/abs/1007.0970}{{\ttfamily 1007.0970}}].

\bibitem{Mannheim:2011ds}
P.~D. Mannheim, \emph{{Making the Case for Conformal Gravity}}, \href{http://dx.doi.org/10.1007/s10701-011-9608-6}{\emph{Found. Phys.} {\bfseries 42} (2012) 388--420}, [\href{https://arxiv.org/abs/1101.2186}{{\ttfamily 1101.2186}}].

\bibitem{Riegert:1984hf}
R.~J. Riegert, \emph{{THE PARTICLE CONTENT OF LINEARIZED CONFORMAL GRAVITY}}, \href{http://dx.doi.org/10.1016/0375-9601(84)90648-0}{\emph{Phys. Lett. A} {\bfseries 105} (1984) 110--112}.

\bibitem{Mannheim:2021oat}
P.~D. Mannheim, \emph{{Solution to the ghost problem in higher-derivative gravity}}, \href{http://dx.doi.org/10.1393/ncc/i2022-22027-6}{\emph{Nuovo Cim. C} {\bfseries 45} (2022) 27}, [\href{https://arxiv.org/abs/2109.12743}{{\ttfamily 2109.12743}}].

\bibitem{Anastasiou:2020mik}
G.~Anastasiou, I.~J. Araya and R.~Olea, \emph{{Einstein Gravity from Conformal Gravity in 6D}}, \href{http://dx.doi.org/10.1007/JHEP01(2021)134}{\emph{JHEP} {\bfseries 01} (2021) 134}, [\href{https://arxiv.org/abs/2010.15146}{{\ttfamily 2010.15146}}].

\bibitem{Miskovic:2009bm}
O.~Miskovic and R.~Olea, \emph{{Topological regularization and self-duality in four-dimensional anti-de Sitter gravity}}, \href{http://dx.doi.org/10.1103/PhysRevD.79.124020}{\emph{Phys. Rev.} {\bfseries D79} (2009) 124020}, [\href{https://arxiv.org/abs/0902.2082}{{\ttfamily 0902.2082}}].

\bibitem{Penrose:1962ij}
R.~Penrose, \emph{{Asymptotic properties of fields and space-times}}, \href{http://dx.doi.org/10.1103/PhysRevLett.10.66}{\emph{Phys. Rev. Lett.} {\bfseries 10} (1963) 66--68}.

\bibitem{Anastasiou:2025dex}
G.~Anastasiou, M.~Bravo and R.~Olea, \emph{{Asymptotic analysis of energy functionals in anti-de Sitter spacetimes}}, \href{http://dx.doi.org/10.1007/JHEP09(2025)093}{\emph{JHEP} {\bfseries 09} (2025) 093}, [\href{https://arxiv.org/abs/2504.06382}{{\ttfamily 2504.06382}}].

\bibitem{Fursaev:2013fta}
D.~V. Fursaev, A.~Patrushev and S.~N. Solodukhin, \emph{{Distributional Geometry of Squashed Cones}}, \href{http://dx.doi.org/10.1103/PhysRevD.88.044054}{\emph{Phys. Rev.} {\bfseries D88} (2013) 044054}, [\href{https://arxiv.org/abs/1306.4000}{{\ttfamily 1306.4000}}].

\bibitem{Dowker:2010yj}
J.~S. Dowker, \emph{{Entanglement entropy for odd spheres}},  \href{https://arxiv.org/abs/1012.1548}{{\ttfamily 1012.1548}}.

\bibitem{Casini:2011kv}
H.~Casini, M.~Huerta and R.~C. Myers, \emph{{Towards a derivation of holographic entanglement entropy}}, \href{http://dx.doi.org/10.1007/JHEP05(2011)036}{\emph{JHEP} {\bfseries 05} (2011) 036}, [\href{https://arxiv.org/abs/1102.0440}{{\ttfamily 1102.0440}}].

\bibitem{Duff:1977ay}
M.~J. Duff, \emph{{Observations on Conformal Anomalies}}, \href{http://dx.doi.org/10.1016/0550-3213(77)90410-2}{\emph{Nucl. Phys. B} {\bfseries 125} (1977) 334--348}.

\bibitem{Bonora:1985cq}
L.~Bonora, P.~Pasti and M.~Bregola, \emph{{Weyl cocycles}}, \href{http://dx.doi.org/10.1088/0264-9381/3/4/018}{\emph{Class. Quant. Grav.} {\bfseries 3} (1986) 635}.

\bibitem{Deser:1993yx}
S.~Deser and A.~Schwimmer, \emph{{Geometric classification of conformal anomalies in arbitrary dimensions}}, \href{http://dx.doi.org/10.1016/0370-2693(93)90934-A}{\emph{Phys. Lett. B} {\bfseries 309} (1993) 279--284}, [\href{https://arxiv.org/abs/hep-th/9302047}{{\ttfamily hep-th/9302047}}].

\bibitem{Faulkner:2015csl}
T.~Faulkner, R.~G. Leigh and O.~Parrikar, \emph{{Shape Dependence of Entanglement Entropy in Conformal Field Theories}}, \href{http://dx.doi.org/10.1007/JHEP04(2016)088}{\emph{JHEP} {\bfseries 04} (2016) 088}, [\href{https://arxiv.org/abs/1511.05179}{{\ttfamily 1511.05179}}].

\bibitem{Fonda:2015nma}
P.~Fonda, D.~Seminara and E.~Tonni, \emph{{On shape dependence of holographic entanglement entropy in AdS$_{4}$/CFT$_{3}$}}, \href{http://dx.doi.org/10.1007/JHEP12(2015)037}{\emph{JHEP} {\bfseries 12} (2015) 037}, [\href{https://arxiv.org/abs/1510.03664}{{\ttfamily 1510.03664}}].

\bibitem{Anastasiou:2020smm}
G.~Anastasiou, J.~Moreno, R.~Olea and D.~Rivera-Betancour, \emph{{Shape dependence of renormalized holographic entanglement entropy}}, \href{http://dx.doi.org/10.1007/JHEP09(2020)173}{\emph{JHEP} {\bfseries 09} (2020) 173}, [\href{https://arxiv.org/abs/2002.06111}{{\ttfamily 2002.06111}}].

\bibitem{Anastasiou:2022pzm}
G.~Anastasiou, I.~J. Araya, A.~Argando\~na and R.~Olea, \emph{{CFT correlators from shape deformations in Cubic Curvature Gravity}}, \href{http://dx.doi.org/10.1007/JHEP11(2022)031}{\emph{JHEP} {\bfseries 11} (2022) 031}, [\href{https://arxiv.org/abs/2208.00093}{{\ttfamily 2208.00093}}].

\bibitem{Bueno:2021fxb}
P.~Bueno, H.~Casini, O.~L. Andino and J.~Moreno, \emph{{Disks globally maximize the entanglement entropy in 2 + 1 dimensions}}, \href{http://dx.doi.org/10.1007/JHEP10(2021)179}{\emph{JHEP} {\bfseries 10} (2021) 179}, [\href{https://arxiv.org/abs/2107.12394}{{\ttfamily 2107.12394}}].

\bibitem{Bueno:2023gey}
P.~Bueno, H.~Casini, O.~L. Andino and J.~Moreno, \emph{{Conformal Bounds in Three Dimensions from Entanglement Entropy}}, \href{http://dx.doi.org/10.1103/PhysRevLett.131.171601}{\emph{Phys. Rev. Lett.} {\bfseries 131} (2023) 171601}, [\href{https://arxiv.org/abs/2307.05164}{{\ttfamily 2307.05164}}].

\bibitem{Erdmenger:1997gy}
J.~Erdmenger, \emph{{Conformally covariant differential operators: Properties and applications}}, \href{http://dx.doi.org/10.1088/0264-9381/14/8/008}{\emph{Class. Quant. Grav.} {\bfseries 14} (1997) 2061--2084}, [\href{https://arxiv.org/abs/hep-th/9704108}{{\ttfamily hep-th/9704108}}].

\bibitem{Miao:2020oey}
R.-X. Miao, \emph{{An Exact Construction of Codimension two Holography}}, \href{http://dx.doi.org/10.1007/JHEP01(2021)150}{\emph{JHEP} {\bfseries 01} (2021) 150}, [\href{https://arxiv.org/abs/2009.06263}{{\ttfamily 2009.06263}}].

\bibitem{Fursaev:1995ef}
D.~V. Fursaev and S.~N. Solodukhin, \emph{{On the description of the Riemannian geometry in the presence of conical defects}}, \href{http://dx.doi.org/10.1103/PhysRevD.52.2133}{\emph{Phys. Rev.} {\bfseries D52} (1995) 2133--2143}, [\href{https://arxiv.org/abs/hep-th/9501127}{{\ttfamily hep-th/9501127}}].

\bibitem{guven2005conformally}
J.~Guven, \emph{Conformally invariant bending energy for hypersurfaces}, {\emph{Journal of Physics A: Mathematical and General} {\bfseries 38} (2005) 7943}.

\bibitem{Gover:2016buc}
A.~R. Gover and A.~Waldron, \emph{{A Calculus for Conformal Hypersurfaces and new higher Willmore energy functionals}},  \href{https://arxiv.org/abs/1611.04055}{{\ttfamily 1611.04055}}.

\bibitem{zhang2017graham}
Y.~Zhang, \emph{Graham-{W}itten's conformal invariant for closed four dimensional submanifolds},  \href{https://arxiv.org/abs/1703.08611}{{\ttfamily 1703.08611}}.

\bibitem{Blitz:2021qbp}
S.~Blitz, A.~R. Gover and A.~Waldron, \emph{{Generalized Willmore energies, Q-curvatures, extrinsic Paneitz operators, and extrinsic Laplacian powers}}, \href{http://dx.doi.org/10.1142/S0219199723500141}{\emph{Commun. Contemp. Math.} {\bfseries 26} (2024) 2350014}, [\href{https://arxiv.org/abs/2111.00179}{{\ttfamily 2111.00179}}].

\bibitem{Olanipekun:2021htq}
P.~O. Olanipekun, \emph{{Study of a Four Dimensional Willmore Energy}}.
\newblock PhD thesis, Monash U., 2021.
\newblock \href{https://arxiv.org/abs/2210.05924}{{\ttfamily 2210.05924}}.

\bibitem{2023arXiv230811433M}
D.~Martino, \emph{{A duality theorem for a four dimensional Willmore energy}},  \href{https://arxiv.org/abs/2308.11433}{{\ttfamily 2308.11433}}.

\bibitem{bernard2024analysis}
{Bernard, Yann}, \emph{{Analysis of Conformally Invariant Energies on four-Dimensional Hypersurfaces}},  \href{https://arxiv.org/abs/2402.15032}{{\ttfamily 2402.15032}}.

\bibitem{Boulanger:2025oli}
N.~Boulanger and D.~Rovere, \emph{{8D conformal gravity with Einstein sector, and its relation to the Q-curvature}},  \href{https://arxiv.org/abs/2511.01368}{{\ttfamily 2511.01368}}.

\bibitem{Wu:2025zbf}
N.~Wu and Z.~Yan, \emph{{Energy reduction for Fourth order Willmore energy}},  \href{https://arxiv.org/abs/2504.08105}{{\ttfamily 2504.08105}}.

\bibitem{Lan:2025evd}
T.~Lan, D.~Martino and T.~Rivi{\`e}re, \emph{{The Analysis of Willmore Surfaces and its Generalizations in Higher Dimensions}},  \href{https://arxiv.org/abs/2511.01777}{{\ttfamily 2511.01777}}.

\bibitem{Lu:2013hx}
H.~Lü, Y.~Pang and C.~N. Pope, \emph{{Black Holes in Six-dimensional Conformal Gravity}}, \href{http://dx.doi.org/10.1103/PhysRevD.87.104013}{\emph{Phys. Rev.} {\bfseries D87} (2013) 104013}, [\href{https://arxiv.org/abs/1301.7083}{{\ttfamily 1301.7083}}].

\bibitem{Gurarie:2004ce}
V.~Gurarie and A.~W.~W. Ludwig, \emph{{Conformal field theory at central charge c=0 and two-dimensional critical systems with quenched disorder}},  in \emph{{From Fields to Strings: Circumnavigating Theoretical Physics: A Conference in Tribute to Ian Kogan}}, pp.~1384--1440, 9, 2004.
\newblock \href{https://arxiv.org/abs/hep-th/0409105}{{\ttfamily hep-th/0409105}}.
\newblock \href{http://dx.doi.org/10.1142/9789812775344_0032}{DOI}.

\bibitem{dubail2010conformal}
J.~Dubail, J.~L. Jacobsen and H.~Saleur, \emph{Conformal field theory at central charge c= 0: A measure of the indecomposability (b) parameters}, {\emph{Nuclear Physics B} {\bfseries 834} (2010) 399--422}.

\bibitem{Miao:2015iba}
R.-X. Miao, \emph{{Universal Terms of Entanglement Entropy for 6d CFTs}}, \href{http://dx.doi.org/10.1007/JHEP10(2015)049}{\emph{JHEP} {\bfseries 10} (2015) 049}, [\href{https://arxiv.org/abs/1503.05538}{{\ttfamily 1503.05538}}].

\bibitem{Camps:2013zua}
J.~Camps, \emph{{Generalized entropy and higher derivative Gravity}}, \href{http://dx.doi.org/10.1007/JHEP03(2014)070}{\emph{JHEP} {\bfseries 03} (2014) 070}, [\href{https://arxiv.org/abs/1310.6659}{{\ttfamily 1310.6659}}].

\bibitem{Hung:2011xb}
L.-Y. Hung, R.~C. Myers and M.~Smolkin, \emph{{On Holographic Entanglement Entropy and Higher Curvature Gravity}}, \href{http://dx.doi.org/10.1007/JHEP04(2011)025}{\emph{JHEP} {\bfseries 04} (2011) 025}, [\href{https://arxiv.org/abs/1101.5813}{{\ttfamily 1101.5813}}].

\bibitem{deBoer:2011wk}
J.~de~Boer, M.~Kulaxizi and A.~Parnachev, \emph{{Holographic Entanglement Entropy in Lovelock Gravities}}, \href{http://dx.doi.org/10.1007/JHEP07(2011)109}{\emph{JHEP} {\bfseries 07} (2011) 109}, [\href{https://arxiv.org/abs/1101.5781}{{\ttfamily 1101.5781}}].

\end{thebibliography}\endgroup

\end{document}